%% file: main.tex
\documentclass[twocolumn,twocolappendix]{aastex631}

\newcommand{\customsig}{$\sigma_{r=10\,{\rm cMpc}}$}

\graphicspath{{./}{figures/}} 

\newcommand\gamornet{G\textsc{a}M\textsc{or}N\textsc{et}}
\newcommand\gampen{GaMPEN}

\newcommand\grizy{\textit{grizy}}
\newcommand\uband{\textit{u}}
\newcommand\gb{\textit{g}}
\newcommand\rb{\textit{r}}
\newcommand\ib{\textit{i}}
\newcommand\zb{\textit{z}}
\newcommand\yb{\textit{y}}
\newcommand\om{\sim \mathcal{O}}

\shorttitle{Denser Environments Cultivate Larger Galaxies}
\shortauthors{Ghosh et al.}

\usepackage{soul}
\usepackage{enumitem}
\usepackage{bm}
\usepackage{amsmath}
\usepackage{gensymb}
\usepackage{subfigure}
\usepackage{multirow}
\usepackage{array}
\newcolumntype{P}[1]{>{\centering\arraybackslash}p{#1}}
\newcolumntype{M}[1]{>{\centering\arraybackslash}m{#1}}
\usepackage{hhline}
\usepackage{xcolor}
\usepackage{verbatim}
\PassOptionsToPackage{hyphens}{url}
\usepackage{hyperref}

\usepackage{xcolor} 
\definecolor{LightGray}{gray}{0.95}

\input{bangla_commands}
\usepackage{CJK}

\begin{document}
\begin{CJK*}{UTF8}{gbsn}

\title{Denser Environments Cultivate Larger Galaxies: A Comprehensive Study beyond the Local Universe with 3 Million Hyper Suprime-Cam Galaxies}

\author[0000-0002-2525-9647]{Aritra Ghosh ({\bngxi Airt/r \*gh*eaSh})}
\altaffiliation{LSST-DA Catalyst Fellow}
\affil{DiRAC Institute and the Department of Astronomy, University of Washington, Seattle, WA, U.S.A}
\affil{Department of Astronomy, Yale University, New Haven, CT, U.S.A.}
\affil{eScience Institute, University of Washington, Seattle, WA, U.S.A.}
\email{aritrag@uw.edu; aritraghsh09@gmail.com}

\author[0000-0002-0745-9792]{C. Megan Urry}
\affil{Department of Physics, Yale University, New Haven, CT, U.S.A.}
\affil{Yale Center for Astronomy and Astrophysics, New Haven, CT, U.S.A.}

\author[0000-0003-2284-8603]{Meredith C. Powell}
\affil{Leibniz-Institut fur Astrophysik Potsdam (AIP), An der Sternwarte 16, D-14482 Potsdam, Germany}

\author[0000-0003-4442-2750]{Rhythm Shimakawa}
\affil{Waseda Institute for Advanced Study (WIAS), Waseda University, Shinjuku, Tokyo, Japan}
\affil{Center for Data Science, Waseda University, Shinjuku, Tokyo, Japan}

\author[0000-0003-3236-2068]{Frank C. van den Bosch}
\affil{Department of Astronomy, Yale University, New Haven, CT, U.S.A.}
\affil{Yale Center for Astronomy and Astrophysics, New Haven, CT, U.S.A.}

\author[0000-0002-6766-5942]{Daisuke Nagai}
\affil{Department of Physics, Yale University, New Haven, CT, U.S.A.}
\affil{Yale Center for Astronomy and Astrophysics, New Haven, CT, U.S.A.}

\author[0000-0001-8073-4554]{Kaustav Mitra}
\affil{Department of Astronomy, Yale University, New Haven, CT, U.S.A.}

\author[0000-0001-5576-8189]{Andrew J. Connolly}
\affil{DiRAC Institute and the Department of Astronomy, University of Washington, Seattle, WA, U.S.A}
\affil{eScience Institute, University of Washington, Seattle, WA,  U.S.A.}

\begin{abstract}
The relationship between galaxy size and environment has remained enigmatic, with over a decade of conflicting results. We present one of the first comprehensive studies of the variation of galaxy radius with environment beyond the local Universe and demonstrate that large-scale environmental density is correlated with galaxy radius independent of stellar mass and galaxy morphology. We confirm with $>5\sigma$ confidence that galaxies in denser environments are up to $\sim25\%$ larger than their equally massive counterparts with similar morphology in less dense regions of the Universe. We achieve this result by correlating projected two-dimensional densities over $\sim360$ deg$^2$ with the structural parameters of $\sim3$ million Hyper Suprime-Cam galaxies at $0.3 \leq z < 0.7$ with $\log M/M_{\odot} \geq 8.9$. Compared to most previous studies, this sample is $\sim100-10,000$ times larger and goes $\sim1$ dex deeper in mass-completeness. We demonstrate that past conflicting results have been driven by small sample sizes and a lack of robust measurement uncertainties. We verify the presence of the above correlation separately for disk-dominated, bulge-dominated, star-forming, and quiescent subpopulations. We find the strength of the correlation to be dependent on redshift, stellar mass, and morphology. The correlation is strongest at lower redshifts and systematically weakens or disappears beyond $z \geq 0.5$. At $z\geq0.5$, more massive galaxies still display a statistically significant correlation. Although some existing theoretical frameworks can be selectively invoked to explain some of the observed correlations, our work demonstrates the need for more comprehensive theoretical investigations of the correlation between galaxy size and environment.
\end{abstract}


\section{Introduction} \label{sec:intro}
The morphological features and structural parameters of galaxies serve as a cornerstone in comprehending their attributes and have played an instrumental role in our understanding of galaxy formation and evolution. Since the mid-20$^{th}$ century, astronomers have studied the intricate correlations between morphology and other fundamental properties of galaxies, leveraging these relationships as powerful tools to further our understanding of galaxy evolution.

Some early pioneering studies, such as \citet{holmberg_58}, unveiled a correlation between the morphology of galaxies and their stellar populations. They found that massive elliptical galaxies typically house older stars and exhibit limited star formation. In contrast, spiral galaxies mostly harbor younger stars and are actively engaged in star formation. This observed correlation has been used to suggest that early-type galaxies may halt star formation due to the scarcity of cold gas, possibly caused by stellar or active galactic nuclei feedback or environmental effects. Conversely, late-type galaxies' ongoing star formation may stem from their ability to keep or acquire gas, potentially due to their lower masses or more isolated settings. We refer the interested reader to \citet{morph_review} for a detailed review.

Another early insight was the relationship between a galaxy's morphology and its environment. \citet{dressler_84} found that elliptical and lenticular galaxies are more common in high-density regions of the Universe, such as the centers of galaxy clusters. In contrast, spiral galaxies are more common in low-density regions, like the outskirts of clusters or the field. This observation was one of the first suggestions of a strong link between a galaxy's environment and its evolution --- a concept that remains an active area of research to this day.  

Subsequent studies have substantiated this relationship, and there is now considerable evidence showing that whether a galaxy is elliptical or spiral depends to a large degree on the density of that particular galaxy's environment \citep[e.g.,][]{gomez_03, vdw08, blanton_09, Tasca09, Cappelari11, Fogarty14, hsc_morph_den}. Although the evidence gets sparser at higher redshifts, the effect has also been shown to exist out to $z\sim2-3$. We note that some studies suggest that this trend nearly vanishes at a fixed stellar mass, implying that mass could be the predominant factor controlling this observational trend \citep[e.g.,][]{Holden07, Bamford09, Brough17, Greene17}.

\begin{deluxetable*}{cP{2.25cm}cccP{2cm}P{3cm}}[htbp]
\tablecaption{Previous Observational Studies of Environmental Dependence of $R_e$ at $z \geq 0.2$ \label{tab:lit_survey}}
\tablecolumns{7}
\tablehead{
\colhead{Reference} & \colhead{Survey/} & \colhead{Redshift} & \colhead{Stellar Mass} & \colhead{Sample} & \colhead{Density} & \colhead{Radius - Env.}\\
\colhead{} & \colhead{Instrument\tablenotemark{a}} & \colhead{} & \colhead{$ (\log M / M_{\odot}) $} & \colhead{Size} & \colhead{Measure\tablenotemark{b}} & \colhead{Correlation\tablenotemark{c}}}
\startdata
    \hline
    \hline
    \citet{afanasiev_23} & CARLA (passive only) & 1.4 - 2.8 & $> 10.5$ & $\sim 300$ & Cluster/ Field & $+$ve for passive galaxies\\
    \hline
    \citet{Siudek22} & VIPERS & $0.5 - 0.9$ & $9 - 11.5$ & $\sim 30,000$ & Density contrast & $+$ve \textit{only} for blue galaxies \\
    \hline
    \citet{Gu21} & CANDELS & 0.5 - 2.5 & $\geq 9.2$ & $\sim13,500$ & Density contrast & No correlation \\
    \hline
    \citet{Afonso19} & VIMOS & 0.8 - 0.9 & $\geq 10$ & $\sim 500$ & Density contrast & $+$ve \textit{only} for galaxies with $\log M/M_{\odot} > 11$ \\
    \hline
    \citet{Matharu19} & HST & $\sim1$ & $\geq10$ & $\sim600$ & Cluster/ Field & $-$ve correlation \\
    \hline
    \citet{Chan18} & KMOS/HST (passive only) & 1.39 - 1.61 & $\geq10.2$ & $\sim3000$& Cluster/ Field & $+$ve in two clusters; $-$ve in one\\
    \hline
    \citet{Kelkar15} & ESO Distant Cluster Survey & 0.4 - 0.8 & $\geq 10.2$ & $\sim400$ & Cluster/ Field & No correlation\\
    \hline
    \citet{Allen15} & ZFOURGE/ CANDELS & $ \sim2.1$ & $\geq 9$ & $\sim500$ & Cluster/ Field & $+$ve \textit{only} for star-forming galaxies\\ 
    \hline
    \citet{Bassett13} & CANDELS & $\sim1.6$ & $ \geq 10.3$ & $\sim 500$ & Cluster/ Field & $+$ve \textit{only} for quiescent galaxies\\
    \hline
    \citet{Huertas-Company13}  &  COSMOS (passive only) & 0.2 - 1.0 & $\geq 10.5$ & $\sim700$ & Cluster/ Field & No correlation for passive galaxies\\
    \hline
    \citet{Lani13} & UKIDSS & 1.0 - 2.0 & $\geq 10.4$ & $\sim96,000$ & Density contrast & $+$ve \textit{only} for passive galaxies\\
    \hline
    \citet{Cooper12} & DEEP2/3 (passive only) & 0.4 - 1.2 & 10 - 11 & $\sim600$ & Density contrast & $+$ve for passive galaxies\\
\enddata
\tablenotetext{a}{Note that some studies only included passive galaxies in their sample and have been denoted accordingly.}
\tablenotetext{b}{``Density contrast" refers to a numerical measurement of overdensity using either the number of neighbors within a certain volume or the distance to the $n^{th}$ nearest neighbor. ``Cluster/Field" refers to categorically separating galaxies into cluster and field galaxies without any quantitative measurement of overdensity.}
\tablenotetext{c}{$+$ve correlation refers to larger galaxies being present in denser environments. Note that for some of the rows, a $+$ve correlation is observed for \textit{only} certain subpopulations; this means that the authors investigated other subpopulations but did not find evidence for correlation.}
\end{deluxetable*}

The relatively well-established morphology-environment correlation is in stark contrast to the relationship between galaxy size (half-light radius; $R_e$) and environment. The latter relationship has remained enigmatic despite being studied for more than a decade, with no broad consensus in the field. Different studies have reported wildly conflicting results beyond the local Universe ($z \geq 0.2$), as summarized in Table \ref{tab:lit_survey}. While some studies have reported a positive correlation\footnote{i.e., larger galaxies in denser environments} of radius with the environment for certain subpopulations of galaxies \citep[e.g.,][]{Cooper12,Lani13,Bassett13,Afonso19,Siudek22}; others have reported no correlation \citep[e.g.,][]{Huertas-Company13, Kelkar15,Gu21}; and yet others have reported a negative correlation \citep[e.g.,][]{Matharu19,Chan18}. Although not a focus of this study, the disagreement also extends to the local Universe \citep[e.g.,][]{Blanton05, Park07, Cappellari13, Huertas-Company13_local, Yoon17, hearin_19}. These observational results are even more surprising given that hierarchical models of galaxy formation in the Lambda cold dark matter ($\Lambda$CDM) paradigm have consistently predicted galaxy sizes to be linked to the properties of their dark matter halos \citep[e.g.,][]{dutton_07, Kravtsov13, somerville18, jiang19}, and some have attempted to model the environmental dependence as well \citep[e.g.,][]{shankar13}. We will discuss later in \S\ref{sec:theory} how our results, while broadly agreeing with the standard galaxy formation framework, demonstrate the need for further theoretical investigation. 

The conflicting results discussed above can be attributed primarily to several challenges faced in previous studies. These include (a) the absence of sufficiently large sample sizes, (b) the absence of robust uncertainty estimates on the measurements of galaxy radius, (c) not using statistically robust frameworks to assess the presence of correlations, (d) the presence of systematic discrepancies between measurements obtained from different surveys employed in the same study (e.g., between cluster and field galaxies).

It is important to note that galaxy sizes are primarily related to stellar mass through the size-mass relationship \citep[e.g.,][]{kormendy_77}, and any correlation of size with the environment is secondary and weaker. Measuring the effect of environment therefore requires a large sample size, as well as accurate uncertainties on $R_e$ measurements.

Additionally, we should also note that more massive galaxies and bulge-dominated/elliptical galaxies preferentially reside in overdense regions of the Universe. Given that both stellar mass and morphology play an important role in determining a galaxy's size, it is extremely important to control for both these parameters to determine whether environment plays a role independent of stellar mass and morphology. We elaborate on this in \S \ref{sec:results}.

The smaller samples used in prior studies also made it difficult to properly account for: i) other galaxy properties that might impact galaxy size (e.g., state of star formation, etc.); ii) the great diversity in the nature of dense environments (e.g., different galaxy clusters); and iii) rare objects (e.g., massive galaxies) that might play an important role in controlling the environmental dependence.

One needs large uniform samples of galaxies with excellent imaging quality and robust statistical frameworks to accurately measure galaxy sizes and examine their correlation with environment over a broad range of stellar mass and redshift. The excellent imaging quality of the Hyper Suprime-Cam Subaru Strategic Program \citep[HSC-SSP; ][]{ssp_2,ssp_1}, in conjunction with state-of-the-art Bayesian machine learning (ML) frameworks like the Galaxy Morphology Posterior Estimation Network \citep[\gampen{};][]{gampen_software_paper}, offers an unprecedented opportunity to address this challenge.  

For the first time, we have access to radius measurements with robust uncertainties for millions of galaxies beyond the local Universe \citep{hsc_wide_morphs} and measurements of environmental density covering $\sim360$ deg$^2$ \citep{hsc_den}. Using these, we present the first comprehensive study of the variation of galaxy radius with environment beyond the local Universe. 

The robustness of \gampen{} coupled with the depth of the HSC data enables us to assemble a large uniform sample of $\sim3$ million galaxies in the redshift range $0.3 \leq z < 0.7$ down to $\sim23$ AB mag. This sample size is $\sim100-10,000$ times larger than most previous studies and complete to stellar masses that are $\gtrapprox1$ dex lower than most previous studies. Using the full Bayesian posteriors predicted by \gampen{} combined with a Monte Carlo analysis framework, we can incorporate the uncertainties in $R_e$ measurements into our correlation analysis. This allows us to confirm/reject the presence of correlations with very high statistical significance --- in most cases with $>5\sigma$ confidence.

We use the comprehensive galaxy morphology catalog for HSC galaxies that was obtained using \gampen{} \citep{hsc_wide_morphs}; and correlate these structural parameter measurements with the wide-field projected density map available from HSC-SSP \citep{hsc_den}. We study the variation of effective radius on a $10$ comoving Mpc (cMpc) scale ---  thus focusing on the large-scale environmental effect by taking advantage of the wide-field coverage of HSC ($\sim 360$ deg$^2$). We investigate the above correlations for the overall sample of galaxies and for four subpopulations derived from the main sample -- disk-dominated, bulge-dominated, star-forming, and quiescent galaxies.

In \S\ref{sec:data}, we describe our sample selection, as well as the determination of structural parameters, large-scale environmental density, and galaxy stellar mass. We also detail our criteria for separating galaxies into the various subpopulations. \S\ref{sec:results} outlines the results of the abovementioned correlation analysis between $R_e$ and environmental density. We summarize our findings in \S\ref{sec:discussion} and discuss the observed correlations in light of existing theoretical frameworks. We end with the primary takeaways of this study as well as future directions in \S\ref{sec:conclusions}. This work uses AB magnitudes and Planck18 cosmology ($H_0=67.7$ km/s/Mpc, $\Omega_{\mathrm{m}}=0.311$, $\Omega_\Lambda=0.689$, \citealp{planck18}).

\section{Data} \label{sec:data}

We use data from the HSC-SSP Public Data Release 2 \citep[PDR2;][]{hsc_pdr2}. Specifically, we use the Wide layer of PDR2, which covers $\sim360$ deg$^2$ in all five broadband filters (\gb\rb\ib\zb\yb) and reaches a depth of $\sim26$ AB mag. In the following subsections, we will outline our sample selection and the various external catalogs used for this study. 

\subsection{Sample Selection} \label{sec:sample_selection}
\begin{figure}[htbp]
    \centering
    \includegraphics[width = 0.47\textwidth]{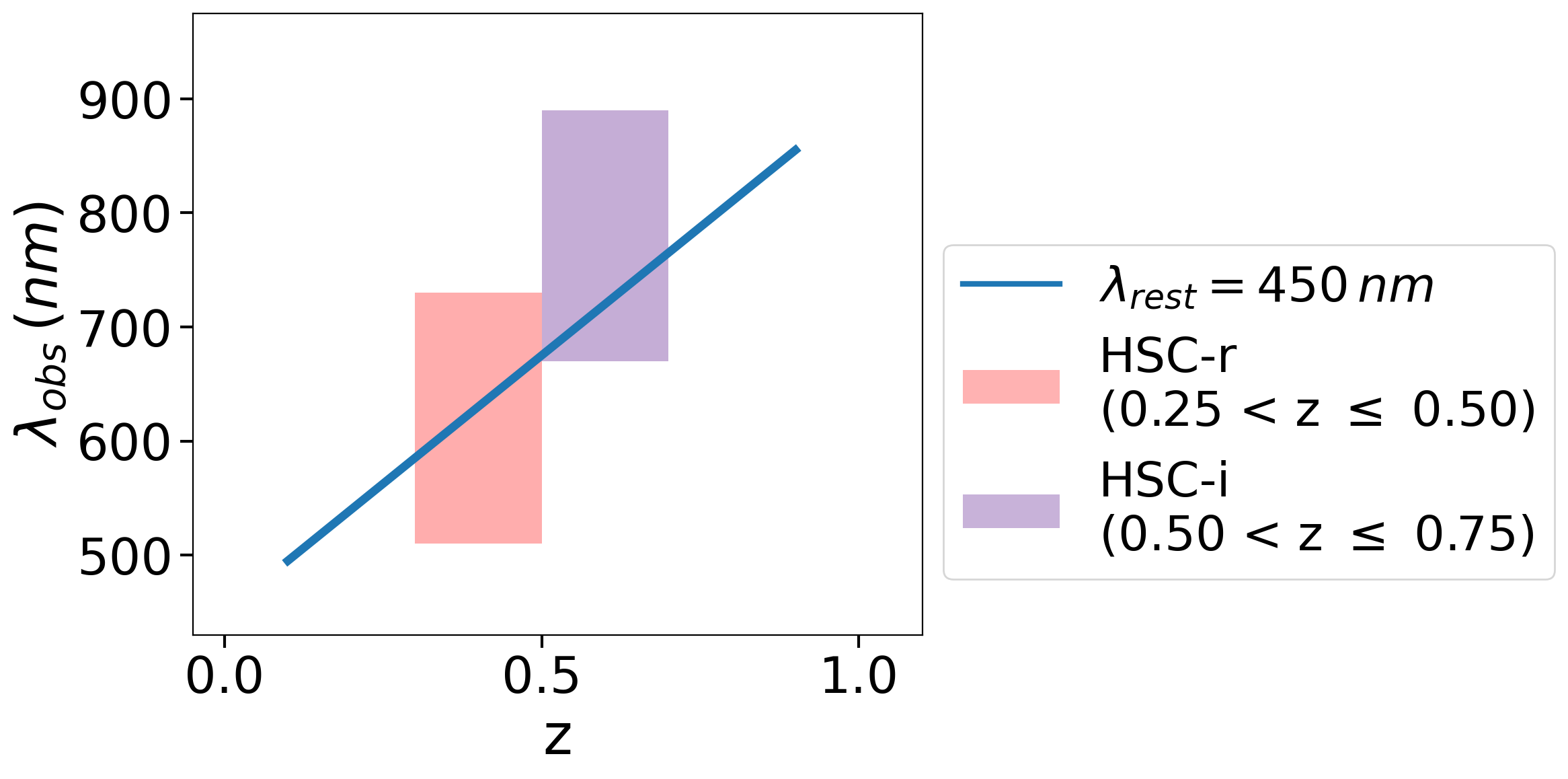}
    \caption{The filter used for morphological determination in \citet{hsc_wide_morphs} for each redshift bin is shown along with the wavelength range sampled by each filter. The blue line shows where rest-frame $450\,\,nm$ emission falls for redshifts labeled on the x-axis. As this figure shows, the chosen filters allow us to consistently perform morphology determination at a rest-frame wavelength of $\sim450\,\,nm$ (i.e., in the rest-frame \gb{}-band).}
    \label{fig:band_comp}
\end{figure}

The starting point for our sample selection is galaxies with $0.3 \leq z < 0.7$ from the \citet{hsc_wide_morphs} morphological catalog (see \S\ref{sec:morph_measurements} for more details). This catalog contains $\sim 8$ million galaxies and includes most HSC-Wide extended sources with $m < 23$ (in the relevant band; see Figure \ref{fig:band_comp}). We only include sources with more than $5\sigma$ detection (in the \texttt{cmodel} measurement) in all HSC bands and exclude galaxies near bright stars or with significant imaging issues (e.g., cosmic-ray hits, saturated pixels) identified using the \texttt{mask\_s18a\_bright\_objectcenter} and \texttt{cleanflags\_any} parameters in PDR2. 

We use high-quality photometric redshifts from the second data release of the HSC photo-z catalog \citep{photoz_hsc_pdr2} for all galaxies in our sample. These redshifts were calculated from HSC five-band photometry using Mizuki \citep{mizuki}, a Bayesian template fitting code, and we refer the interested reader to \citet{photoz_hsc_pdr2} for more details. For $m < 23$ HSC-Wide galaxies, the biweight dispersion of $\Delta z=\left(z_{\text {photo }}-z_{\text {spec }}\right) /\left(1+z_{\text {spec }}\right)$ is $\leq 0.05$. To further isolate a subsample with secure photometric redshifts, we only include galaxies for which the \texttt{photoz\_risk\_best} parameter $<0.1$. This parameter controls the risk of the photometric redshift being outside the range $z_{\mathrm{true}} \pm 0.15(1+z_{\mathrm{true}})$, with 0 being extremely safe to 1 being extremely risky. We also exclude galaxies with reduced chi-square $\chi_{\nu}^2 > 5$ for the best-fitting model, as recommended by \citet{photoz_hsc_pdr2}. Following \citet{hsc_den}, we employ a further selection cut for galaxies in $0.4 \leq z \leq 0.5$, with a strong Balmer-Lyman break degeneracy. To minimize potential contamination from Lyman-break galaxies at $z\sim3$, we exclude galaxies that have specific star-formation rates (sSFRs)  $> 1$ Gyr$^{-1}$. This is because misclassified objects at $z\sim3$ mostly show very young ages ($\sim 1$ Gyr) with very high sSFRs. Note that the contamination of Lyman-break galaxies for higher redshift bins, including \ib-band dropouts, is negligible given our detection criteria in all HSC bands, including the \gb-band. 

\begin{deluxetable*}{c|c}[htbp]
\tablecaption{Sample Sizes at Different Stages of the Sample Selection Process \label{tab:sample_sel}}
\tablecolumns{2}
\tablehead{Selection Cuts & Number of Galaxies} 
\startdata
    \hline
    \hline
    \citet{hsc_wide_morphs} HSC-Wide morphological catalog \tablenotemark{a} & 7,805,186\\
    \hline
    Galaxies with more than $5\sigma$ detection in all HSC-Wide bands & 7,706,785\\
    \hline
    After excluding galaxies near bright stars & 7,475,782\\
    \hline\
    After photometric redshift cuts based on $\chi_{\nu}^2$ and sSFR & 7,194,030\\
    \hline
    Galaxies with $0.3 \leq z < 0.7$ & 5,200,228 \\
    \hline
    After positional  crossmatch with the \citet{hsc_den} catalog\tablenotemark{b} & 2,894,716 \\
    \hline
\enddata
\tablenotetext{a}{The cuts based on magnitude, \texttt{photoz\_risk\_best}, and \texttt{cleanflags\_any} are already incorporated in this catalog.}
\tablenotetext{b}{The \citet{hsc_den} catalog measures environmental density in five of the seven HSC-Wide fields, while the \citet{hsc_wide_morphs} catalog measures structural parameters in all seven fields. }
\end{deluxetable*}

We split our sample into four redshift slices: $0.3 \leq z < 0.4$, $0.4 \leq z < 0.5$, $0.5 \leq z < 0.6$, and $0.6 \leq z < 0.7$. Within each redshift slice, we match the positions of galaxies in our sample to the fine grid of HSC PDR2 survey locations at which density measurements are available from the \citet{hsc_den} catalog (see \S\ref{sec:density_measurements} for more details). Based on the grid size used in \citet{hsc_den}, we allow a maximum positional error of $64\arcsec$ and discard all unmatched sources. This step ensures that the two survey areas (morphology and density estimation) are completely consistent. Note that the \citet{hsc_den} catalog also included $m < 23$ sources with more than $>5\sigma$ detection in all HSC bands and followed a data-cleaning procedure very similar to ours (see \S2 of \citealp{hsc_den}). This leaves us with a final sample of $\sim2.9$ million galaxies, with $0.49$ million, $0.70$ million, $0.65$ million, and $1.05$ million in the four redshift bins, respectively. The total sample size at different stages of the sample selection process is outlined in Table \ref{tab:sample_sel}.

Our decision to use four redshift slices is primarily driven by the precision of the photometric redshifts used in this work. We have set the width of the slices such that it is $\geq2\times$ the biweight dispersion of $\Delta z=\left(z_{\text {photo }}-z_{\text {spec }}\right) /\left(1+z_{\text {spec }}\right)$. Using finer redshift bins might be feasible for a spectroscopic subsample, but is unfortunately not possible for the entire sample presented in this work.

\subsection{Morphology and Size Measurements} \label{sec:morph_measurements}
For this work, the effective radius ($R_e$) is defined as a measure of the radius that encloses half of the total flux of the galaxy; and the bulge-to-total light ratio ($L_B/L_T$) is the fraction of flux contained in the bulge component of the galaxy obtained from a bulge + disk decomposition. Throughout this article, we will use effective radius and `size' interchangeably, but we note that in the recent past, alternative definitions of galaxy size have also been proposed \citep[e.g.,][]{graham_19, miller_19, chamba_20}.

The $R_e$ and $L_B/L_T$ measurements used in this work are from the \citet{hsc_wide_morphs} morphological catalog and were obtained using \gampen{}. \gampen{} is an ML framework that can estimate full Bayesian posteriors for different structural parameters of galaxies and has been extensively tested on both simulations and HSC-Wide data \citep{gampen_software_paper,hsc_wide_morphs}. \gampen{}'s design is inspired by the successful morphology classification framework, Galaxy Morphology Network \citep[\gamornet;][]{gamornet_software_paper}, and various Visual Geometry Group \citep[VGG;][]{vgg} network variants that are highly effective at large-scale visual recognition. We refer the interested reader to \citet{gampen_software_paper} for more details.

\gampen{}'s predicted posterior distributions are extremely well calibrated ($\lesssim 5\%$ deviation) and outperform uncertainty estimates from light-profile fitting codes by $\sim15-60\%$ \citep[Figure 21 of ][]{hsc_wide_morphs}. Specifically for $R_e$ measurements of $0.3 \leq z <0.7$ HSC-Wide galaxies, the dispersion in \gampen{}'s prediction error ($\left(R_{\text {pred.}}-R_{\text {true}}\right) /R_{\text {true}}$) is $\leq0.07$. \gampen{}'s $R_e$ estimates of these galaxies have also been tested against measurements obtained using GALFIT \citep{galfit}. \gampen{}'s measurements are consistent overall and outperform those of GALFIT for galaxies with $R_e \leq 2\arcsec$ ($\sim 12.6$ kpc at $z=0.5$) \citep[Figure 31 of ][]{hsc_wide_morphs}.

As shown in Figure \ref{fig:band_comp}, we use different imaging bands for galaxies at different redshifts --- this allows us to consistently perform morphology determination in the rest-frame \gb-band across our entire sample. We use \rb-band for $0.3 \leq z < 0.50$, and \ib-band for $0.50 \leq z < 0.7$. Additionally, as outlined in Appendix \ref{sec:ap:size_corr}, we further investigated the effect of color gradients on our measured $R_e$ values by correcting them to a rest-frame wavelength of 450\,\,nm. We found the applied size corrections to be $\lesssim5\%$, with no significant impact on the correlation coefficients reported in this study. 

For all structural parameters used in this study, we use the most probable value (mode) of the \gampen{}-predicted posterior distributions as the measured value and the width of the $68\%$ confidence interval as the error bar. Note that all measurements used in this study and the ML models used to obtain them are publicly available \citep{hsc_wide_morphs}.

\subsection{Density Measurements} \label{sec:density_measurements}

\begin{figure*}[htbp]
    \centering
    \includegraphics[width = 0.7\textwidth]{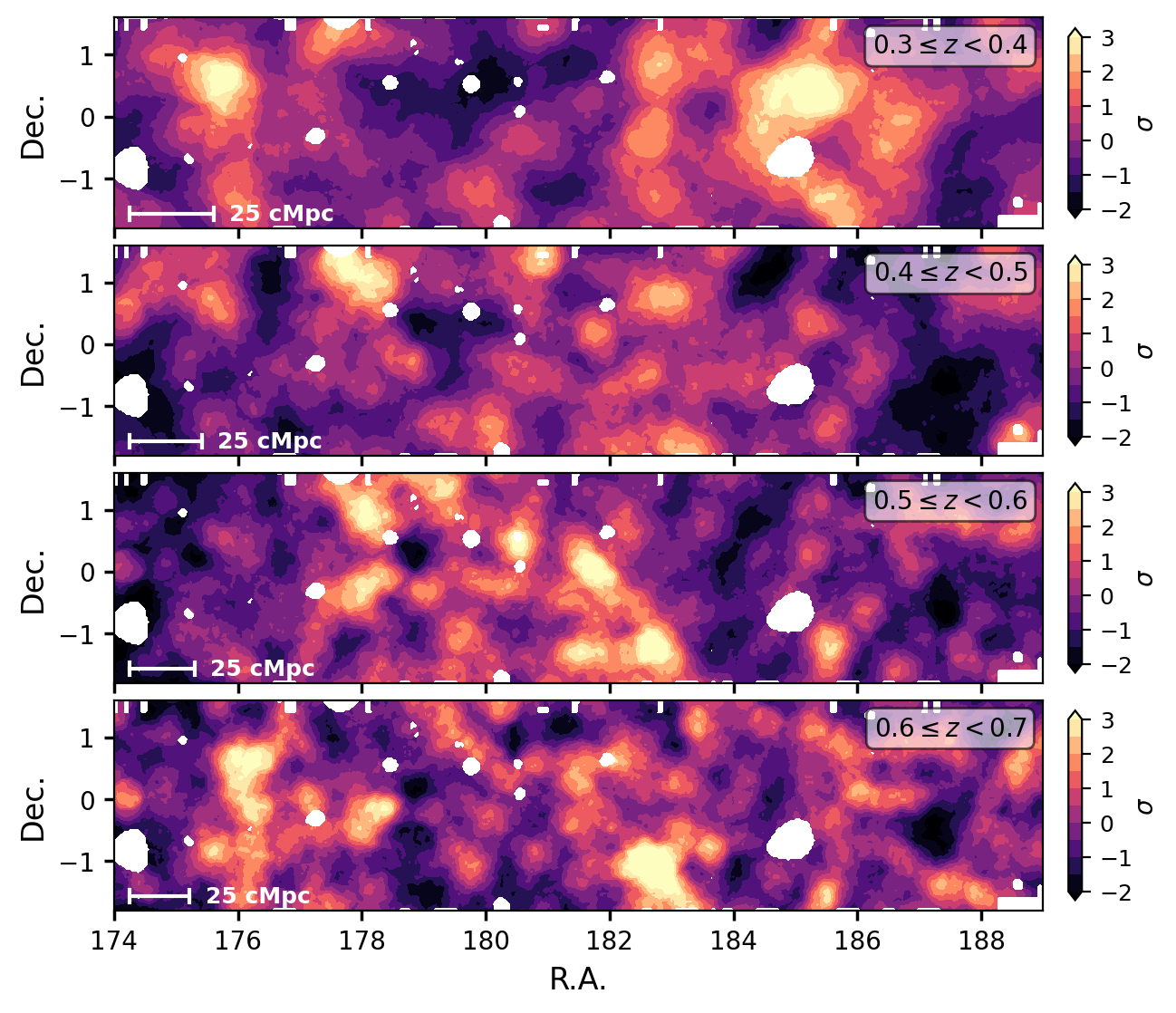}
    \caption{Projected two-dimensional densities are shown for \texttt{W04 GAMA12H}, one of the five HSC-Wide fields used in this study. Each row corresponds to a different redshift slice within which the densities are measured. Colors indicate the density excess in standard deviation within $r=10$ comoving Mpc as denoted by the color bars on the right. For reference, a fixed scale of 25 comoving Mpc is shown in each redshift slice. The white regions demarcate bright-star masks used during density measurement. Note that mask corrections are already incorporated into the density measurements used in this work, and we refer the interested reader to \citet{hsc_den} for more details.}
    \label{fig:hsc_den_map}
\end{figure*}

The density measurements used in this study are from the \citet{hsc_den} catalog, which measured the projected two-dimensional density in five of the seven HSC-Wide fields, covering an area of $\sim 360$ deg$^2$. \citet{hsc_den} chose the $\sim 360$ deg$^2$ survey area by identifying regions in which the $>5\sigma$ limiting magnitude of the point-spread function is deeper than $i=26$; this criterion was chosen to prevent the need to correct the variance of number densities due to depth variation across the survey field. Note that both this study and \citet{hsc_den} use the same underlying catalog (HSC PDR2), and thus all catalog quantities (e.g., photometric redshifts) are consistent among the two studies. 

The projected density map has a spatial resolution of $\sim1.5\arcmin$, and the redshift slices constructed in this study mirror the ones used in \citet{hsc_den}. We set the density estimate at each grid point of the survey area to be the density contrast in standard deviation measured using a top-hat aperture of $r=10$\,cMpc within the specific redshift slice. This is denoted by $\sigma_{r=10\,{\rm cMpc}}$ throughout this article and represents entries 128, 130, 132, and 134 in Table A1 of \citet{hsc_den}. $\sigma_{r=10\,{\rm cMpc}}$ represents $(n_{\mathrm{r}} - n_{\mathrm{r, mean}}) / \Sigma_{\mathrm{r}}$, where $ n_\mathrm{r}$ is the measured number density of galaxies within the given aperture; $ n_{\mathrm{r, mean}} $ and $\Sigma_r$ refer to the mean and standard deviation of $ n_\mathrm{r}$ at each redshift in the whole survey area. We refer the interested reader to \S 3.1 and Appendix A of \citet{hsc_den} for more information about the measured densities.

One of the five fields (\texttt{W04 GAMA12H}) is shown as an example in Figure \ref{fig:hsc_den_map}. The brightest structures in the figure show regions with greater than $3\sigma$ excess at their density peaks in the projected density distributions. They are part of the $\sim50$ large-scale overdensities (comparable to supercluster-embedded regions)  that \citet{hsc_den} detect. We refer the interested reader to \S 3.2 and Figure 9 of \citet{hsc_den} for more details.
 
Note that \citet{hsc_den} tested their pipeline on a mock galaxy catalog and found that their projected overdensities well trace the total masses of embedded massive dark matter halos at each redshift slice. They also confirmed the consistency of their measurements at $z \leq 0.6$ against mean lens shear signals from a weak-lensing stacking analysis.

\vspace{0.1in}

\subsection{Stellar Masses} \label{sec:mass_completeness}
The stellar masses used in this study are from the same Mizuki catalog that was used for photometric redshift estimates. We remind the reader that our study only includes sources with more than $5\sigma$ detection in all HSC bands, and we have also discarded sources with reduced chi-square $\chi_{\nu}^2 > 5$ for the best-fitting model. \citet{hsc_photoz_pdr1} and \citet{hsc_morph_den} have compared these stellar masses from Mizuki, based on HSC \gb{}\rb\ib\zb\yb{} photometry, against stellar masses obtained from the NEWFIRM Medium Band Survey \citep{newfirm} and COSMOS2015 \citep{cosmos_2015}; both of which are based on $>30$ band photometry. The above studies found the Mizuki measurements to be overestimates, especially at higher redshifts. However, at $z \leq 0.7$, this discrepancy is $\lesssim0.1$ dex. These differences can be attributed to the systematic differences in the data and the adopted template error functions and priors. Note that when comparing two independent surveys, stellar mass offsets of $\sim0.2-0.3$ dex can be expected even when both data sets have deep photometry in multiple filters \citep[e.g.,][]{dokkum_14}.

\begin{figure}[htbp]
    \centering
    \includegraphics[width = 0.46\textwidth]{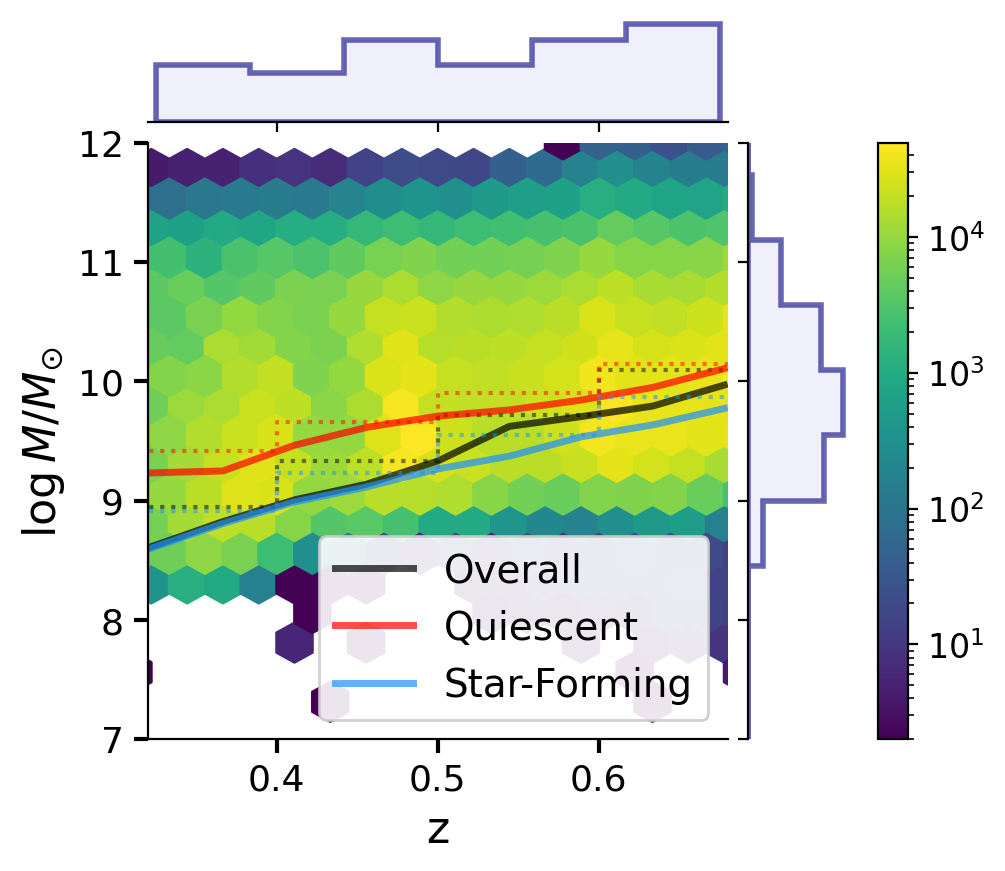}
    \caption{Distribution of stellar mass as a function of redshift for all galaxies in our sample. The mass-redshift space is divided into hexagonal bins of roughly equal size, with the number of galaxies in each bin represented according to the color bar on the right. Marginal histograms are also plotted along each axis. The solid black line shows the $90\%$ stellar mass-completeness limit for the entire sample, while the red and blue solid lines show the mass-completeness limits for the quiescent and star-forming subpopulations, respectively. The faint dotted lines demarcate each redshift slice's adopted lower mass limits ($M_{\rm c}$).}
    \label{fig:mass_comp}
\end{figure}

To estimate the $90\%$ stellar mass-completeness limit of our sample, we follow the approach outlined in \citet{Pozzetti10} and \citet{Weigel16}. We select galaxies in narrow redshift bins and calculate their limiting stellar mass ($M_{\rm lim}$); i.e., the mass they would have if their apparent magnitude were equal to $23$, the magnitude limit we chose for our sample. Thereafter, we selected $20\%$ of the faintest galaxies and defined the stellar mass-completeness limit as the $90^{th}$ quantile of $M_{lim}$ values in each redshift bin. Figure \ref{fig:mass_comp} shows the determined mass-completeness limit for all galaxies, as well as separate completeness limits for star-forming and quiescent galaxies (refer to \S\ref{sec:sep_into_subsamples} for a description of how we isolate these subpopulations). Our sample is complete down to $\log (M/M_\odot)\sim8.5$ at $z=0.3$ and $\sim10$ at $z=0.7$. The completeness limits at the lowest and highest redshifts are $\log (M/M_\odot)\sim$ 8.6 (9.2) and 9.9 (10.1), respectively, for star-forming (quiescent) galaxies. To be conservative, we implement a step function on top of the stellar mass-completeness limit obtained, as shown in Figure \ref{fig:mass_comp}. The values of this step function are used as the lower mass limit ($M_{\rm c}$) for each redshift slice and are shown in Table \ref{tab:mass_comp}.

\begin{deluxetable}{c|ccc}[htbp]
\tablecaption{Lower Mass Limits ($M_{\rm c}$) Used at Different Redshifts\label{tab:mass_comp}}
\tablecolumns{4}
\tablehead{
Redshift Slice & \multicolumn{3}{c}{Lower Mass Limit ($\log M/M_\odot$)}\\
 & Overall & Quiescent & Star Forming} 
\startdata
    \hline
    \hline
    $0.3 \leq z < 0.4$ & 8.95 & 9.41 & 8.91 \\
    $0.4 \leq z < 0.5$ & 9.33 & 9.66 & 9.23 \\
    $0.5 \leq z < 0.6$ & 9.72 & 9.90 & 9.55 \\
    $0.6 \leq z < 0.7$ & 10.09 & 10.15 & 9.87 \\
\enddata
\end{deluxetable}

\subsection{Separating into Subpopulations} \label{sec:sep_into_subsamples}
In \S\ref{sec:results}, we will explore the variation of structural parameters with environmental density for various subpopulations of galaxies derived from our main sample. To divide galaxies into disk- and bulge-dominated samples, we classify all galaxies with $L_B/L_T \leq 0.4$ as disk-dominated and all galaxies with $L_B/L_T \geq 0.6$ as bulge-dominated. 

To separate galaxies into star-forming and quiescent subsamples, we use their rest-frame Sloan Digital Sky Survey (SDSS) \uband{}-\rb{} versus \rb{}-\zb{} colors, calculated using spectral energy distribution (SED) fitting. The use of the \uband{}-\rb{} versus \rb{}-\zb{} color-color diagram has already been demonstrated to be an effective way to separate these subpopulations \citep[e.g.,][]{Holden12, Chang15, Lopes16, hsc_mass_size}. Note that for our sample, we are unable to use \textit{UVJ} selection because SED fitting using HSC \grizy{} photometry does not allow us to obtain robust estimates for rest-frame \textit{J} magnitudes. We obtain the SDSS rest-frame magnitudes from the same Mizuki catalog that we used for redshift and stellar mass estimation. 

\citet{hsc_mass_size} has demonstrated that applying the \citet{Holden12} \textit{urz} color-color selection directly to HSC data does not optimally distinguish quiescent and star-forming galaxies; instead, the boundaries need to be adjusted. Thus, we follow \citet{Kawin16, hsc_mass_size} and use our sample to self-calibrate the region separating quiescent from star-forming galaxies. Appendix \ref{sec:ap:cc_sel} shows the final separation boundaries obtained and an extended description of the procedure.

The number of galaxies in each subpopulation, along with the overlap between disk-dominated/star-forming and bulge-dominated/quiescent samples, is shown in Table \ref{tab:sample_overlap}. Note that the disk-dominated $+$ bulge-dominated combination always has fewer galaxies than the star-forming $+$ quiescent combination. This is because galaxies with intermediate values of $L_B/L_T$ ($0.4 < L_B/L_T < 0.6$) are not classified as either disk-dominated or bulge-dominated.

\begin{deluxetable}{c|r|c}[htbp]
\tablecaption{Number of Galaxies in Each Subpopulation  \label{tab:sample_overlap}}
\tablecolumns{3}
\tablehead{
\colhead{Redshift Slice} & \colhead{Sample} & \colhead{Overlap}}
\startdata
    \hline
    \hline
    \multirow{4}{*}{$0.3 \leq z < 0.4$} & Disk-dominated: 281,448 & \multirow{2}{*}{256,415} \\
    & Star-forming: 369,868 & \\
    \cline{2-3}
     & Bulge-dominated:  83,014 & \multirow{2}{*}{58,589} \\
     & Quiescent: 117,889 & \\
     \hline
     \multirow{4}{*}{$0.4 \leq z < 0.5$} & Disk-dominated: 393,942 & \multirow{2}{*}{347,015} \\
    & Star-forming: 525,928 & \\
    \cline{2-3}
     & Bulge-dominated:  110,862 & \multirow{2}{*}{73,765} \\
     & Quiescent: 177,423 & \\
     \hline
     \multirow{4}{*}{$0.5 \leq z < 0.6$} & Disk-dominated: 403,170 & \multirow{2}{*}{340,764} \\
    & Star-forming: 447,095 & \\
    \cline{2-3}
     & Bulge-dominated:  107,656 & \multirow{2}{*}{80,405} \\
     & Quiescent: 202,217 & \\
     \hline
     \multirow{4}{*}{$0.6 \leq z < 0.7$} & Disk-dominated: 653,682 & \multirow{2}{*}{573,062} \\
    & Star-forming: 756,135 & \\
    \cline{2-3}
     & Bulge-dominated:  158,144 & \multirow{2}{*}{117,099} \\
     & Quiescent: 298,161 & \\
     \hline
\enddata
\end{deluxetable}

\section{Results} \label{sec:results}
The central focus of this work is to study how the structural parameters of galaxies vary as a function of their environmental density. In \S\ref{sec:data}, we described how a measurement of environmental density is assigned to galaxies from the \citet{hsc_wide_morphs} morphological catalog. We now use this combined information to study how the environment affects galaxy radius. 

In \S\ref{sec:rad_den_all}, we will outline our results for all galaxies in our sample. In \S\ref{sec:rad_den_sub}, we will focus on four subpopulations of galaxies: disk-dominated, bulge-dominated, star-forming, and quiescent. 

\subsection{All Galaxies} \label{sec:rad_den_all}
\begin{figure*}[htbp]
    \centering
    \includegraphics[width = 0.95\textwidth]{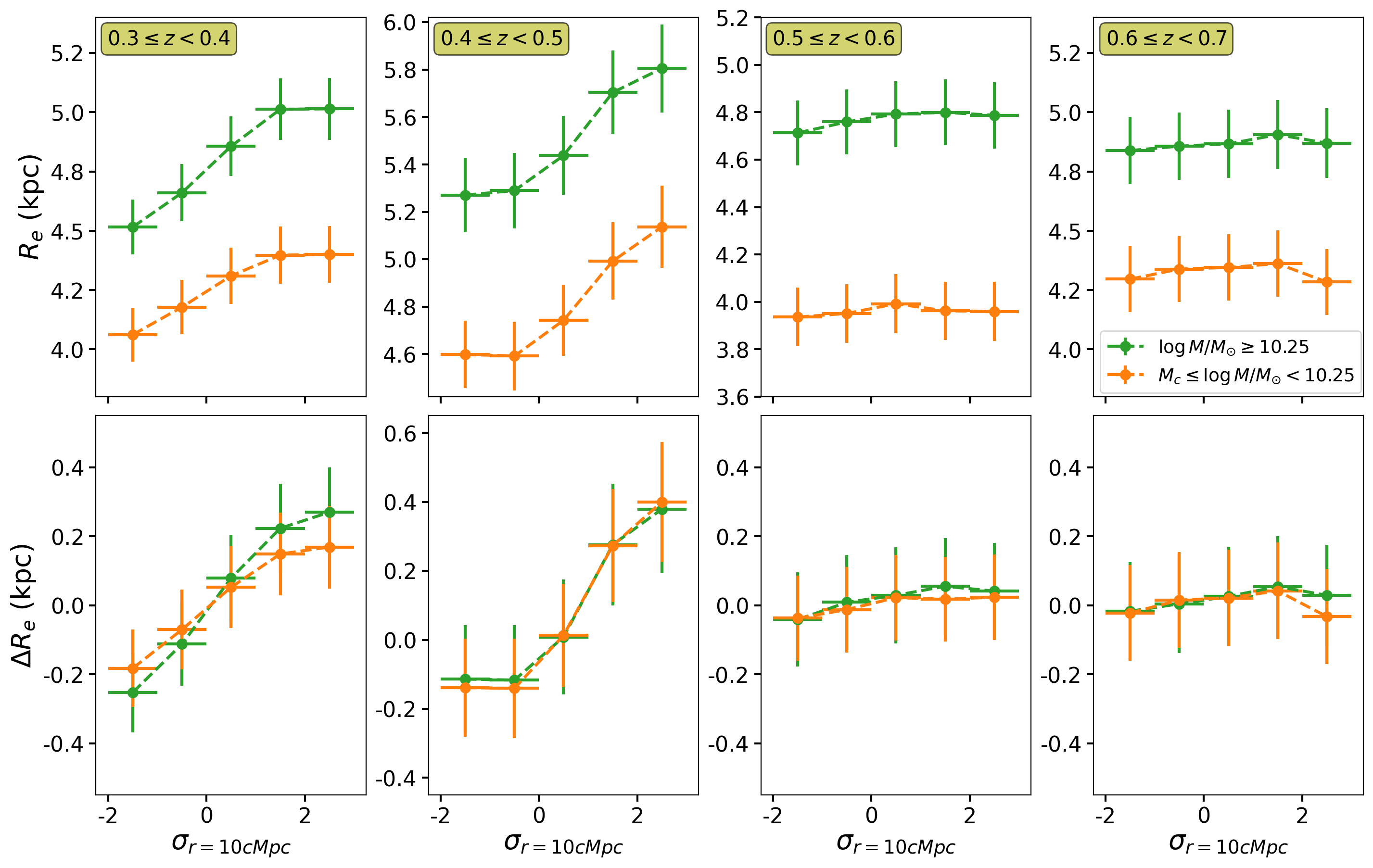}
    \caption{Effective radius ($R_e$) and deviation in effective radius from the average size, $\Delta R_e = R_e - \overline{R_e}(M,L_B/L_T)$, are plotted against density excess ($\sigma_{r=10\,{\rm cMpc}}$). Each column shows the relationship at a different redshift slice. The different colored lines represent different stellar mass ranges, as denoted in the figure legend, with $M_{\rm c}$ referring to the mass-completeness limit at the corresponding redshift. The range of densities ($-2 \leq \sigma_{r=10\,{\rm cMpc}} < 3$) is split into five equal bins, and each point shows the median value of $R_e$ or $\Delta R_e$ within the corresponding bin. The vertical error bars depict the typical uncertainty (i.e., median $68\%$ confidence interval) predicted by \gampen{} for all galaxies in that bin. Although each panel shows a different range of values, the minimum $R_e$ and $\Delta R_e$ scales of $0.2$ kpc are consistent in each row.}
    \label{fig:rad_den_all}
\end{figure*}

We separate the galaxies in each redshift slice into five bins based on their environmental density. The bins are linearly spaced and span the range $\sigma_{r=10\,{\rm cMpc}} = [-2,3)$. We investigate the distribution of half-light radius ($R_e$) within each bin. The top row of Figure \ref{fig:rad_den_all} shows the median value of the effective radius in each density bin. The vertical error bar at each point reflects the median $1\sigma$ width of the $R_e$ posterior distribution predicted by \gampen{} for all galaxies in that bin. To investigate the prevalence of the correlation across different stellar mass regimes, we also split the sample into two mass bins: $M_{\rm c} \leq \log M/M_{\odot} < 10.25$ and $\log M/M_{\odot} \geq 10.25$. This chosen threshold ($\sim10.25M_{\odot}$) corresponds to the $75^{th}$ quantile of the overall stellar mass distribution. 

As noted in \S \ref{sec:intro}, the stellar mass and morphology of a galaxy play an important role in determining its $R_e$ through the size-mass relationship. Given that overdense regions of the Universe contain preferentially more massive and bulge-dominated/elliptical galaxies, it is extremely important to control for these two parameters while judging the presence of a correlation. We demonstrate in Appendix \ref{sec:ap:no_bin} that for the density scales considered in this work, the distribution of stellar mass is reasonably well matched across the different density environments when using the two mass bins mentioned above. 

However, in order to further rule out any chance that our results are driven by variations in either stellar mass or morphology across the different density environments, we define the parameter 

\begin{equation}
    \Delta R_e = R_e - \overline{R_e}(M,L_B/L_T),
\label{eq:delta_re}
\end{equation}

\noindent
to quantify the deviation in size of a galaxy from the (stellar-mass- and $L_B/L_T$-dependent) average size. This quantity allows us to investigate whether the environment of a galaxy plays a role in setting its size independent of stellar mass and morphology. To determine $\Delta R_e$, we first calculate the average size of galaxies ($\overline{R_e}$) over a fine grid of $100\times100$ points in the $(M, L_B/L_T)$ space within each redshift slice. At each grid point, we calculate the median $R_e$ for all galaxies with $\log M/M_{\odot} \pm 0.125$ and $L_B/L_T \pm 0.1$. We choose these ranges to approximate the typical error in the prediction of $M$, $L_B/L_T$ and to ensure the presence of sufficient samples across the entire grid. 

Once we have calculated $\overline{R_e}(M,L_B/L_T)$ over the entire grid, we map every galaxy in our catalog to the closest grid point and calculate $\Delta R_e$ for each galaxy according to Eq. \ref{eq:delta_re}. The bottom row of Figure \ref{fig:rad_den_all} shows the median value of $\Delta R_e$ across the five environmental density bins. The vertical error bar at each point reflects the median $1\sigma$ width of the $\Delta R_e$ posterior distribution for all galaxies in that bin.

\begin{deluxetable*}{c|c|cccc}[htbp]
\tablecaption{Measured Correlation Coefficients for $\Delta R_e$ vs. $\sigma_{r=10\,{\rm cMpc}}$ \label{tab:corr_all}}
\tablecolumns{5}
\tablehead{Mass Range & & $0.3 \leq z < 0.4$ & $0.4 \leq z < 0.5$ & $0.5 \leq z < 0.6$ & $0.6 \leq z < 0.7$ \\ 
          ($\log M/M_{\odot}$) & & & &}
\startdata
    \hline
    \hline
    \multirow[c]{3}{*}{$\left[M_{\rm c},10.25\right)$\tablenotemark{a}} & $\rho$   & $8.5\times10^{-2} \pm 3.8\times10^{-4}$ & $1.1\times10^{-1} \pm 4.0\times10^{-4}$ & $2.3\times10^{-2} \pm 8.9\times10^{-4}$ & $9.5\times10^{-3} \pm 9.1\times10^{-4}$ \\
                                    & $p$      & $(<10^{-300}) \pm (<10^{-300})$ & $(<10^{-300}) \pm (<10^{-300})$ & $2.4\times10^{-25} \pm 5.1\times10^{-20}$ &  $2.6\times10^{-3} \pm 3.8\times10^{-3}$   \\
                                    & $\alpha$ & $225\sigma_{\rho}$ & $269\sigma_{\rho}$ & $25\sigma_{\rho}$ & $10\sigma_{\rho}$  \\
                                    & $>5\sigma$ & \checkmark & \checkmark &  \checkmark &   \\
    \hline
    \multirow{3}{*}{$\geq10.25$} & $\rho$ & $1.3\times10^{-1} \pm 8.0\times10^{-4}$   & $1.0\times10^{-1} \pm 6.1\times10^{-4}$ & $2.3\times10^{-2} \pm 6.6\times10^{-4}$ & $1.9\times10^{-2} \pm 4.5\times10^{-4}$ \\
                                                & $p$  & $(<10^{-300}) \pm (<10^{-300})$ & $(<10^{-300}) \pm (<10^{-300})$ & $6.8\times10^{-27} \pm 5.5\times10^{-24}$ &  $1.4\times10^{-29} \pm 8.6\times10^{-27}$   \\
                                             & $\alpha$ & $165\sigma_{\rho}$ & $168\sigma_{\rho}$ & $35\sigma_{\rho}$ & $42\sigma_{\rho}$  \\
                                             & $>5\sigma$ & \checkmark & \checkmark & \checkmark & \checkmark \\
    \hline
    \hline
\enddata
\tablenotetext{a}{$M_{\rm c}$ refers to the $90\%$ stellar mass-completeness limit for each corresponding redshift slice}
\tablecomments{The above table shows correlation coefficients measured using a Monte Carlo-based Spearman rank correlation test. $\rho$ refers to the Spearman correlation coefficient, and $p$ refers to the probability of a spurious correlation. For both $\rho$ and $p$, median $\pm$ standard deviation values are reported above. $\alpha$ refers to the distance between the median value of $\rho$ and $\rho=0$ (signifying no correlation). It is reported above in terms of the standard deviation of $\rho$ ($\sigma_\rho$). A \checkmark indicates that we can confirm a positive correlation between $R_e$ and $\sigma_{r=10\,{\rm cMpc}}$ with $>5\sigma$ confidence.}
\end{deluxetable*}

Figure \ref{fig:rad_den_all} shows that both $R_e$ and $\Delta R_e$ display similar overall trends across the four redshift slices. However, the trends are not identical, and there are small differences (e.g., the rightmost two points in $0.3 \leq z < 0.4$, the lower mass bin in $0.5 \leq z < 0.6$). Due to the reasons elucidated earlier, we will use $\Delta R_e$ as the primary parameter from here onward to judge the presence of correlations for the rest of this paper. We should also note that using $\Delta R_e$ allows us to make a fairer comparison across the different redshift slices, as the distribution of $L_B/L_T$ is not identical across the four different redshift bins given the varying lower mass limits adopted in Table \ref{tab:mass_comp}. We refer the interested reader to Appendix \ref{sec:ap:lb_lt_var} for more details.

Note that the median $R_e$ values in the second redshift slice of Figure \ref{fig:rad_den_all} are larger than those in the other three slices. This difference is $<5\%$ when considering the full range of $R_e$ values. We have verified that this difference is not an artifact of our analysis pipeline but instead happens to be driven by the combination of mass-completeness and other cuts summarized in \S \ref{sec:data}. We refer the interested reader to Appendix \ref{sec:ap:rad_distr} for more details. This also demonstrates why we use $\Delta R_e$ to investigate the presence of correlations. $\Delta R_e$ is calculated independently in each redshift slice and accounts for the fact that the median $R_e$ is not identical in each redshift slice.

Although Figure \ref{fig:rad_den_all} accurately represents the available data, it does not fully capture all available information. \gampen{} predicts the full posterior distribution of structural parameters; thus, for every galaxy in our sample, we have access to the full probability distribution function of $R_e$. Therefore, we can embark on a more statistically robust analysis than judging the existence of correlations by visual inspection alone. Rather, we use a Monte Carlo sampling technique to incorporate each of these predicted distributions into judging whether a correlation exists or not and estimate with what statistical significance we can reject the null hypothesis of no correlation. 

\begin{figure*}[htbp]
    \centering
    \includegraphics[width = 0.98\textwidth]{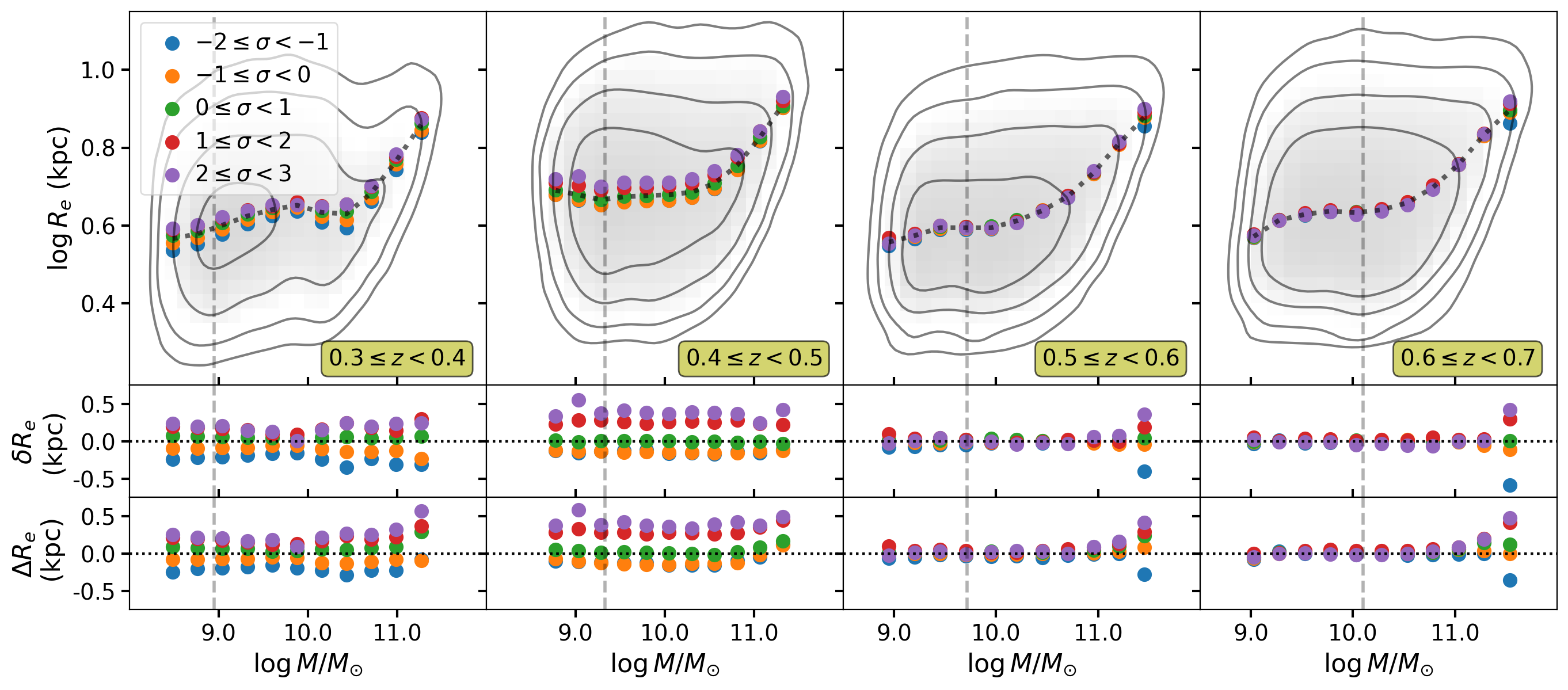}
    \caption{The size-mass relationship for the four different redshift slices. The colored points in the top panels depict the median effective radius at a given stellar mass for all galaxies within a given density excess bin (as shown in the figure legend). The contours and shading in the top panels delineate denser regions of the size-mass plane, and the black dotted lines running alongside the colored points show the overall trend. The second row shows how the colored points deviate from the overall median (black dotted) trend line in the top row, i.e., $\delta R_e = R_e - \overline{R_e}(M)$. The bottom panels show the median values of $\Delta R_e = R_e - \overline{R_e}(M,L_B/L_T)$ for galaxies in different environments at a given stellar mass. The gray dashed vertical lines throughout the figure show the overall mass-completeness in each redshift slice.}
    \label{fig:r_m_all}
\end{figure*}

Specifically, for every galaxy in our sample, we use the predicted posterior distribution to draw 5000 samples of $R_e$, effectively creating 5000 stochastic copies of our dataset. For each of these datasets, we use the Spearman rank correlation test \citep{spearman_original} to judge the existence of a correlation between $\Delta R_e$ and $\sigma_{r=10\,{\rm cMpc}}$. Spearman test is a nonparametric method to assess how well the relationship between two variables can be described using a monotonic function. The test estimates two variables, $\rho$ and $p$. $\rho$ is the correlation coefficient that can take values between $-1$ and $+1$, with the end limits signifying the variables being perfect monotonic functions of each other. $\rho=0$ signifies no correlation between the variables. The value of $p$ roughly indicates the probability of an uncorrelated system producing datasets with a Spearman correlation at least as extreme as the computed $\rho$. The above Monte Carlo procedure results in distributions of $\rho$ and $p$ values for each mass bin at the four redshift slices. We refer an interested reader to Appendix \ref{sec:ap:corr_coeff} for an extended description of the Monte Carlo procedure. 

The values of $p$ and $\rho$ obtained using the above procedure are reported in Table \ref{tab:corr_all}. Note that Table \ref{tab:corr_all} also reports the value of $\alpha$. We define $\alpha$ as the distance between the median value of $\rho$ and $\rho=0$, and report it in terms of the standard deviation ($\sigma_{\rho}$) of the $\rho$ distribution. Larger values of $\alpha$ indicate a stronger and more statistically significant correlation. The \checkmark{s} in Table \ref{tab:corr_all} signify whether we can reject the null hypothesis of noncorrelation at a significance level of $5\sigma$ or more. To assign a \checkmark, we verify whether $(\overline{p} + \sigma_p) < 3\times10^{-7} $ \texttt{AND} $\alpha>5\sigma_{\rho}$. Note that $3\times10^{-7}$ corresponds to the fractional probability threshold of $\overline{p}$ for a five-sigma detection. Cases where $10^{-12} \leq (\overline{p} + \sigma_p) < 3\times10^{-7} $ are demarcated with ``Borderline \checkmark". The choice of $10^{-12}$ as the upper limit represents a way to delineate these cases from situations where $(\overline{p} + \sigma_p) \sim \mathcal{O}(10^{-100})$.

Table \ref{tab:corr_all} and the lower row of Figure \ref{fig:rad_den_all} show that galaxies in denser environments are consistently larger than galaxies of similar mass and morphology in less dense environments. The relationship is present across the entire mass range probed by this study, but the strength of the correlation varies significantly with redshift --- the effect is strongest at lower redshifts and weakens/disappears at higher redshifts. For the two lowest-redshift bins ($0.3 \leq z <0.5 $), we can confirm the existence of a strong positive correlation between $\Delta R_e$ and $\sigma_{r=10\,{\rm cMpc}}$ with more than $5\sigma$ confidence. At these redshifts, galaxies in denser environments are $\sim10-20\%$ larger than their counterparts with similar mass and morphology in less dense environments. For $0.5 \leq z < 0.6$, the correlation is much weaker (as can be seen from the value of $\alpha$ in Table \ref{tab:corr_all}), but we have enough statistics to confirm the presence of this weak correlation with more than $5\sigma$ confidence. For $0.6 \leq z < 0.7$, the correlation can be confirmed with more than $5\sigma$ confidence only for the higher mass bin.

To investigate the variation of $R_e$ and $\Delta R_e$ with $\sigma_{r=10\,{\rm cMpc}}$ in finer mass bins, we plot the size-mass relationship for our sample in Figure \ref{fig:r_m_all}. The top row shows how galaxies with different environments are located on the size-mass plane by plotting the median $\log(R_e)$ of galaxies in different environments at a given stellar mass. The second row ($\delta R_e$ panels) shows how the colored points deviate from the median trend line of the size-mass relationship (i.e., $R_e - \overline{R_e}(M)$). The bottom panels display $\Delta R_e$ for galaxies in different environments at a given stellar mass.

For $z < 0.5$, Figure \ref{fig:r_m_all} confirms what we saw previously: we consistently find that galaxies in denser environments are larger (by $\sim 0.5-0.7$ kpc) than galaxies of similar mass and morphology in less dense environments. For the two higher-z bins, the overall effect is much weaker or absent, consistent with what we found earlier. However, Figure \ref{fig:r_m_all} adds a new dimension to our results in the $z \geq 0.5$ bins. There appears to be a ``critical stellar mass," above which we observe a strong correlation between radius and environment. The effect is significantly stronger than what can be seen for less massive galaxies at these higher redshifts. In Figure \ref{fig:r_m_all}, although this effect is visible over primarily one mass bin, we have verified that when we use even finer mass bins, we observe a steady gradual increase in this effect for $\log M/M_{\odot} > 10^{11.25} \sim 2\times10^{11}$. We ran the Monte Carlo correlation analysis for galaxies above this critical stellar mass. We observed the following values: [$\rho = 8.2\times10^{-2} \pm 4.6\times10^{-3}$, $p = 1.0\times10^{-12} \pm 3.5\times10^{-10}$, $\alpha=17\sigma_{\rho}$] for the $0.5 \leq z < 0.6$ slice; and [$\rho = 6.6\times10^{-2} \pm 2.9\times10^{-3}$, $p = 1.3\times10^{-14} \pm 1.5\times10^{-12}$, $\alpha=22\sigma_{\rho}$] for the $0.6 \leq z < 0.7$ slice. Therefore, we have more than enough statistics at these higher masses to confirm the correlation with $>5\sigma$ confidence. Note that this high critical stellar mass might explain why some of the studies in Table \ref{tab:lit_survey} did not observe any correlation at these higher redshifts --- many of them probably did not have enough statistics at these very high masses.

In the top row of Figure \ref{fig:r_m_all}, we also observe the existence of a mass-dependant slope of the size-mass relationship across all four redshift slices. The slope of the size-mass relationship is shallower at lower masses, with a pivotal stellar mass above which the slope becomes much steeper. This effect has been observed previously across a wide redshift range, and it has been posited that the pivotal stellar mass represents a mass above which both the stellar mass growth and the size growth of galaxies transition from being star-formation-dominated to dry-merger-dominated \citep[e.g.,][]{mowla19,hsc_mass_size}. The observed pivotal stellar masses ($\log M/M_{\odot} \sim 10.3-10.6$) in Figure \ref{fig:r_m_all} are consistent with what has been observed previously, and we can additionally confirm that the pivotal stellar mass does not appear to depend on environmental density.

\subsection{Specific Subpopulations} \label{sec:rad_den_sub}
\begin{figure*}[htbp]
    \centering
    \includegraphics[width = 0.9\textwidth]{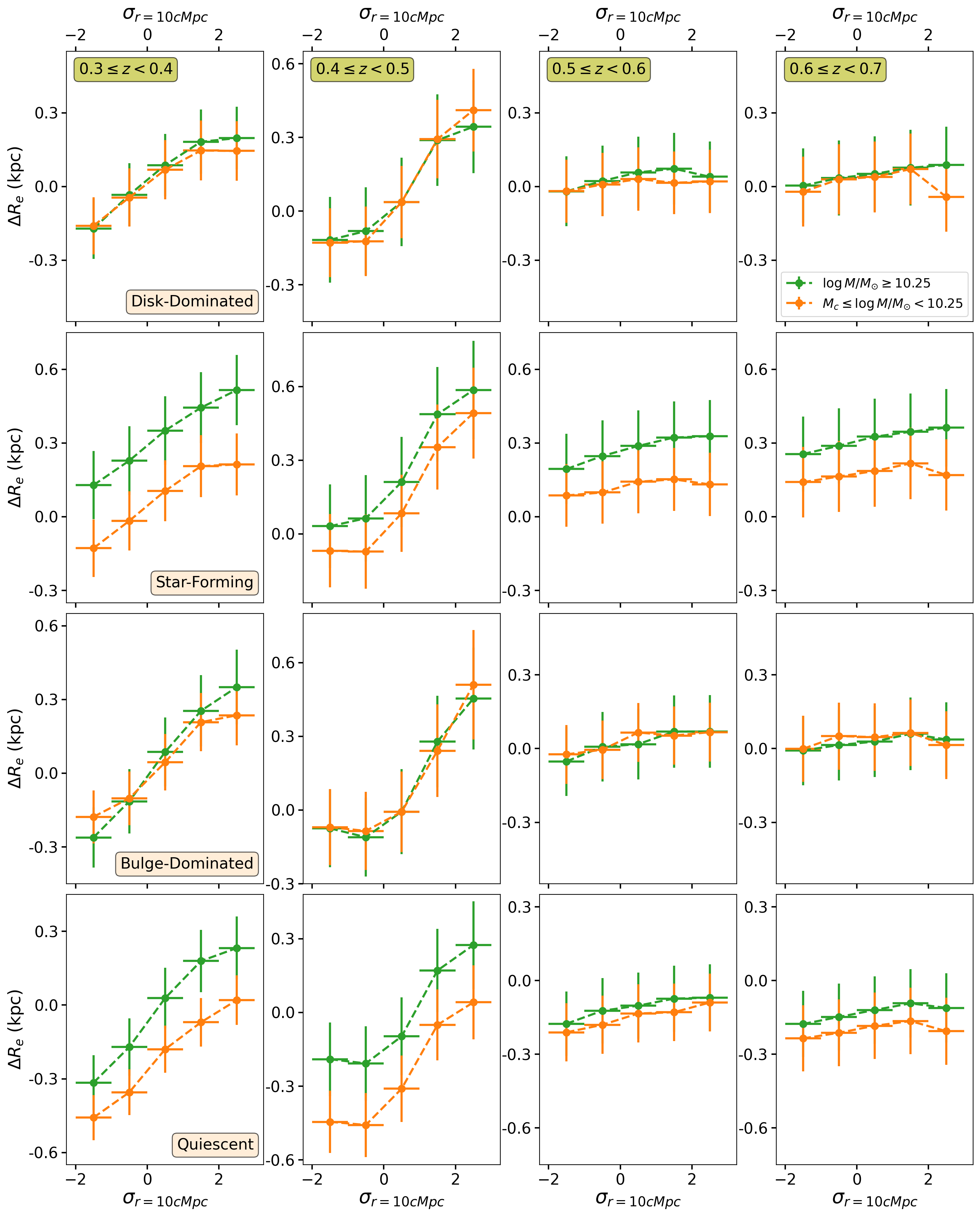}
    \caption{Variation of $\Delta R_e$ with density excess for different subpopulations of galaxies (four rows) at different redshifts (four columns). The different colored lines represent different stellar mass ranges, as denoted in the figure legend, with $M_{\rm c}$ referring to the mass-completeness limit at the corresponding redshift. The points depict the median $\Delta R_e$ in each density bin. Vertical error bars show the typical uncertainty (i.e., median $68\%$ confidence interval) predicted by \gampen{} for all galaxies in that bin. Although each panel shows a different $\Delta R_e$ range, the minimum $\Delta R_e$ scale of $0.3$ kpc is equal across all panels.}
    \label{fig:rad_den_sub}
\end{figure*}

\begin{figure*}[htbp]
    \centering
    \includegraphics[width = 0.86\textwidth]{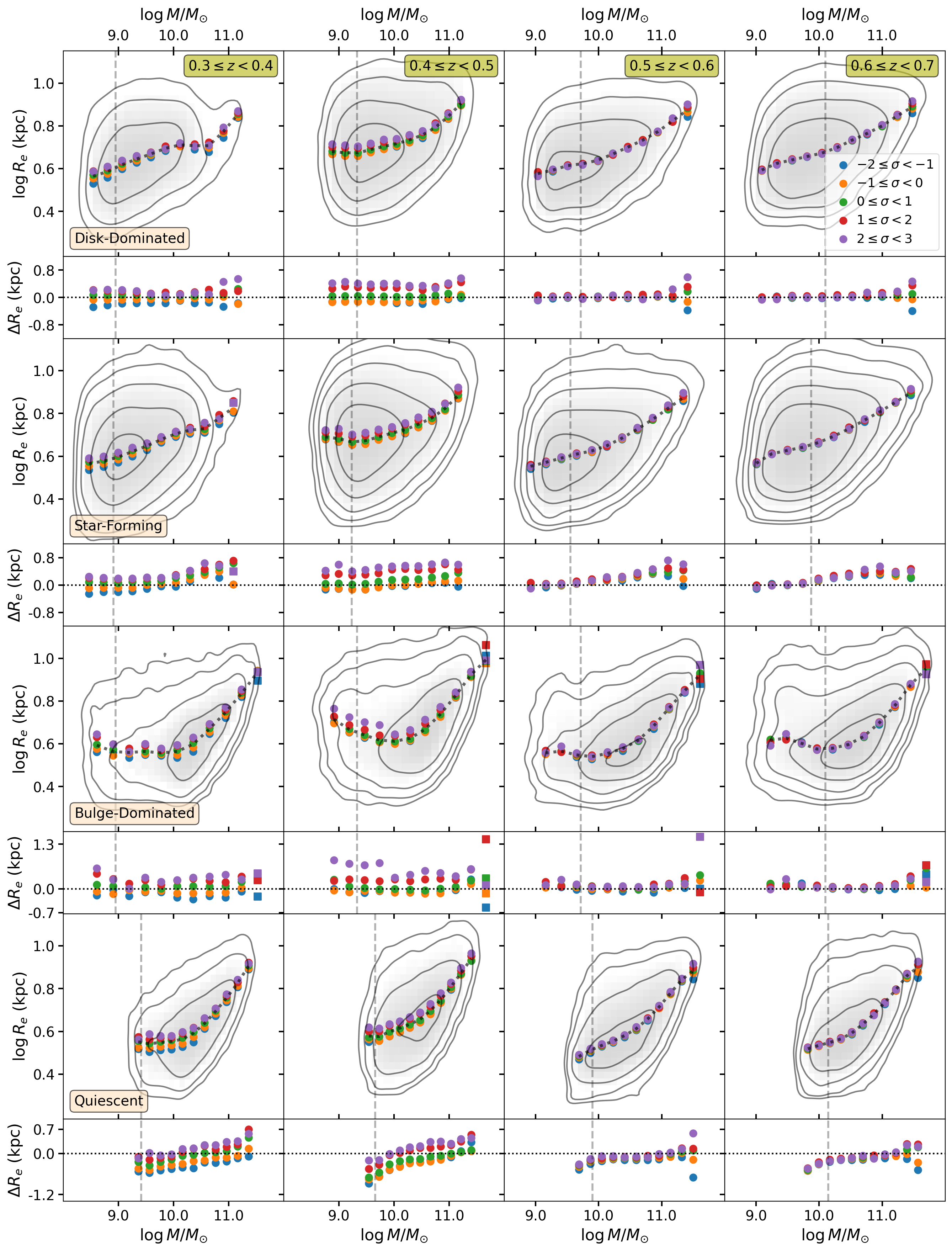}
    \caption{The size-mass relationship is shown (top panels) in the four different redshift slices (four columns) for different subpopulations of galaxies (four rows). The colored points in the top panels depict the median effective radius at a given stellar mass for all galaxies within a given density excess bin. The black dotted lines running alongside the colored points show the overall trend. The panels underneath each contour plot show the median values of $\Delta R_e = R_e - \overline{R_e}(M,L_B/L_T)$ for galaxies in different environments at a given stellar mass. The gray dashed vertical lines show the corresponding mass-completeness limit in each redshift slice. Note that some of the colored points are squares (instead of circles), indicating fewer than 100 galaxies in these bins.}
    \label{fig:r_m_sub}
\end{figure*}

In \S\ref{sec:sep_into_subsamples}, we described how we split our total sample into subpopulations of disk-dominated, bulge-dominated, star-forming, and quiescent galaxies. This section describes the results of running the same analysis we did in \S\ref{sec:rad_den_all} for each subpopulation. The $\Delta R_e$ versus $\sigma_{r=10\,{\rm cMpc}}$ plot for each subpopulation is shown in Figure \ref{fig:rad_den_sub}, and the results of the correlation analysis are reported in Table \ref{tab:corr_subpop_ab}. To be concise, we only include order-of-magnitude values in Table \ref{tab:corr_subpop_ab}, and the interested reader can refer to Appendix \ref{sec:ap:corr_coeff} for the unabridged version of this table. Finally, the size-mass diagram, plotted individually for each subpopulation, is shown in Figure \ref{fig:r_m_sub}. The format of these figures and the table is similar to what was used in the last section for the overall analysis. 

\begin{deluxetable*}{c|c|c|cccc}
\tablecaption{$\Delta R_e$ vs. $\sigma_{r=10\,{\rm cMpc}}$ Correlation Coefficients for Different Subpopulations \label{tab:corr_subpop_ab}}
\tablecolumns{7}
\tablehead{Subpopulation & Mass Range & & $0.3 \leq z < 0.4$ & $0.4 \leq z < 0.5$ & $0.5 \leq z < 0.6$ & $0.6 \leq z < 0.7$ \\ 
          & ($\log M/M_{\odot}$) & & & &}
\startdata
    \hline
    \hline
    \multirow{8}{*}{Disk-dominated} & \multirow[c]{4}{*}{$\left[M_{\rm c},10.25\right)$\tablenotemark{a}} & $\rho$   & $\om(10^{-2})$ & $\om(10^{-1})$ & $\om(10^{-2})$ & $\om(10^{-2})$ \\
                                    &                                  & $p$      & $\om(10^{-263})$ & $(<10^{-300})$ & $\om(10^{-6})$ &  $\om(10^{-3})$   \\
                                    &                                  & $\alpha$ & $172\sigma_{\rho}$ & $241\sigma_{\rho}$ & $17\sigma_{\rho}$ & $11\sigma_{\rho}$  \\
                                    & & $>5\sigma$ & \checkmark & \checkmark & &   \\
                 \cline{2-7}
                 & \multirow[c]{4}{*}{$\geq10.25$} & $\rho$   & $\om(10^{-2})$ & $\om(10^{-2})$ & $\om(10^{-2})$ & $\om(10^{-2})$ \\
                                    &             & $p$                    & $\om(10^{-43})$ & $\om(10^{-148})$ & $\om(10^{-7})$ &  $\om(10^{-11})$   \\
                                    & & $\alpha$                           & $67\sigma_{\rho}$ & $100\sigma_{\rho}$ & $21\sigma_{\rho}$ & $23\sigma_{\rho}$  \\
                                    & & $>5\sigma$ & \checkmark & \checkmark & Borderline \checkmark & Borderline \checkmark \\
    \hline
    \hline
    \multirow{8}{*}{Star-forming} & \multirow[c]{4}{*}{$\left[M_{\rm c},10.25\right)$}       & $\rho$   & $\om(10^{-2})$ & $\om(10^{-1})$ & $\om(10^{-2})$ & $\om(10^{-2})$ \\
                                    &                                     & $p$      & $(<10^{-300})$ & $(<10^{-300})$ & $\om(10^{-21})$ &  $\om(10^{-7})$   \\
                                    &                                     & $\alpha$ & $206\sigma_{\rho}$ & $265\sigma_{\rho}$ & $29\sigma_{\rho}$ & $19\sigma_{\rho}$  \\
                                    & & $>5\sigma$ & \checkmark & \checkmark & \checkmark &  Borderline \checkmark \\
                 \cline{2-7}
                 & \multirow[c]{4}{*}{$\geq10.25$} & $\rho$   & $\om(10^{-2})$ & $\om(10^{-1})$ & $\om(10^{-2})$ & $\om(10^{-2})$ \\
                                    &             & $p$                    & $\om(10^{-22})$ & $\om(10^{-177})$ & $\om(10^{-15})$ &  $\om(10^{-15})$   \\
                                    & & $\alpha$                           & $46\sigma_{\rho}$ & $105\sigma_{\rho}$ & $28\sigma_{\rho}$ & $33\sigma_{\rho}$  \\
                                    & & $>5\sigma$ & \checkmark & \checkmark & \checkmark & \checkmark \\
    \hline
    \hline
    \multirow{8}{*}{Bulge-dominated} & \multirow[c]{4}{*}{$\left[M_{\rm c},10.25\right)$} & $\rho$   & $\om(10^{-1})$ & $\om(10^{-2})$ & $\om(10^{-2})$ & $\om(10^{-3})$ \\
                                    &                                     & $p$   & $\om(10^{-94})$ & $\om(10^{-31})$ & $\om(10^{-19})$ &  $\om(10^{-1})$   \\
                                    & & $\alpha$                                  & $86\sigma_{\rho}$ & $42\sigma_{\rho}$ & $26\sigma_{\rho}$ & $2\sigma_{\rho}$  \\
                                    & & $>5\sigma$ & \checkmark & \checkmark &  \checkmark &  \\
                 \cline{2-7}
                 & \multirow[c]{4}{*}{$\geq10.25$} & $\rho$    & $\om(10^{-1})$ & $\om(10^{-2})$ & $\om(10^{-2})$ & $\om(10^{-2})$ \\
                                    &             & $p$                     & $\om(10^{-196})$ & $\om(10^{-131})$ & $\om(10^{-19})$ &  $\om(10^{-7})$   \\
                                    & & $\alpha$                            & $117\sigma_{\rho}$ & $93\sigma_{\rho}$ & $24\sigma_{\rho}$ & $14\sigma_{\rho}$  \\
                                    & & $>5\sigma$ & \checkmark & \checkmark & \checkmark & Borderline \checkmark \\
    \hline
    \hline
    \multirow{8}{*}{Quiescent} & \multirow[c]{4}{*}{$\left[M_{\rm c},10.25\right)$} & $\rho$   & $\om(10^{-1})$ & $\om(10^{-1})$ & $\om(10^{-2})$ & $\om(10^{-2})$ \\
                                    &                            & $p$      & $\om(10^{-210})$ & $\om(10^{-191})$ & $\om(10^{-17})$ &  $\om(10^{-5})$   \\
                                    &                            & $\alpha$ & $143\sigma_{\rho}$ & $119\sigma_{\rho}$ & $21\sigma_{\rho}$ & $15\sigma_{\rho}$  \\
                                    & & $>5\sigma$  & \checkmark & \checkmark &  \checkmark &  \\
                 \cline{2-7}
                 & \multirow[c]{4}{*}{$\geq10.25$} & $\rho$   & $\om(10^{-1})$ & $\om(10^{-1})$ & $\om(10^{-2})$ & $\om(10^{-2})$ \\
                                    &             & $p$                    & $\om(10^{-299})$ & $\om(10^{-273})$ & $\om(10^{-22})$ &  $\om(10^{-29})$   \\
                                    &     & $\alpha$                       & $159\sigma_{\rho}$ & $134\sigma_{\rho}$ & $29\sigma_{\rho}$ & 35$\sigma_{\rho}$  \\
                                    & & $>5\sigma$  & \checkmark & \checkmark & \checkmark & \checkmark\\
    \hline
    \hline
\enddata
\tablenotetext{a}{$M_{\rm c}$ refers to the $90\%$ stellar mass-completeness limit for each corresponding redshift slice}.
\tablecomments{The above table shows correlation coefficients measured using a Monte Carlo-based Spearman rank correlation test for different subpopulations of galaxies. The format of the table and the variables used are identical to that of Table \ref{tab:corr_all}. For formatting constraints, we only report median order-of-magnitude values for $\rho$ and $p$. The unabridged version of this table, presented in Appendix \ref{sec:ap:corr_coeff}, reports the full median and standard deviation values. A \checkmark indicates that we can confirm a positive correlation between $R_e$ and $\sigma_{r=10\,{\rm cMpc}}$ with $>5\sigma$ confidence.}
\end{deluxetable*}

As can be seen from Table \ref{tab:corr_subpop_ab} and Figure \ref{fig:rad_den_sub}, for $z < 0.5$, the positive correlation between $\Delta R_e$ and $\sigma_{r=10\,{\rm cMpc}}$ can be confirmed with more than $5\sigma$ confidence for each subpopulation for the entire mass range probed. For disk-dominated and star-forming galaxies, the effect appears to be marginally stronger in the lower mass bin compared to the higher mass bin (see $\alpha$ values in Table \ref{tab:corr_subpop_ab}). On the other hand, for bulge-dominated and quiescent systems, the effect seems to be marginally stronger at the higher-mass end compared to the lower-mass end. 

In examining the two higher-redshift slices, the correlation is weaker or absent when contrasted with the lower-redshift slices across each subpopulation. However, for all subpopulations, galaxies with $\log M/M_{\odot} > 10.25$ demonstrate a positive correlation that can be confirmed with greater than $5\sigma$ confidence. At the lower-mass end, the correlation is either nonexistent or diminished for these galaxies at higher redshifts.


In line with expectations from previous studies \citep[e.g.,][]{mowla19, hsc_mass_size}, Figure \ref{fig:r_m_sub} shows that the steeper slope of the size-mass relation at higher masses is primarily driven by bulge-dominated and quiescent systems. The change in the slope of the power law is significantly less prevalent for disk-dominated and star-forming galaxies. Even when broken down into subpopulations, we do not observe any environmental density dependence of the pivotal stellar mass at which the slope steepens in the size-mass relation.

For the $z \geq 0.5$ slices, Figure \ref{fig:r_m_sub} confirms the existence of the same critical stellar mass of $\sim 10^{11.25}M_{\odot}$ for each subpopulation, above which $\Delta R_e$ displays sensitivity to environmental density. However, we must note that because the sample was split into different groups, we can confirm this result at $>5\sigma$ confidence only for the disk-dominated and quiescent systems using our correlation analysis. This effect can only be confirmed at $\gtrapprox3\sigma$ confidence for the bulge-dominated and star-forming systems. It is also interesting to note that for galaxies above the critical mass, the $\Delta R_e$ is significantly higher for bulge-dominated and quiescent systems ($\sim1.2-1.6$ kpc) compared to disk-dominated and star-forming galaxies ($\sim0.6-1.0$ kpc). 

As Figures \ref{fig:rad_den_sub} and \ref{fig:r_m_sub} show, $\Delta R_e$ can be entirely positive/negative for certain subpopulations of galaxies and/or stellar mass ranges. Note that this is not unexpected. The definition of $\Delta R_e$ [$= R_e - \overline{R_e}(M,L_B/L_T)$] does not impose any restrictions on the variation of $\Delta R_e$ with stellar mass. The only condition enforced by the definition is that average $\Delta R_e$ should be zero when considering the \textit{entire sample} of galaxies. However, even this requirement is not true for each subpopulation individually, as $\overline{R_e}(M,L_B/L_T)$ is calculated over the entire sample and not each subpopulation separately.

We also note that the results reported for the disk-dominated (bulge-dominated) and star-forming (quiescent) subpopulations are not identical to each other. This is expected given the completely independent way these subsamples were chosen --- there is considerable overlap between the samples, but they are not identical, as shown in Table \ref{tab:sample_overlap}.

\section{Summary and Discussion} \label{sec:discussion}

In this section, we summarize the main findings from \S\ref{sec:results}, compare our results to previous studies, and reflect on possible explanations for the observed correlations. 

\subsection{Summary of Observed Correlations} \label{sec:summary}

For $0.3 \leq z < 0.5$:

\begin{itemize}
    \item We can confirm with $>5 \sigma$ confidence that galaxies in denser environments are larger than their counterparts with similar mass and morphology in less dense environments. 
    \begin{itemize}
        \item The above result holds for the entire mass range probed in this study. We cannot detect any significant monotonic change in the strength of the correlation with mass (see lower panels of Figure \ref{fig:r_m_all}). 
        \item However, the effect appears to be marginally stronger in the lower mass bin (see $\alpha$ in Table \ref{tab:corr_all}).
    \end{itemize}
    \item The $>5\sigma$ positive correlation between $\Delta R_e$ and $\sigma_{r=10\,{\rm cMpc}}$ also holds individually for disk-dominated, bulge-dominated, star-forming, and quiescent subpopulations drawn from our total sample. 
    \begin{itemize}
        \item The above result also holds for the entire mass range probed in this study. We cannot detect any significant monotonic change in the strength of the correlation with mass (see the lower panels of Figure \ref{fig:r_m_sub}).
        \item However, the effect does appear to be marginally stronger in the lower mass bin for disk-dominated/star-forming galaxies and in the higher mass bin for bulge-dominated/quiescent galaxies (see $\alpha$ in Table \ref{tab:corr_subpop_ab}).
    \end{itemize}  
\end{itemize}

For $0.5 \leq z < 0.7$:
\begin{itemize}
    \item The overall positive trend between galaxy radius and large-scale structure density is much weaker/absent compared to the lower redshifts. 
    \begin{itemize}
        \item  For $0.5 \leq z < 0.6$, we detect a $>5\sigma$ positive correlation in both the mass bins, although the strength of these correlations is much weaker than what was seen for $z < 0.5$. For $0.6 \leq z < 0.7$, we cannot detect any statistically significant correlation for the lower mass bin.
        \item There appears to be a critical stellar mass ($\sim2\times10^{11} M_\odot$) above which there is a strong preference for galaxies in denser environments to be larger. Although at the tail end of the mass distribution, we have enough statistics to confirm this effect with $>5\sigma$ confidence. 
    \end{itemize}
    \item When the sample is divided into disk-dominated, star-forming, bulge-dominated, and quiescent galaxies, we correspondingly find that the overall positive correlation seen at lower z disappears/weakens when compared within each subpopulation.
    \begin{itemize}
        \item However, for $\log M/M_{\odot} > 10.25$, we can detect a positive correlation with $>5\sigma$ confidence for both the higher-redshift slices. This effect weakens/disappears in the lower mass bin. 
        \item The existence of the previously mentioned critical stellar mass ($\sim2\times10^{11} M_\odot$) can be observed for each subpopulation, above which we see a clear correlation of radius and environmental density. The magnitude of $\Delta R_e$ above the critical stellar mass is higher for bulge-dominated and quiescent systems compared to disk-dominated and star-forming systems. 
    \end{itemize}
\end{itemize}

Given that $z=0.5$ emerges as a ``pivotal redshift'', where the nature of the correlation changes, it would be interesting to probe the $0.45 < z < 0.55$ redshift slice. However, we currently do not have \gampen{} models that work on i-band images at $z < 0.5$ and on r-band images at $z \geq 0.5$ --- making it impossible to perform the analysis consistently in a single band at $0.45 < z < 0.55$. We are currently training \gampen{} models that work on all HSC filters at all redshifts. Once those models are ready, we will present our analysis of the $0.45 < z < 0.55$ redshift slice in a future publication.

\subsection{Comparison to Previous Studies of Radius and Environment} \label{sec:comp}

The correlations detailed in \S\ref{sec:summary} help us to understand the contradictory results previously reported in the literature, as outlined in Table \ref{tab:lit_survey}. Since a galaxy's radius is primarily determined by stellar mass and morphology, examining its secondary dependence on the environment requires a large, uniform sample. Our findings emphasize that this substantial sample must be supplemented by a robust analytical framework capable of predicting and accounting for uncertainties when assessing the existence of correlations.

All studies referenced in Table \ref{tab:lit_survey} used sample sizes that are smaller by a factor of $\sim100-10,000$ compared to this work. Furthermore, none of the previously mentioned studies utilized an end-to-end Bayesian framework like \gampen{}, thus lacking a robust handle on the uncertainty in their $R_e$ measurement\footnote{As shown in \citet{hsc_wide_morphs}, light-profile fitting tools underestimate uncertainties by $15-60\%$.} nor did these studies incorporate errors into their correlation analysis in a statistically robust manner.

Therefore, given the low values of the observed correlation coefficients and their dependence on redshift, stellar mass, morphology, and state of star formation, it is not surprising that prior studies documented a range of positive, negative, and no correlations, as shown in Table \ref{tab:lit_survey}. These earlier works were limited in scope, examining a narrow segment of all galaxies rather than a comprehensive overview. If we restrict our total sample size to $\sim1000$ galaxies and repeat our analysis without taking into account $R_e$ errors, we cannot reproduce our own results with high statistical significance. Every time we draw a different sample from all galaxies and the $R_e$ distribution of each galaxy, we observe markedly varying trends.

Some of the previous studies probing $z > 0.5$ have observed the correlation solely for massive passive/quiescent/early-type galaxies \citep[e.g.,][]{Cooper12,Lani13,Bassett13}. This agrees with the fact that in our study at $z > 0.5$, the statistical significance of the observed correlation is stronger for quiescent and bulge-dominated galaxies in the higher mass bin. Additionally, \citet{Afonso19} observed the correlation only for galaxies with $\log M/M_{\odot} > 11$, which coincides with the critical stellar mass we observed for $z > 0.5$ galaxies. 

\subsection{Possible Explanations of Observed Correlations} \label{sec:theory}
Given our current understanding of the hierarchical model of galaxy formation in the $\Lambda$CDM paradigm, galaxy sizes are affected by various factors. There is no straightforward theoretical explanation for why the sizes of galaxies across different morphological types and star-formation states should depend on environmental density (at a fixed stellar mass and morphology). In fact, there is considerable disagreement among theoretical studies regarding what underlying factors could be controlling a potential correlation between galaxy size and environment. This demonstrates that comprehensive follow-up work using N-body and cosmological simulations is needed to determine the cause of the observed correlations. However, given the existing literature, it is possible to point to a few different reasons that may partially explain some of the correlations observed in \S\ref{sec:results}.

For disk-dominated systems, classical models of galaxy formation \citep[e.g.,][]{fall80,mo98} based their radial sizes on the spin of their dark matter halos and predicted that galaxy sizes should be proportional to the spin and size ($R_{vir}$) of the halo. Since then, there have been several follow-up studies using the standard abundance-matching model in N-body simulations as well as large-scale cosmological simulations \citep[e.g.,][]{Kravtsov13, somerville18, hearin_19, jiang19, yang21}, all of which have converged on the general prediction

\begin{equation}
    R_e = A R_{vir},
    \label{eq:r_e_vir}
\end{equation}

\noindent although we should note that the predicted nature of A has varied. Different studies have predicted A to be either constant or dependent on halo concentration and/or halo spin (see \citealt{wechsler_tinker} for an extended discussion). 

It is well known that in the $\Lambda$CDM paradigm, the clustering of dark matter halos depends on various halo properties in addition to halo mass \citep[e.g.,][]{wechsler02,wechsler06,gao07} -- this is commonly referred to as `assembly bias' (see \citealt{mao18} for a recent overview). As a consequence of this effect, halos (of the same mass) with higher concentration and spin are known to cluster more strongly, and strongly clustered halos typically coincide with higher-density environments.

Therefore, if the parameter $A$ in Eq. \ref{eq:r_e_vir} is indeed positively correlated with halo spin for disk-dominated galaxies, as some studies suggest \citep[e.g.,][]{yang21}, it can be expected from Eq. \ref{eq:r_e_vir} that disk-dominated galaxies in denser environments will be preferentially larger compared to equally massive counterparts in less dense environments. However, we should also note that some studies \citep[e.g.,][]{jiang19} have suggested that the sizes of disk-dominated galaxies are only weakly correlated with halo spin; but show significant anticorrelation with halo concentration (i.e., at a fixed halo mass, smaller galaxies tend to live in more concentrated halos). Note that this suggests the opposite of what we find -- that disk-dominated galaxies should be smaller in denser environments (as halos with higher concentration cluster more strongly).

\citet{somerville18} studied the redshift evolution of $R_e/(\lambda R_{vir}) \equiv A/\lambda$, where $\lambda$ is the dimensionless halo spin parameter. Note that for a given $R_{vir}$ and $\Delta \lambda$, higher values of $R_e/(\lambda R_{vir})$ will lead to higher values of $\Delta R_e$. For more massive galaxies ($\log M/M_{\odot}\sim 11$), \citet{somerville18} report that $R_e/(\lambda R_{vir})$ increases by $\sim50\%$ from $z\sim2$ to $z\sim0.3$. This is consistent with what we find overall: for disk-dominated galaxies in the higher mass bin, for a given $\Delta \sigma_{r=10\,{\rm cMpc}}$, $\Delta R_e$ is the largest at lower redshifts. Although we report a similar trend for lower-mass galaxies as well, \citet{somerville18} do not. For lower-mass galaxies, ($\log M/M_{\odot}\sim 10$), they find $R_e/(\lambda R_{vir})$ to decrease by $\sim30\%$ over the same redshift range when using the \citet{peebles_spin} definition of spin and to be constant when using the \citet{bullock_spin} definition of spin.

For bulge-dominated systems, the above arguments do not hold. However, the correlation observed in these systems may be explained by the correlation between size growth in these systems and mergers. It has been proposed by many that for bulge-dominated/quiescent systems, size growth at higher masses is dominated by minor, dry mergers, and the same conclusion has also been drawn from observational results \citep[e.g.,][]{shen03, vdw_14, mowla19, hsc_mass_size}. Some studies based on semianalytic hierarchical galaxy formation models \citep[e.g.,][]{shankar13} have suggested that bulge-dominated galaxies in denser environments should appear larger (at a fixed stellar mass) due to mergers being more efficient in denser environments. Since mergers do not always dominate size growth at lower masses \citep[e.g.,][]{yang13}, it is not surprising that the observed correlations for bulge-dominated/quiescent systems are weaker in the lower mass bin.

Mergers could also explain the existence of the critical stellar mass observed at $z \geq 0.5$. $\log M/M_{\odot} \sim 11.25$ may represent a critical stellar mass beyond which mergers play an increased role in the size and mass growth of bulge-dominated/quiescent systems at $z \geq 0.5$. This would explain why we always see a strong correlation between $\Delta R_e$ and $\sigma_{r=10\,{\rm cMpc}}$ for bulge-dominated/quiescent galaxies with $\log M/M_{\odot} \geq 11.25$. However, the same argument cannot be applied to justify the existence of the critical stellar mass we observe for disk-dominated/star-forming systems.

The observed correlations between galaxy properties (e.g., size) and environment may not indicate a causal environmental dependence. For example, it has been shown that the environmental dependence of the galaxy luminosity function is a natural outcome of an environment-independent galaxy-halo mass relation \citep{mo04}. In a similar vein, it could be argued that the existence of the critical stellar mass and/or the stronger correlations for more massive bulge-dominated/quiescent systems could have its origins in the varied scatter of the stellar mass-halo mass relation. The scatter in halo mass (for a given stellar mass) is larger at higher stellar masses than at lower stellar masses (for the stellar mass range of our sample). This large scatter in halo masses (and hence $R_{vir}$) for a given stellar mass could be causing the galaxy size and environment correlation to be stronger beyond a mass threshold for certain cases.

As the above arguments show, although some models/simulations can be selectively invoked to explain different parts of the observed size-environment correlations, there is no uniform theoretical framework to explain all of the correlations reported in this work. This demonstrates the need for more comprehensive theoretical follow-up work that can robustly investigate the correlations reported here using directly comparable simulation data sets.


\vspace{-0.05in}
\section{Conclusions} \label{sec:conclusions}

In this work, we presented one of the first comprehensive studies of the variation of galaxy radius with large-scale environmental density beyond $z \geq 0.2$. We used a comprehensive sample of $\sim3$ million HSC galaxies in the redshift range $0.3 \leq z < 0.7$ with masses down to $\log M/M_{\odot}\sim8.9$. This represents a $\sim100-10,000$ times increase in sample size and a $\sim1$ dex improvement in mass-completeness compared to most previous studies. We combined this extensive sample with robust estimates of structural parameters and associated uncertainties from \gampen{}, allowing us to perform a statistically robust correlation analysis using Monte Carlo sampling. The following are the primary takeaways. 

\begin{itemize}
    \item We confirm with $>5\sigma$ confidence that galaxy half-light radius is positively correlated with large-scale environmental density, even when controlling for stellar mass and morphology. At the scales considered in this study (10 comoving Mpc), galaxies in denser environments are up to $\sim25\%$ larger than equally massive counterparts with similar morphology in less dense environments. 
    \item We find that the strength of the above correlation depends on redshift, stellar mass, and galaxy morphology. The correlation is strongest at $z < 0.5$ and becomes systematically weaker/disappears at $z \geq 0.5$.
    \item At $z < 0.5$, we find the overall correlation to be persistent (at $>5\sigma$ confidence) across the entire stellar mass range covered in this study and individually across all subpopulations: disk-dominated, bulge-dominated, star-forming, and quiescent galaxies. For disk-dominated/star-forming galaxies, the correlation is marginally stronger at lower masses, while for bulge-dominated/quiescent galaxies, it is marginally stronger at higher masses. 
    \item At $z\geq0.5$, although the overall correlation is weaker/absent compared to lower redshifts, some noteworthy deviations exist. For galaxies with $\log M/M_{\odot} > 10.25$, we can consistently confirm the presence of a correlation with $>5\sigma$ confidence, although the strength of this correlation is much weaker compared to $z < 0.5$. We observe this correlation in the overall sample of galaxies as well as across all subpopulations: disk-dominated, bulge-dominated, star-forming, and quiescent galaxies.
    \item At $z\geq0.5$, we also find the existence of a critical stellar mass ($\log M/M_{\odot} \sim 11.25 \sim 2\times 10^{11}$), beyond which we find the existence of a strong and significant correlation. The critical stellar mass is observed to be present in the overall sample as well as individually for all studied subpopulations of galaxies. 
\end{itemize}

This comprehensive study represents an important step in resolving decades of contradictory results on the correlation between galaxy size and environmental density beyond $z \geq 0.2$. Our findings emphasize that earlier works were limited in their scope, examining a narrow segment of all galaxies rather than a comprehensive overview. These previous conflicting results were largely driven by the use of small nonuniform samples, the failure to properly account for measurement uncertainties while assessing the presence of correlations, and not properly controlling for additional factors that affect galaxy radius (e.g., stellar mass, morphology).

There is no straightforward theoretical understanding of why the sizes of galaxies across different morphological types and star-formation states should depend on environmental density (at a fixed stellar mass and morphology). That being said, we note that the reported correlations in disk-dominated galaxies could be driven by the galaxy-halo connection and halo assembly bias, while the correlations in bulge-dominated/quiescent galaxies could be driven by the varying efficiency of mergers in denser environments. The existence of the critical stellar mass could be driven by (a) the varied scatter of halo masses in the stellar mass-halo mass relation; and/or (b) the varying prevalence of mergers across different stellar mass regimes. Although the above arguments along with relevant models/simulations can be selectively invoked to explain different parts of the observed size-environment correlation, there is no uniform theoretical framework to explain all of the correlations reported in this work. 



Comparative theoretical follow-up work using N-body and cosmological simulations is needed to expand on the theoretical framework we presented above and conclusively establish the reason for the observed correlations. Such future efforts will also be greatly assisted by current/upcoming large spectroscopic programs such as the Dark Energy Spectroscopic Instrument \citep{desi}, the Subaru Prime Focus Spectrograph \citep{pfs}, and the Spectro-Photometer for the History of the Universe, Epoch of Reionization, and Ices Explorer \citep{spherex}. These spectroscopic surveys will enable us to investigate the correlation between structural parameters and environment over a comparable volume at much finer scales. They will eliminate the need to use redshift slices and enable the study of redshift-space clustering of galaxies of different sizes, masses, and morphological types. Finally, upcoming space-based wide-field surveys, such as Euclid \citep{euclid} and the Nancy Grace Roman Space Telescope \citep{ngrst} will allow us to perform similar comprehensive analyses at higher redshifts.

\section*{acknowledgments}
C.M.U. and A.G. would like to acknowledge support from the National Aeronautics and Space Administration via grant 80NSSC23K0488. 

A.G. would like to acknowledge support from the Yale Graduate School of Arts \& Sciences through the Dean's Emerging Scholars Research Award.

A.G. is supported by an LSST-DA Catalyst Fellowship; this publication was thus made possible through the support of grant 62192 from the John Templeton Foundation to LSST-DA. The opinions expressed in this publication are those of the author(s) and do not necessarily reflect the views of LSST-DA or the John Templeton Foundation.

A.G. acknowledges support from the DiRAC Institute in the Department of Astronomy at the University of Washington. The DiRAC Institute is supported through generous gifts from the Charles and Lisa Simonyi Fund for Arts and Sciences and the Washington Research Foundation.

A.G. acknowledges support from the University of Washington eScience Institute for the UW Data Science Postdoctoral Fellowship.

The Hyper Suprime-Cam (HSC) collaboration includes the astronomical communities of Japan and Taiwan and Princeton University. The HSC instrumentation and software were developed by the National Astronomical Observatory of Japan (NAOJ), the Kavli Institute for the Physics and Mathematics of the Universe (Kavli IPMU), the University of Tokyo, the High Energy Accelerator Research Organization (KEK), the Academia Sinica Institute for Astronomy and Astrophysics in Taiwan (ASIAA), and Princeton University. Funding was contributed by the FIRST program from Japanese Cabinet Office, the Ministry of Education, Culture, Sports, Science and Technology (MEXT), the Japan Society for the Promotion of Science (JSPS), Japan Science and Technology Agency (JST), the Toray Science Foundation, NAOJ, Kavli IPMU, KEK, ASIAA, and Princeton University. 

This paper makes use of software developed for the Large Synoptic Survey Telescope. We thank the LSST Project for making their code available as free software at \href{http://dm.lsst.org}{http://dm.lsst.org}.

The Pan-STARRS1 Surveys (PS1) have been made possible through contributions of the Institute for Astronomy, the University of Hawaii, the Pan-STARRS Project Office, the Max-Planck Society and its participating institutes, the Max Planck Institute for Astronomy, Heidelberg and the Max Planck Institute for Extraterrestrial Physics, Garching, The Johns Hopkins University, Durham University, the University of Edinburgh, Queen’s University Belfast, the Harvard-Smithsonian Center for Astrophysics, the Las Cumbres Observatory Global Telescope Network Incorporated, the National Central University of Taiwan, the Space Telescope Science Institute, the National Aeronautics and Space Administration under grant No. NNX08AR22G issued through the Planetary Science Division of the NASA Science Mission Directorate, the National Science Foundation under grant No. AST-1238877, the University of Maryland, and Eotvos Lorand University (ELTE), and the Los Alamos National Laboratory.

Based in part on data collected at the Subaru Telescope and retrieved from the HSC data archive system, which is operated by Subaru Telescope and Astronomy Data Center at National Astronomical Observatory of Japan.

\appendix
\section{Effect of Color Gradients on Our Results} \label{sec:ap:size_corr}
Multiple previous studies have reported that observed galaxy half-light radii are typically smaller when measured using longer-wavelength bands \citep[e.g.,][]{barbera_10,kelvin_12,vdw_14,lange_15,hsc_mass_size}. This effect has been reported across a wide range of redshifts and demonstrates that color gradients need to be accounted for while using half-light radius measurements in investigating galaxy evolution. We follow \citet{vdw_14} to quantify the relationship between galaxy sizes and the wavelength at which they were imaged. We then use this relation to correct the $R_e$ measurements to a common rest-frame wavelength of $450\,\,nm$ and investigate the impact the corrected radii have on our results. 

The general procedure for incorporating color gradients involves measuring the $R_e$ of galaxies in multiple imaging bands and then fitting a linear relationship of the form 

\begin{equation}
    \log R_{e;\mathrm{obs}} = A_{\lambda} \log \lambda_{\mathrm{rest}} + B_{\lambda}
    \label{eq:color_corr}
\end{equation}

\noindent where $A_{\lambda}$ and $B_{\lambda}$ represent the slope and intercept, respectively. $R_{e;\mathrm{obs}}$ is the observed effective radius, and $\lambda_{\mathrm{rest}}$ is the rest-frame wavelength of a galaxy imaged at redshift $z$ and corresponds to $\lambda_{\mathrm{rest}} = \lambda_{\mathrm{obs}}/(1+z)$. $\lambda_{\mathrm{obs}}$ is the effective wavelength of observation and corresponds to $621.84\,\,nm$ for HSC \rb-band imaging and $772.70\,\,nm$ for HSC \ib-band imaging. 

Note that the linear fit outlined in Eq. \ref{eq:color_corr} needs to be performed separately for quiescent and star-forming galaxies in small bins of stellar mass ($\sim 0.4$ dex) within each redshift slice. The fitted value of the slope $A_{\lambda}$ can then be used as the color gradient $\Delta \log R_e/\Delta \log \lambda$ at the given redshift and stellar mass. \citet{hsc_mass_size} extensively studied the impact of color gradients on $R_e$ measurements derived using HSC imaging. Since the mass and redshift range covered in this work is a subset of what was covered in \citet{hsc_mass_size}, we can use the best-fit values of $A_{\lambda}$ directly from Table 2 of \citet{hsc_mass_size}. 

We use these $A_{\lambda}$ values to obtain the corrected half-light radii ($R_{e;\mathrm{corr}}$) at a rest-frame of $450\,\,nm$ using the following equation:

\begin{equation}
    R_{e;\mathrm{corr}} = R_{e;\mathrm{obs}}\left( \frac{1 + z}{1+z_p} \right)^{A_{\lambda}} ,
    \label{eq:rad_corr}
\end{equation}

\noindent where $z_p$ is the pivotal redshift at which the effective wavelength of the imaging band corresponds to the rest-frame wavelength of $450\,\,nm$.

Using Eq. \ref{eq:rad_corr}, we corrected our $R_e$ measurements for all four redshift slices to a common rest-frame wavelength of $450\,\,nm$. We found the applied corrections to be pretty mild, with fractional correction $(R_{e;\mathrm{corr}} - R_{e;\mathrm{obs}})/R_{e;\mathrm{obs}}$ $\lesssim5\%$ across our entire sample. As the above section shows, the color gradient corrections applied to $R_e$ depend primarily on stellar mass and redshift, while this study addresses the variation of $R_e$ with environment within narrow bins of redshift and stellar mass. Therefore, prima facie, one should not expect the above corrections to impact our results significantly. However, to be certain, we redid the calculations for Tables \ref{tab:corr_all} and \ref{tab:corr_subpop_ab} using the corrected $R_e$ and found no change in the presence of statistically significant correlations denoted by $\checkmark$ in both these tables.

\section{Separation into Star-forming and Quiescent Subpopulations} \label{sec:ap:cc_sel}

\begin{figure*}[htbp]
    \centering
    \includegraphics[width = 0.93\textwidth]{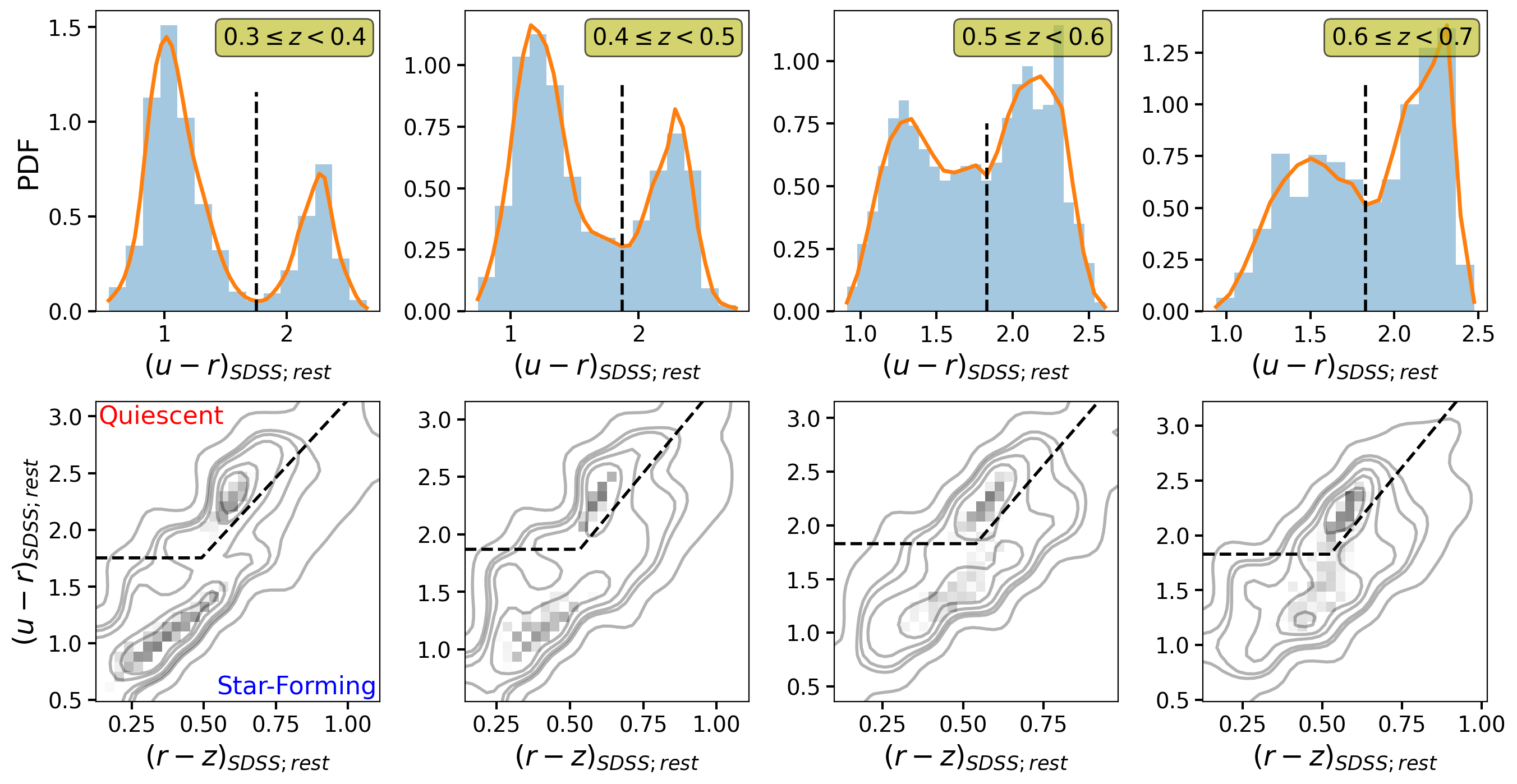}
    \caption{ (\textit{Top row}): Distribution of SDSS rest-frame \uband{}-\rb{} colors for the four redshift slices. Solid orange lines depict kernel density estimation (KDE) fits to ascertain the \uband{}-\rb{} probability density function. Dashed vertical lines mark the local minima (obtained using the KDE estimate) between the two peaks of the \uband{}-\rb{} distribution. (\textit{Bottom row}): Distribution of galaxies on the SDSS rest-frame \uband-\rb{} versus \rb{}-\zb{} color-color plane for the four redshift slices. Dashed lines delineate the boundaries between quiescent and star-forming subpopulations. See text for details on how the separation is defined.}
    \label{fig:color_sep}
\end{figure*}

To separate the galaxies in our sample into star-forming and quiescent subpopulations, we follow \citet{Kawin16, hsc_mass_size}. We begin by defining a generic region on the \textit{urz} color-color diagram for quiescent galaxies:

\begin{equation}
    \begin{aligned}
        u-r & >A \times(r-z)+\mathrm{ZP} , \\
        u-r & >(u-r)^{\prime} , \\
        r-z & <1.15 ,
\end{aligned}
\label{eq:color-color}
\end{equation}

\noindent where A, ZP, and $(u-r)^{\prime}$ are variables that we derive using galaxies in each redshift slice with stellar masses above the overall stellar mass-completeness limit. 

\begin{deluxetable}{cccc}[htbp]
\tablecaption{Best-fit Values For Eq. \ref{eq:color-color}  \label{tab:color-color}}
\tablecolumns{4}
\tablehead{
\colhead{Redshift Slice} & \colhead{$(u-r)^{\prime}$} & \colhead{A} & \colhead{ZP}}
\startdata
    \hline
    \hline
    $0.3 \leq z < 0.4$ & 1.75 & 2.75 & 0.40 \\
    $0.4 \leq z < 0.5$ & 1.87 & 3.07 & 0.23 \\
    $0.5 \leq z < 0.6$ & 1.83 & 3.48 & -0.05 \\
    $0.6 \leq z < 0.7$ & 1.83 & 3.52 & -0.02 \\
\enddata
\end{deluxetable}

First, we measure $(u-r)^{\prime}$ as the local minimum between the two peaks observed in the \uband-\rb{} color distribution (see the top panels of Figure \ref{fig:color_sep}). Subsequently, we derive the value of $A$ in Eq. \ref{eq:color-color} by fitting the slope of the red sequence in the \textit{urz} diagram. Once the slope of the line has been fixed, we need to place it in an optimal location on the diagram. For this, we measure the \textit{urz} color distance of all galaxies from the line defined by the slope $A$ in Eq. \ref{eq:color-color}. Finally, we set the zero-point $ZP$ as the local minimum between the two peaks in the measured \textit{urz} color distance distribution. 

The bottom panels of Figure \ref{fig:color_sep} show the final separation boundaries obtained for each redshift slice. The final fitted values for all the parameters in Eq. \ref{eq:color-color} are also shown in Table \ref{tab:color-color}. We refer the interested reader to \citet{Kawin16} for a more detailed description of the abovementioned procedure.

\section{Analyzing the effect of Stellar Mass Variations on Our Correlation Analysis \label{sec:ap:no_bin}}

Here we investigate the impact of stellar mass variations across different density environments on the correlations reported in \S \ref{sec:results}. It is well known that the size of a galaxy is mainly determined by its stellar mass through the size-mass relationship \citep[e.g.,][]{kormendy_77}. Therefore, while determining the variation of $R_e$ with environment, it is important to control for stellar mass.

\begin{figure*}[htbp]
    \centering
    \includegraphics[width = 0.9\textwidth]{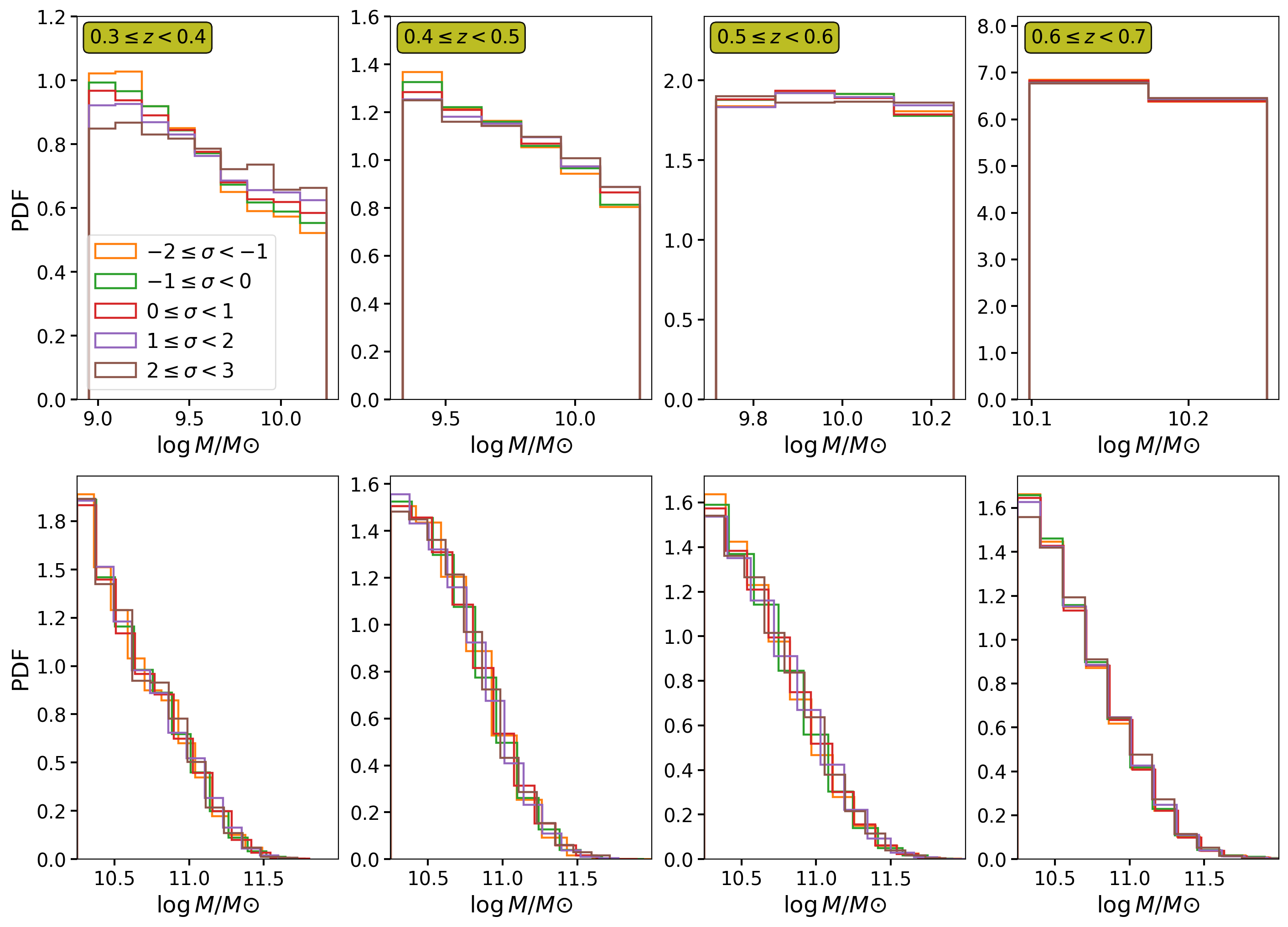}
    \caption{Probability density functions (i.e., normalized distributions) of stellar mass in different density environments. The top row shows galaxies with $M_{\rm c} \leq \log M/M_{\odot} < 10.25$  and the bottom row shows galaxies with $\log M/M_{\odot} \geq 10.25$. Each row shows the distributions on a different redshift slice.}
    \label{fig:mass_distr}
\end{figure*}

In Figure \ref{fig:rad_den_all}, we divide the sample into two mass bins while investigating the variation of $R_e$ with \customsig. Figure \ref{fig:mass_distr} shows the normalized distribution (probability density function) of the stellar mass in the different density environments for the two mass bins used in \ref{fig:rad_den_all}. The top row shows galaxies with $M_{\rm c} \leq \log M/M_{\odot} < 10.25$  and the bottom row shows galaxies with $\log M/M_{\odot} \geq 10.25$. As can be seen from Figure \ref{fig:mass_distr}, the stellar mass distributions are reasonably well matched, although there are noticeable variations between the highest- and lowest-density environments in some panels (e.g., the lower mass bin of $0.3 \leq z < 0.4$). Below, we describe a test we carried out to assess the impact of the minute differences in the stellar mass distributions shown in Figure \ref{fig:mass_distr}.

\begin{figure}[htb]
    \centering
    \includegraphics[width = 0.4\textwidth]{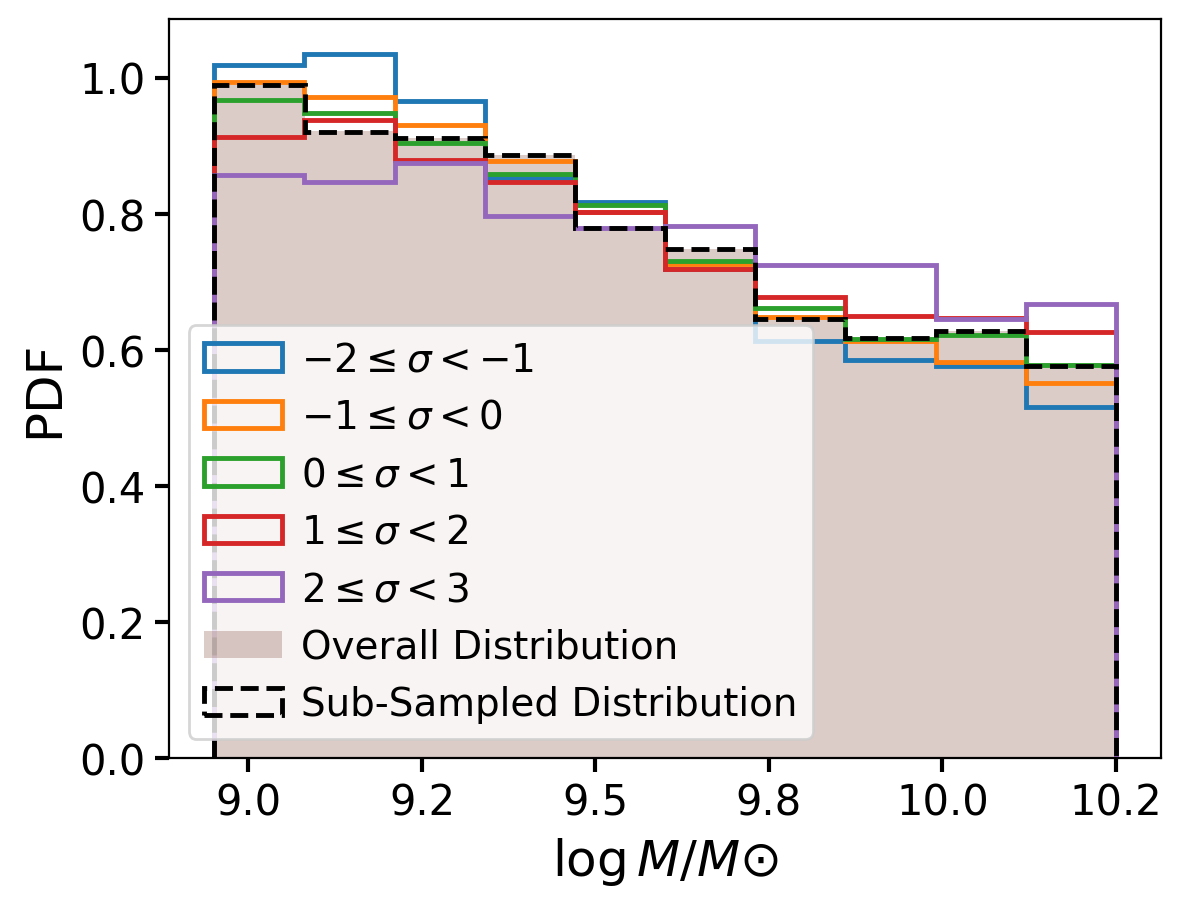}
    \caption{The colored lines show the probability density functions (i.e., normalized distributions) of galaxies in different environments for $0.3 \leq z < 0.4$ galaxies with $8.95 \leq \log M/M_{\odot} < 10.25$. The shaded histogram shows the overall distribution of galaxies in the above sample. We draw samples from each of the colored histograms to produce corresponding subsamples for each density environment such that all subsamples have a common distribution (denoted by the dotted histogram) that follows the overall distribution.}
    \label{fig:sub_sampling}
\end{figure}

We first calculate the overall distribution of galaxies in each panel in Figure \ref{fig:mass_distr}. Thereafter, we subsample galaxies from each of the individual density environments such that the mass distribution of galaxies in each density environment matches the overall distribution and is thus identical. As a demonstration of this procedure, we show the overall distribution of galaxies and the subsampled distribution for the lower mass bin at $0.3 \leq z < 0.4$ in Figure \ref{fig:sub_sampling}.

\begin{figure*}[htbp]
    \centering
    \includegraphics[width = 0.9\textwidth]{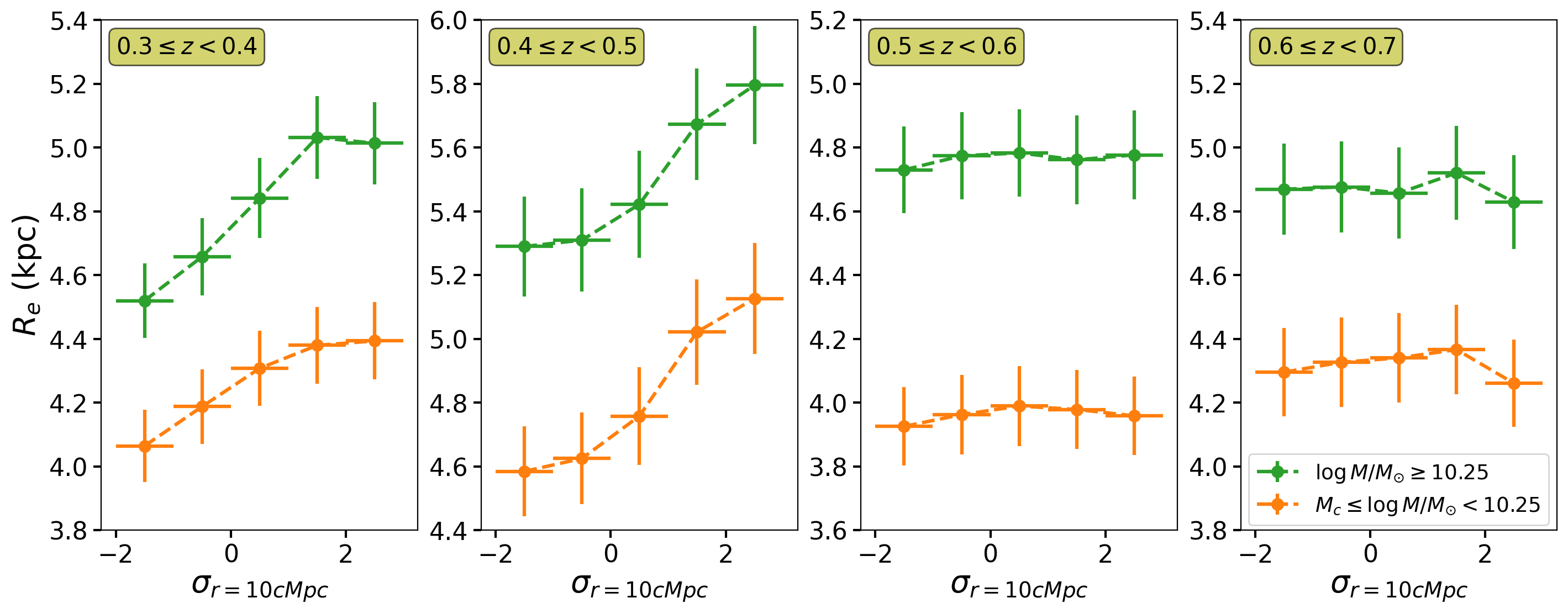}
    \caption{Variation of effective radius with density excess for a subsample of galaxies with identical mass distributions across the different density environments (see text for more details). Each panel shows the relation at a different redshift slice, and the different colored lines represent different stellar mass ranges, as denoted in the figure legend, with $M_{\rm c}$ referring to the mass-completeness limit at the corresponding redshift. The vertical error bars depict the typical uncertainty (i.e., median $68\%$ confidence interval) predicted by \gampen{} for all galaxies in that bin.}
    \label{fig:rad_den_all_matched}
\end{figure*}

Thereafter, using these new subsampled collections of galaxies, we recreate the top row of Figure \ref{fig:rad_den_all}, and this is shown as Figure \ref{fig:rad_den_all_matched}. As can be seen, this is almost identical to the top row of Figure \ref{fig:rad_den_all}. We also redid the calculations for Table \ref{tab:corr_all} using the subsampled galaxies and found no change to the presence of statistically significant correlations denoted by $\checkmark$s in this table.

The results of the above test demonstrate that stellar mass is well controlled in our correlation analysis, and the results reported in \S \ref{sec:rad_den_all} cannot be an artifact of mass binning or driven by variation in stellar mass distributions across the different density environments.

\section{Distribution of $L_B/L_T$ across the Redshift Slices \label{sec:ap:lb_lt_var}}
In \S \ref{sec:results}, we introduced the $\Delta R_e = R_e - \overline{R_e}(M,L_B/L_T)$ parameter to properly control for stellar mass and morphology in our correlation analysis. Controlling for $L_B/L_T$ is important given that morphology affects the size of a galaxy and denser regions of the Universe contain more elliptical galaxies. An additional reason that $L_B/L_T$ should be controlled for is that the distribution of $L_B/L_T$ is not identical across the four different redshift slices considered in our sample. 

\begin{figure*}[htbp]
    \centering
    \includegraphics[width = 0.95\textwidth]{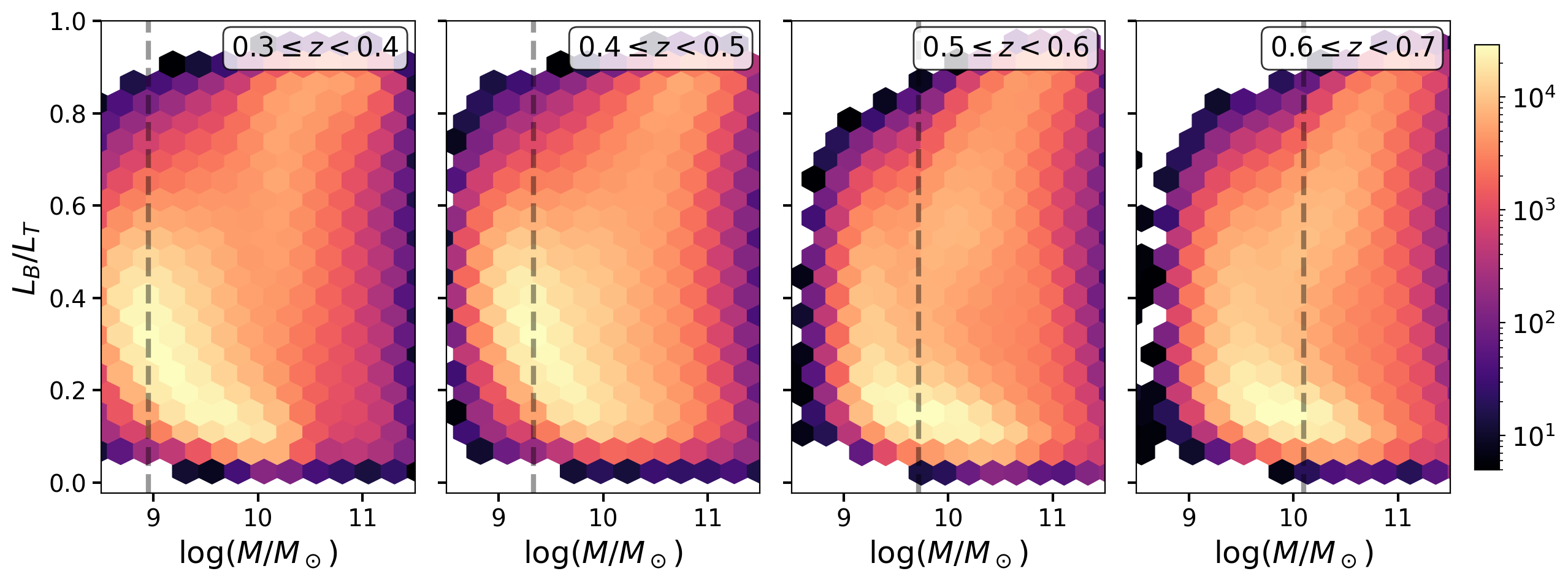}
    \caption{Distribution of bulge-to-total light ratio against stellar mass across the four different redshift slices. The figure shows galaxies in hexagonal bins of roughly equal size, with the number of galaxies in each bin represented according to the color bar on the right. Note that we are using a logarithmic color bar to explore the full distribution of galaxies, down to five galaxies per bin. The black dashed vertical lines show the overall mass-completeness in each redshift slice.}
    \label{fig:lb_lt_mass}
\end{figure*}

Figure \ref{fig:lb_lt_mass} shows the distribution of $L_B/L_T$ against stellar mass across the four redshift slices. Since we use a mass-complete sample for our correlation analysis, we only include galaxies that are to the right of the dashed vertical line in each panel. Therefore, the marginalized distribution of $L_B/L_T$ differs across the four redshift slices. Especially, as we push to higher redshifts, the relative number of galaxies with intermediate $L_B/L_T$ values (e.g., $0.25 < L_B/L_T < 0.65$) decreases. Therefore, to make a fair comparison across the different redshift slices, it is important to control for $L_B/L_T$ by using $\Delta R_e$ instead of $R_e$ for the correlation analysis.  

\color{black}

\section{Distribution of Effective Radii} \label{sec:ap:rad_distr}
In Figure \ref{fig:rad_den_all} of \S \ref{sec:rad_den_all}, the average $R_e$ for the $0.4 \leq z < 0.5$ redshift slice appears to be significantly different compared to the other slices. We demonstrate below that this is primarily because we do not plot the full range of $R_e$ values in Figure \ref{fig:rad_den_all} and zoom in on the region around median $R_e$ for each redshift slice. 

\begin{figure*}[htbp]
    \centering
    \includegraphics[width = 0.6\textwidth]{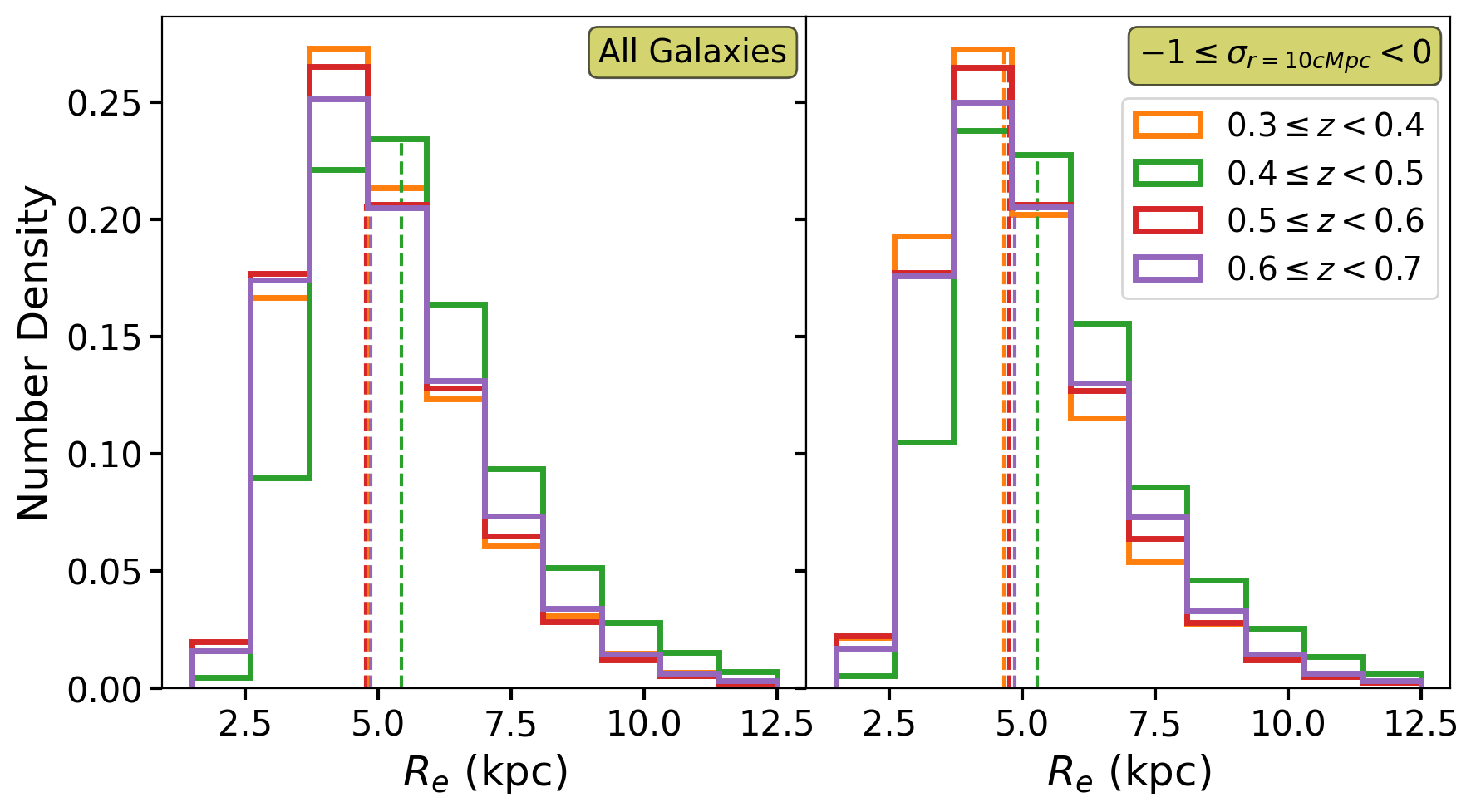}
    \caption{(\textit{Left}): Distribution of effective radii for the overall sample in each redshift slice. (\textit{Right}): The same distribution for galaxies in regions where $-1 \leq \sigma_{r=10cMpc} < 0$. Vertical dashed lines show the median of each distribution.}
    \label{fig:size_dist}
\end{figure*}

As shown in Figure \ref{fig:size_dist}, the full range of $R_e$ values spans a larger portion of the parameter space, and this discrepancy is $<5\%$ when considering the full range of $R_e$. As we have also shown in Figure \ref{fig:size_dist}, the discrepancy in median $R_e$ for a specific environmental bin (e.g., $-1 \leq \sigma_{r=10cMpc} < 0$) largely tracks the discrepancy in the overall sample. Therefore, this is not an artifact of our analysis pipeline but rather a by-product of our sample selection process. 

It just happens that a combination of the sample selection cuts summarized in \S 2.1 along with the mass-completeness cuts leads to a dataset where the median $R_e$ in the first, third, and fourth redshift slices are almost equal, and the median $R_e$ in the second redshift slice is slightly larger. We verified that this is not a by-product of a single sample selection cut -- in fact, the median $R_e$ values shown in Figure \ref{fig:size_dist} change gradually throughout the sample selection process. The above also shows why we chose to use $\Delta R_e$ (instead of $R_e$) as the primary parameter to investigate the presence of correlations. $\Delta R_e$ is calculated independently in each redshift slice and takes into account the fact that the median $R_e$ is not identical in each redshift slice.

\section{Calculating Correlation Coefficients \label{sec:ap:corr_coeff}}

As outlined previously in \S\ref{sec:rad_den_all}, we use the Spearman rank correlation test \citep{spearman_original} to judge the existence of a correlation between $R_e$ and $\sigma_{r=10\,{\rm cMpc}}$. Spearman's test, widely used in astronomy, is a nonparametric method to assess how well the relationship between two variables can be described using a monotonic function. The test estimates two variables, $\rho$ and $p$. The correlation coefficient, $\rho$, has a value between $-1$ and $+1$, with the end limits signifying the variables being perfect monotonic functions of each other; $\rho=0$ signifies no correlation between the variables. The value of $p$ roughly indicates the probability of an uncorrelated system producing datasets with a Spearman correlation at least as extreme as the computed $\rho$.

In astronomical studies, the Spearman rank correlation coefficient is often quoted without any estimate of the error in its value. The $R_e$ measurements used in this study are accompanied by robust uncertainties predicted by \gampen{}; therefore, we use a Monte Carlo-based method to incorporate these uncertainties into our calculation of correlation coefficients. This provides us with a robust statistical framework to assess the level of correlation present in the data and the level of statistical certainty with which we can reject the null hypothesis of no correlation being present in the data. Note that Monte Carlo-based methods have been used previously in the literature to augment Spearman's rank correlation test, and our method below closely follows the method outlined in \citet{curran_14}.

For every galaxy in our sample, we can access the posterior distribution of $R_e$ predicted by \gampen{}. From this posterior distribution, we draw 5000 samples of $R_e$ for every galaxy. This effectively creates 5000 different data sets, each containing $\sim3$ million galaxies. In each of these 5000 data sets, we partition the data identically as outlined in \S\ref{sec:rad_den_all} based on their redshift and stellar mass. Within each redshift and stellar mass bin, we measure $\rho$ and $p$ using Spearman's rank correlation test. Therefore, for each redshift and stellar mass partition, we have a distribution of 5000 different values of the correlation coefficient $\rho$, each accompanied by the statistical significance parameter $p$. We visually inspect these distributions to ensure that they are well approximated by Gaussian distributions and report the median and standard deviations of $\rho$ and $p$ in Tables \ref{tab:corr_all} \& \ref{tab:corr_subpop_ab}. Table \ref{tab:corr_subpop} shows the full version of Table \ref{tab:corr_subpop_ab}.

\movetabledown=2in
\movetableright=-2in
\begin{rotatetable*}
\begin{deluxetable*}{c|c|c|cccc}
\tablecaption{$\Delta R_e$ vs. Density Correlation Coefficients for Different Subpopulations. Unabridged Version of Table \ref{tab:corr_subpop_ab} \label{tab:corr_subpop}}
\tablecolumns{7}
\tablehead{Subpopulation & Mass Range & & $0.3 \leq z < 0.4$ & $0.4 \leq z < 0.5$ & $0.5 \leq z < 0.6$ & $0.6 \leq z < 0.7$ \\ 
          & ($\log M/M_{\odot}$) & & & &}
\startdata
    \hline
    \hline
    \multirow{8}{*}{Disk-dominated} & \multirow[c]{4}{*}{$\left[M_{\rm c},10.25\right)$} & $\rho$   & $7.6\times10^{-2} \pm 4.4\times10^{-4}$ & $1.2\times10^{-1} \pm 4.8\times10^{-4}$ & $1.3\times10^{-2} \pm 7.5\times10^{-4}$ & $1.2\times10^{-2} \pm 1.0\times10^{-3}$ \\
                                    &                                     & $p$  & $6.8\times10^{-263} \pm (<10^{-300})$ & 
                                    $(<10^{-300}) \pm (<10^{-300})$ & 
                                    $2.6\times10^{-6} \pm 9.8\times10^{-6}$ &  
                                    $3.9\times10^{-3} \pm 4.5\times10^{-3}$   \\
                                    &                                 & $\alpha$ & $172\sigma_{\rho}$ & $241\sigma_{\rho}$ & $17\sigma_{\rho}$ & $11\sigma_{\rho}$  \\
                                    & & $>5\sigma$ & \checkmark & \checkmark &  & \\
                 \cline{2-7}
                 & \multirow[c]{4}{*}{$\geq10.25$} & $\rho$   & $9.1\times10^{-2} \pm 1.3\times10^{-3}$ & $9.6\times10^{-2} \pm 9.6\times10^{-4}$ & $1.7\times10^{-2} \pm 8.1\times10^{-4}$ & $1.8\times10^{-2} \pm 7.7\times10^{-4}$ \\
                                    &             & $p$                    & $6.3\times10^{-46} \pm 2.9\times10^{-43}$ & $2.1\times10^{-158} \pm 1.0\times10^{-148}$ & $2.5\times10^{-8} \pm 1.5\times10^{-7}$ &  $1.1\times10^{-13} \pm 1.8\times10^{-11}$   \\
                                    & & $\alpha$                           & $67\sigma_{\rho}$ & $100\sigma_{\rho}$ & $21\sigma_{\rho}$ & $23\sigma_{\rho}$  \\
                                    & & $>5\sigma$ & \checkmark & \checkmark &  Borderline \checkmark & Borderline \checkmark \\
    \hline
    \hline
    \multirow{8}{*}{Star-forming} & \multirow[c]{4}{*}{$\left[M_{\rm c},10.25\right)$}       & $\rho$   & $8.2\times10^{-2} \pm 3.9\times10^{-4}$ & $1.1\times10^{-1} \pm 4.2\times10^{-4}$ & $2.4\times10^{-2} \pm 7.9\times10^{-4}$ & $1.3\times10^{-2} \pm 6.9\times10^{-4}$ \\
                                    &                                     & $p$      & $(<10^{-300}) \pm (<10^{-300})$ & 
                                    $(<10^{-300}) \pm (<10^{-300})$ & 
                                    $3.1\times10^{-25} \pm 9.8\times10^{-21}$ &  $1.1\times10^{-8} \pm 1.2\times10^{-7}$   \\
                                    &                                     & $\alpha$ & $206\sigma_{\rho}$ & $265\sigma_{\rho}$ & $29\sigma_{\rho}$ & $19\sigma_{\rho}$  \\
                                    & & $>5\sigma$ & \checkmark & \checkmark & \checkmark &  Borderline \checkmark \\
                 \cline{2-7}
                 & \multirow[c]{4}{*}{$\geq10.25$} & $\rho$   & $8.4\times10^{-2} \pm 1.8\times10^{-3}$ & $1.1\times10^{-1} \pm 1.0\times10^{-3}$ & $2.9\times10^{-2} \pm 1.0\times10^{-3}$ & $2.2\times10^{-2} \pm 6.7\times10^{-4}$ \\
                                    &             & $p$                    & $3.5\times10^{-24} \pm 1.8\times10^{-22}$ & $6.5\times10^{-177} \pm (<10^{-300})$ & $3.3\times10^{-17} \pm 4.4\times10^{-15}$ &  $3.0\times10^{-17} \pm 2.3\times10^{-15}$   \\
                                    & & $\alpha$                          & $46\sigma_{\rho}$ & $105\sigma_{\rho}$ & $28\sigma_{\rho}$ & $33\sigma_{\rho}$  \\
                                    & & $>5\sigma$ & \checkmark & \checkmark & \checkmark & \checkmark \\
    \hline
    \hline
    \multirow{8}{*}{Bulge-dominated} & \multirow[c]{4}{*}{$\left[M_{\rm c},10.25\right)$} & $\rho$   & $1.1\times10^{-1} \pm 1.3\times10^{-3}$ & $6.6\times10^{-2} \pm 1.6\times10^{-3}$ & $5.7\times10^{-2} \pm 2.1\times10^{-3}$ & $7.5\times10^{-3} \pm 2.9\times10^{-3}$ \\
                                    &                                     & $p$   & $5.2\times10^{-101} \pm 8.1\times10^{-94}$ & $3.6\times10^{-36} \pm 2.7\times10^{-31}$ & $7.3\times10^{-22} \pm 8.0\times10^{-19}$ &  $3.7\times10^{-1} \pm 1.8\times10^{-1}$   \\
                                    & & $\alpha$                                  & $86\sigma_{\rho}$ & $42\sigma_{\rho}$ & $26\sigma_{\rho}$ & $2\sigma_{\rho}$  \\
                                    & & $>5\sigma$ & \checkmark & \checkmark &  \checkmark &  \\
                 \cline{2-7}
                 & \multirow[c]{4}{*}{$\geq10.25$} & $\rho$   & $1.5\times10^{-1} \pm 1.3\times10^{-3}$ & $9.7\times10^{-2} \pm 1.0\times10^{-3}$ & $3.8\times10^{-2} \pm 1.5\times10^{-3}$ & $1.8\times10^{-2} \pm 1.2\times10^{-3}$ \\
                                    &             & $p$                    & $1.3\times10^{-196} \pm (<10^{-300})$ & $4.4\times10^{-139} \pm 6.2\times10^{-131}$ & $5.1\times10^{-23} \pm 6.2\times10^{-19}$ &  $4.9\times10^{-9} \pm 3\times10^{-7}$   \\
                                    & & $\alpha$                           & $117\sigma_{\rho}$ & $93\sigma_{\rho}$ & $24\sigma_{\rho}$ & $14\sigma_{\rho}$  \\
                                    & & $>5\sigma$ & \checkmark & \checkmark & \checkmark & Borderline \checkmark\\
    \hline
    \hline
    \multirow{8}{*}{Quiescent} & \multirow[c]{4}{*}{$\left[M_{\rm c},10.25\right)$} & $\rho$   & $1.4\times10^{-1} \pm 9.8\times10^{-4}$ & $1.2\times10^{-1} \pm 1.0\times10^{-3}$ & $4.2\times10^{-2} \pm 2.0\times10^{-3}$ & $2.8\times10^{-2} \pm 1.8\times10^{-3}$ \\
                                    &                            & $p$      & $3.6\times10^{-210} \pm (<10^{-300})$ & $9.4\times10^{-191} \pm (<10^{-300})$ & $1.4\times10^{-21} \pm 3.5\times10^{-17}$ &  $4.1\times10^{-6} \pm 1.9\times10^{-5}$   \\
                                    & & $\alpha$                            & $143\sigma_{\rho}$ & $119\sigma_{\rho}$ & $21\sigma_{\rho}$ & $15\sigma_{\rho}$  \\
                                    & & $>5\sigma$ & \checkmark & \checkmark &  \checkmark &  \\
                 \cline{2-7}
                 & \multirow[c]{4}{*}{$\geq10.25$} & $\rho$   & $1.4\times10^{-1} \pm 9.0\times10^{-4}$ & $1.0\times10^{-1} \pm 7.8\times10^{-4}$ & $3.0\times10^{-2} \pm 1.0\times10^{-3}$ & $2.7\times10^{-2} \pm 7.1\times10^{-4}$ \\
                                    &             & $p$                    & $1.4\times10^{-299} \pm (<10^{-300})$ & $4.9\times10^{-273} \pm (<10^{-300})$ & $1.8\times10^{-27} \pm 1.9\times10^{-22}$ &  $2.6\times10^{-34} \pm 1.1\times10^{-29}$   \\
                                    & & $\alpha$                           & $159\sigma_{\rho}$ & $134\sigma_{\rho}$ & $29\sigma_{\rho}$ & 35$\sigma_{\rho}$  \\
                                    & & $>5\sigma$ & \checkmark & \checkmark & \checkmark & \checkmark \\
    \hline
    \hline
\enddata
\end{deluxetable*}
\end{rotatetable*}

\software{Numpy \citep{numpy},
          Scipy \citep{Virtanen2020SciPyPython},
          Astropy \citep{astropy_1,astropy_2},
          Pandas \citep{pandas},
          Matplotlib \citep{matplotlib},
          }

\bibliographystyle{aasjournal}
\bibliography{references}

\end{CJK*}
\end{document}

%% file: bangla_commands.tex
\font\bngxi=bang10 scaled 1100

\def\*#1*#2{o\null{#2}{#1}}

\def\d#1{\oalign{\smash{#1}\crcr\hidewidth{$\!$\rm.}\hidewidth}}

\def\sh#1{\setbox0=\hbox{#1}%
     \kern-.02em\copy0\kern-\wd0
     \kern.04em\copy0\kern-\wd0
     \kern-.02em\raise.0433em\box0 }

%% file: main.bbl
\begin{thebibliography}{}
\expandafter\ifx\csname natexlab\endcsname\relax\def\natexlab#1{#1}\fi
\providecommand{\url}[1]{\href{#1}{#1}}
\providecommand{\dodoi}[1]{doi:~\href{http://doi.org/#1}{\nolinkurl{#1}}}
\providecommand{\doeprint}[1]{\href{http://ascl.net/#1}{\nolinkurl{http://ascl.net/#1}}}
\providecommand{\doarXiv}[1]{\href{https://arxiv.org/abs/#1}{\nolinkurl{https://arxiv.org/abs/#1}}}

\bibitem[{Afanasiev {et~al.}(2023)Afanasiev, Mei, Fu, Shankar, Amodeo, Stern, Cooke, Gonzalez, Noirot, Rettura, Wylezalek, De~Breuck, Hatch, Stanford, \& Vernet}]{afanasiev_23}
Afanasiev, A.~V., Mei, S., Fu, H., {et~al.} 2023, Astronomy {\&} Astrophysics, 670, A95, \dodoi{10.1051/0004-6361/202244634}

\bibitem[{Aihara {et~al.}(2018)Aihara, Arimoto, Armstrong, Arnouts, Bahcall, Bickerton, Bosch, Bundy, Capak, Chan, Chiba, Coupon, Egami, Enoki, Finet, Fujimori, Fujimoto, Furusawa, Furusawa, Goto, Goulding, Greco, Greene, Gunn, Hamana, Harikane, Hashimoto, Hattori, Hayashi, Hayashi, He{\l}miniak, Higuchi, Hikage, Ho, Hsieh, Huang, Huang, Ikeda, Imanishi, Inoue, Iwasawa, Iwata, Jaelani, Jian, Kamata, Karoji, Kashikawa, Katayama, Kawanomoto, Kayo, Koda, Koike, Kojima, Komiyama, Konno, Koshida, Koyama, Kusakabe, Leauthaud, Lee, Lin, Lin, Lupton, Mandelbaum, Matsuoka, Medezinski, Mineo, Miyama, Miyatake, Miyazaki, Momose, More, More, Moritani, Moriya, Morokuma, Mukae, Murata, Murayama, Nagao, Nakata, Niida, Niikura, Nishizawa, Obuchi, Oguri, Oishi, Okabe, Okamoto, Okura, Ono, Onodera, Onoue, Osato, Ouchi, Price, Pyo, Sako, Sawicki, Shibuya, Shimasaku, Shimono, Shirasaki, Silverman, Simet, Speagle, Spergel, Strauss, Sugahara, Sugiyama, Suto, Suyu, Suzuki, Tait, Takada, Takata, Tamura, Tanaka, Tanaka, Tanaka,
  Tanaka, Terai, Terashima, Toba, Tominaga, Toshikawa, Turner, Uchida, Uchiyama, Umetsu, Uraguchi, Urata, Usuda, Utsumi, Wang, Wang, Wong, Yabe, Yamada, Yamanoi, Yasuda, Yeh, Yonehara, \& Yuma}]{ssp_2}
Aihara, H., Arimoto, N., Armstrong, R., {et~al.} 2018, Publications of the Astronomical Society of Japan, 70, \dodoi{10.1093/pasj/psx066}

\bibitem[{Aihara {et~al.}(2019)Aihara, Alsayyad, Ando, Armstrong, Bosch, Egami, Furusawa, Furusawa, Goulding, Harikane, Hikage, Ho, Hsieh, Huang, Ikeda, Imanishi, Ito, Iwata, Jaelani, Kakuma, Kawana, Kikuta, Kobayashi, Koike, Komiyama, Li, Liang, Lin, Luo, Lupton, Lust, Macarthur, Matsuoka, Mineo, Miyatake, Miyazaki, More, Murata, Namiki, Nishizawa, Oguri, Okabe, Okamoto, Okura, Ono, Onodera, Onoue, Osato, Ouchi, Shibuya, Strauss, Sugiyama, Suto, Takada, Takagi, Takata, Takita, Tanaka, Terai, Toba, Uchiyama, Utsumi, Wang, Wang, \& Yamada}]{hsc_pdr2}
Aihara, H., Alsayyad, Y., Ando, M., {et~al.} 2019, Publications of the Astronomical Society of Japan, 71, 114, \dodoi{10.1093/PASJ/PSZ103}

\bibitem[{Allen {et~al.}(2015)Allen, Kacprzak, Spitler, Glazebrook, Labb{\'{e}}, Tran, Straatman, Nanayakkara, Brammer, Quadri, Cowley, Monson, Papovich, Persson, Rees, Tilvi, \& Tomczak}]{Allen15}
Allen, R.~J., Kacprzak, G.~G., Spitler, L.~R., {et~al.} 2015, The Astrophysical Journal, 806, 3, \dodoi{10.1088/0004-637X/806/1/3}

\bibitem[{Bamford {et~al.}(2009)Bamford, Nichol, Baldry, Land, Lintott, Schawinski, Slosar, Szalay, Thomas, Torki, Andreescu, Edmondson, Miller, Murray, Raddick, \& Vandenberg}]{Bamford09}
Bamford, S.~P., Nichol, R.~C., Baldry, I.~K., {et~al.} 2009, Monthly Notices of the Royal Astronomical Society, 393, 1324, \dodoi{10.1111/J.1365-2966.2008.14252.X}

\bibitem[{Bassett {et~al.}(2013)Bassett, Papovich, Lotz, Bell, Finkelstein, Newman, Tran, Almaini, Lani, Cooper, Croton, Dekel, Ferguson, Kocevski, Koekemoer, Koo, McGrath, McIntosh, \& Wechsler}]{Bassett13}
Bassett, R., Papovich, C., Lotz, J.~M., {et~al.} 2013, The Astrophysical Journal, 770, 58, \dodoi{10.1088/0004-637X/770/1/58}

\bibitem[{Blanton {et~al.}(2005)Blanton, Eisenstein, Hogg, Schlegel, \& Brinkmann}]{Blanton05}
Blanton, M.~R., Eisenstein, D., Hogg, D.~W., Schlegel, D.~J., \& Brinkmann, J. 2005, The Astrophysical Journal, 629, 143, \dodoi{10.1086/422897/FULLTEXT/}

\bibitem[{Blanton \& Moustakas(2009)}]{blanton_09}
Blanton, M.~R., \& Moustakas, J. 2009, Annual Review of Astronomy and Astrophysics, 47, 159, \dodoi{10.1146/annurev-astro-082708-101734}

\bibitem[{Brough {et~al.}(2017)Brough, van~de Sande, Owers, d’Eugenio, Sharp, Cortese, Scott, Croom, Bassett, Bekki, Bland-Hawthorn, Bryant, Davies, Drinkwater, Driver, Foster, Goldstein, L{\'{o}}pez-S{\'{a}}nchez, Medling, Sweet, Taranu, Tonini, Yi, Goodwin, Lawrence, \& Richards}]{Brough17}
Brough, S., van~de Sande, J., Owers, M.~S., {et~al.} 2017, The Astrophysical Journal, 844, 59, \dodoi{10.3847/1538-4357/aa7a11}

\bibitem[{Bullock {et~al.}(2001)Bullock, Dekel, Kolatt, Kravtsov, Klypin, Porciani, \& Primack}]{bullock_spin}
Bullock, J.~S., Dekel, A., Kolatt, T.~S., {et~al.} 2001, The Astrophysical Journal, 555, 240, \dodoi{10.1086/321477}

\bibitem[{Cappellari(2013)}]{Cappellari13}
Cappellari, M. 2013, The Astrophysical Journal Letters, 778, L2, \dodoi{10.1088/2041-8205/778/1/L2}

\bibitem[{Cappellari {et~al.}(2011)Cappellari, Emsellem, Krajnovi{\'{c}}, McDermid, Serra, Alatalo, Blitz, Bois, Bournaud, Bureau, Davies, Davis, de~Zeeuw, Khochfar, Kuntschner, Lablanche, Morganti, Naab, Oosterloo, Sarzi, Scott, Weijmans, \& Young}]{Cappelari11}
Cappellari, M., Emsellem, E., Krajnovi{\'{c}}, D., {et~al.} 2011, Monthly Notices of the Royal Astronomical Society, 416, 1680, \dodoi{10.1111/J.1365-2966.2011.18600.X}

\bibitem[{Chamba {et~al.}(2020)Chamba, Trujillo, \& Knapen}]{chamba_20}
Chamba, N., Trujillo, I., \& Knapen, J.~H. 2020, Astronomy {\&} Astrophysics, 633, L3, \dodoi{10.1051/0004-6361/201936821}

\bibitem[{Chan {et~al.}(2018)Chan, Beifiori, Saglia, Mendel, Stott, Bender, Galametz, Wilman, Cappellari, Davies, Houghton, Prichard, Lewis, Sharples, \& Wegner}]{Chan18}
Chan, J. C.~C., Beifiori, A., Saglia, R.~P., {et~al.} 2018, The Astrophysical Journal, 856, 8, \dodoi{10.3847/1538-4357/AAADB4}

\bibitem[{Chang {et~al.}(2015)Chang, Wel, Cunha, \& Rix}]{Chang15}
Chang, Y.~Y., Wel, A. V.~D., Cunha, E.~D., \& Rix, H.~W. 2015, The Astrophysical Journal Supplement Series, 219, 8, \dodoi{10.1088/0067-0049/219/1/8}

\bibitem[{Conselice(2014)}]{morph_review}
Conselice, C.~J. 2014, Annual Review of Astronomy and Astrophysics, 52, 291, \dodoi{10.1146/annurev-astro-081913-040037}

\bibitem[{Cooper {et~al.}(2012)Cooper, Griffith, Newman, Coil, Davis, Dutton, Faber, Guhathakurta, Koo, Lotz, Weiner, Willmer, \& Yan}]{Cooper12}
Cooper, M.~C., Griffith, R.~L., Newman, J.~A., {et~al.} 2012, Monthly Notices of the Royal Astronomical Society, 419, 3018, \dodoi{10.1111/J.1365-2966.2011.19938.X}

\bibitem[{Curran(2014)}]{curran_14}
Curran, P.~A. 2014.
\newblock \url{http://arxiv.org/abs/1411.3816}

\bibitem[{{DESI Collaboration} {et~al.}(2016){DESI Collaboration}, Aghamousa, Aguilar, Ahlen, Alam, Allen, Prieto, Annis, Bailey, Balland, Ballester, Baltay, Beaufore, Bebek, Beers, Bell, Bernal, Besuner, Beutler, Blake, Bleuler, Blomqvist, Blum, Bolton, Briceno, Brooks, Brownstein, Buckley-Geer, Burden, Burtin, Busca, Cahn, Cai, Cardiel-Sas, Carlberg, Carton, Casas, Castander, Cervantes-Cota, Claybaugh, Close, Coker, Cole, Comparat, Cooper, Cousinou, Crocce, Cuby, Cunningham, Davis, Dawson, de~la Macorra, De~Vicente, Delubac, Derwent, Dey, Dhungana, Ding, Doel, Duan, Ealet, Edelstein, Eftekharzadeh, Eisenstein, Elliott, Escoffier, Evatt, Fagrelius, Fan, Fanning, Farahi, Farihi, Favole, Feng, Fernandez, Findlay, Finkbeiner, Fitzpatrick, Flaugher, Flender, Font-Ribera, Forero-Romero, Fosalba, Frenk, Fumagalli, Gaensicke, Gallo, Garcia-Bellido, Gaztanaga, Fusillo, Gerard, Gershkovich, Giannantonio, Gillet, Gonzalez-de Rivera, Gonzalez-Perez, Gott, Graur, Gutierrez, Guy, Habib, Heetderks, Heetderks, Heitmann,
  Hellwing, Herrera, Ho, Holland, Honscheid, Huff, Hutchinson, Huterer, Hwang, Laguna, Ishikawa, Jacobs, Jeffrey, Jelinsky, Jennings, Jiang, Jimenez, Johnson, Joyce, Jullo, Juneau, Kama, Karcher, Karkar, Kehoe, Kennamer, Kent, Kilbinger, Kim, Kirkby, Kisner, Kitanidis, Kneib, Koposov, Kovacs, Koyama, Kremin, Kron, Kronig, Kueter-Young, Lacey, Lafever, Lahav, Lambert, Lampton, Landriau, Lang, Lauer, Goff, Guillou, Van~Suu, Lee, Lee, Leitner, Lesser, Levi, L'Huillier, Li, Liang, Lin, Linder, Loebman, Luki{\'{c}}, Ma, MacCrann, Magneville, Makarem, Manera, Manser, Marshall, Martini, Massey, Matheson, McCauley, McDonald, McGreer, Meisner, Metcalfe, Miller, Miquel, Moustakas, Myers, Naik, Newman, Nichol, Nicola, da~Costa, Nie, Niz, Norberg, Nord, Norman, Nugent, O'Brien, Oh, Olsen, Padilla, Padmanabhan, Padmanabhan, Palanque-Delabrouille, Palmese, Pappalardo, P{\^{a}}ris, Park, Patej, Peacock, Peiris, Peng, Percival, Perruchot, Pieri, Pogge, Pollack, Poppett, Prada, Prakash, Probst, Rabinowitz, Raichoor, Ree,
  Refregier, Regal, Reid, Reil, Rezaie, Rockosi, Roe, Ronayette, Roodman, Ross, Ross, Rossi, Rozo, Ruhlmann-Kleider, Rykoff, Sabiu, Samushia, Sanchez, Sanchez, Schlegel, Schneider, Schubnell, Secroun, Seljak, Seo, Serrano, Shafieloo, Shan, Sharples, Sholl, Shourt, Silber, Silva, Sirk, Slosar, Smith, Smoot, Som, Song, Sprayberry, Staten, Stefanik, Tarle, Tie, Tinker, Tojeiro, Valdes, Valenzuela, Valluri, Vargas-Magana, Verde, Walker, Wang, Wang, Weaver, Weaverdyck, Wechsler, Weinberg, White, Yang, Yeche, Zhang, Zhao, Zheng, Zhou, Zhou, Zhu, Zou, \& Zu}]{desi}
{DESI Collaboration}, Aghamousa, A., Aguilar, J., {et~al.} 2016.
\newblock \url{http://arxiv.org/abs/1611.00036}

\bibitem[{Dor{\'{e}} {et~al.}(2014)Dor{\'{e}}, Bock, Ashby, Capak, Cooray, de~Putter, Eifler, Flagey, Gong, Habib, Heitmann, Hirata, Jeong, Katti, Korngut, Krause, Lee, Masters, Mauskopf, Melnick, Mennesson, Nguyen, {\"{O}}berg, Pullen, Raccanelli, Smith, Song, Tolls, Unwin, Venumadhav, Viero, Werner, \& Zemcov}]{spherex}
Dor{\'{e}}, O., Bock, J., Ashby, M., {et~al.} 2014.
\newblock \url{http://arxiv.org/abs/1412.4872}

\bibitem[{Dressler(1984)}]{dressler_84}
Dressler, A. 1984, Annual Review of Astronomy and Astrophysics, 22, 185, \dodoi{10.1146/annurev.aa.22.090184.001153}

\bibitem[{Dutton {et~al.}(2007)Dutton, van~den Bosch, Dekel, \& Courteau}]{dutton_07}
Dutton, A.~A., van~den Bosch, F.~C., Dekel, A., \& Courteau, S. 2007, The Astrophysical Journal, 654, 27, \dodoi{10.1086/509314}

\bibitem[{Fall \& Efstathiou(1980)}]{fall80}
Fall, S.~M., \& Efstathiou, G. 1980, Monthly Notices of the Royal Astronomical Society, 193, 189, \dodoi{10.1093/mnras/193.2.189}

\bibitem[{Fogarty {et~al.}(2014)Fogarty, Scott, Owers, Brough, Croom, Pracy, Houghton, Bland-Hawthorn, Colless, Davies, Heath~Jones, Allen, Bryant, Goodwin, Green, Konstantopoulos, Lawrence, Richards, Cortese, \& Sharp}]{Fogarty14}
Fogarty, L.~M., Scott, N., Owers, M.~S., {et~al.} 2014, Monthly Notices of the Royal Astronomical Society, 443, 485, \dodoi{10.1093/MNRAS/STU1165}

\bibitem[{Gao \& White(2007)}]{gao07}
Gao, L., \& White, S. D.~M. 2007, Monthly Notices of the Royal Astronomical Society: Letters, 377, L5, \dodoi{10.1111/j.1745-3933.2007.00292.x}

\bibitem[{Ghosh {et~al.}(2020)Ghosh, Urry, Wang, Schawinski, Turp, \& Powell}]{gamornet_software_paper}
Ghosh, A., Urry, C.~M., Wang, Z., {et~al.} 2020, The Astrophysical Journal, 895, 112, \dodoi{10.3847/1538-4357/ab8a47}

\bibitem[{Ghosh {et~al.}(2022)Ghosh, Urry, Rau, Perreault-Levasseur, Cranmer, Schawinski, Stark, Tian, Ofman, Ananna, Auge, Cappelluti, Sanders, \& Treister}]{gampen_software_paper}
Ghosh, A., Urry, C.~M., Rau, A., {et~al.} 2022, The Astrophysical Journal, 935, 138, \dodoi{10.3847/1538-4357/ac7f9e}

\bibitem[{Ghosh {et~al.}(2023)Ghosh, Urry, Mishra, Perreault-Levasseur, Natarajan, Sanders, Nagai, Tian, Cappelluti, Kartaltepe, Powell, Rau, \& Treister}]{hsc_wide_morphs}
Ghosh, A., Urry, C.~M., Mishra, A., {et~al.} 2023, The Astrophysical Journal, 953, 134, \dodoi{10.3847/1538-4357/acd546}

\bibitem[{Gomez {et~al.}(2003)Gomez, Nichol, Miller, Balogh, Goto, Zabludoff, Romer, Bernardi, Sheth, Hopkins, Castander, Connolly, Schneider, Brinkmann, Lamb, SubbaRao, \& York}]{gomez_03}
Gomez, P.~L., Nichol, R.~C., Miller, C.~J., {et~al.} 2003, The Astrophysical Journal, 584, 210, \dodoi{10.1086/345593/FULLTEXT/}

\bibitem[{Graham(2019)}]{graham_19}
Graham, A.~W. 2019, Publications of the Astronomical Society of Australia, 36, e035, \dodoi{10.1017/PASA.2019.23}

\bibitem[{Greene {et~al.}(2017)Greene, Leauthaud, Emsellem, Goddard, Ge, Andrews, Brinkman, Brownstein, Greco, Law, Lin, Masters, Merrifield, More, Okabe, Schneider, Thomas, Wake, Yan, \& Drory}]{Greene17}
Greene, J.~E., Leauthaud, A., Emsellem, E., {et~al.} 2017, The Astrophysical Journal, 851, L33, \dodoi{10.3847/2041-8213/aa8ace}

\bibitem[{Gu {et~al.}(2021)Gu, Fang, Yuan, Lu, \& Liu}]{Gu21}
Gu, Y., Fang, G., Yuan, Q., Lu, S., \& Liu, S. 2021, The Astrophysical Journal, 921, 60, \dodoi{10.3847/1538-4357/AC1CE0}

\bibitem[{Harris {et~al.}(2020)Harris, Millman, van~der Walt, Gommers, Virtanen, Cournapeau, Wieser, Taylor, Berg, Smith, Kern, Picus, Hoyer, van Kerkwijk, Brett, Haldane, del R{\'{i}}o, Wiebe, Peterson, G{\'{e}}rard-Marchant, Sheppard, Reddy, Weckesser, Abbasi, Gohlke, \& Oliphant}]{numpy}
Harris, C.~R., Millman, K.~J., van~der Walt, S.~J., {et~al.} 2020, Nature 2020 585:7825, 585, 357, \dodoi{10.1038/s41586-020-2649-2}

\bibitem[{Hearin {et~al.}(2019)Hearin, Behroozi, Kravtsov, \& Moster}]{hearin_19}
Hearin, A., Behroozi, P., Kravtsov, A., \& Moster, B. 2019, Monthly Notices of the Royal Astronomical Society, 489, 1805, \dodoi{10.1093/mnras/stz2251}

\bibitem[{Holden {et~al.}(2012)Holden, Van Der~Wel, Rix, \& Franx}]{Holden12}
Holden, B.~P., Van Der~Wel, A., Rix, H.~W., \& Franx, M. 2012, The Astrophysical Journal, 749, 96, \dodoi{10.1088/0004-637X/749/2/96}

\bibitem[{Holden {et~al.}(2007)Holden, Illingworth, Franx, Blakeslee, Postman, Kelson, van~der Wel, Demarco, Magee, Tran, Zirm, Ford, Rosati, \& Homeier}]{Holden07}
Holden, B.~P., Illingworth, G.~D., Franx, M., {et~al.} 2007, The Astrophysical Journal, 670, 190, \dodoi{10.1086/521777}

\bibitem[{Holmberg(1958)}]{holmberg_58}
Holmberg, E. 1958, Meddelanden fran Lunds Astronomiska Observatorium Serie II, 136, 1

\bibitem[{Huertas-Company {et~al.}(2013{\natexlab{a}})Huertas-Company, Shankar, Mei, Bernardi, Aguerri, Meert, \& Vikram}]{Huertas-Company13_local}
Huertas-Company, M., Shankar, F., Mei, S., {et~al.} 2013{\natexlab{a}}, The Astrophysical Journal, 779, 29, \dodoi{10.1088/0004-637X/779/1/29}

\bibitem[{Huertas-Company {et~al.}(2013{\natexlab{b}})Huertas-Company, Mei, Shankar, Delaye, Raichoor, Covone, Finoguenov, Kneib, Le~F'evre, \& Povi'c}]{Huertas-Company13}
Huertas-Company, M., Mei, S., Shankar, F., {et~al.} 2013{\natexlab{b}}, Monthly Notices of the Royal Astronomical Society, 428, 1715, \dodoi{10.1093/MNRAS/STS150}

\bibitem[{Hunter(2007)}]{matplotlib}
Hunter, J.~D. 2007, Computing in Science {\&} Engineering, 9, 90, \dodoi{10.1109/MCSE.2007.55}

\bibitem[{Jiang {et~al.}(2019)Jiang, Dekel, Kneller, Lapiner, Ceverino, Primack, Faber, Macci{\`{o}}, Dutton, Genel, \& Somerville}]{jiang19}
Jiang, F., Dekel, A., Kneller, O., {et~al.} 2019, Monthly Notices of the Royal Astronomical Society, 488, 4801, \dodoi{10.1093/mnras/stz1952}

\bibitem[{Kawinwanichakij {et~al.}(2016)Kawinwanichakij, Quadri, Papovich, Kacprzak, Labb{\'{e}}, Spitler, Straatman, Tran, Allen, Behroozi, Cowley, Dekel, Glazebrook, Hartley, Kelson, Koo, Lee, Lu, Nanayakkara, Persson, Primack, Tilvi, Tomczak, \& van Dokkum}]{Kawin16}
Kawinwanichakij, L., Quadri, R.~F., Papovich, C., {et~al.} 2016, The Astrophysical Journal, 817, 9, \dodoi{10.3847/0004-637X/817/1/9}

\bibitem[{Kawinwanichakij {et~al.}(2021)Kawinwanichakij, Silverman, Ding, George, Damjanov, Sawicki, Tanaka, Taranu, Birrer, Huang, Li, Onodera, Shibuya, \& Yasuda}]{hsc_mass_size}
Kawinwanichakij, L., Silverman, J.~D., Ding, X., {et~al.} 2021, The Astrophysical Journal, 921, 38, \dodoi{10.3847/1538-4357/AC1F21}

\bibitem[{Kelkar {et~al.}(2015)Kelkar, KelkarArag{\'{o}}n-Salamanca, Gray, Maltby, Vulcani, De~Lucia, Poggianti, \& Zaritsky}]{Kelkar15}
Kelkar, K., KelkarArag{\'{o}}n-Salamanca, A., Gray, M.~E., {et~al.} 2015, Monthly Notices of the Royal Astronomical Society, 450, 1246, \dodoi{10.1093/MNRAS/STV670}

\bibitem[{Kelvin {et~al.}(2012)Kelvin, Driver, Robotham, Hill, Alpaslan, Baldry, Bamford, Bland-Hawthorn, Brough, Graham, H{\"{a}}ussler, Hopkins, Liske, Loveday, Norberg, Phillipps, Popescu, Prescott, Taylor, \& Tuffs}]{kelvin_12}
Kelvin, L.~S., Driver, S.~P., Robotham, A.~S., {et~al.} 2012, Monthly Notices of the Royal Astronomical Society, 421, 1007, \dodoi{10.1111/J.1365-2966.2012.20355.X/2/M{\_}MNRAS0421-1007-MU8.GIF}

\bibitem[{Kormendy(1977)}]{kormendy_77}
Kormendy, J. 1977, The Astrophysical Journal, 218, 333, \dodoi{10.1086/155687}

\bibitem[{Kravtsov(2013)}]{Kravtsov13}
Kravtsov, A.~V. 2013, The Astrophysical Journal, 764, L31, \dodoi{10.1088/2041-8205/764/2/L31}

\bibitem[{La~Barbera {et~al.}(2010)La~Barbera, De~Carvalho, De~La~Rosa, Lopes, Kohl-Moreira, \& Capelato}]{barbera_10}
La~Barbera, F., De~Carvalho, R.~R., De~La~Rosa, I.~G., {et~al.} 2010, Monthly Notices of the Royal Astronomical Society, 408, 1313, \dodoi{10.1111/J.1365-2966.2010.16850.X}

\bibitem[{Laigle {et~al.}(2016)Laigle, McCracken, Ilbert, Hsieh, Davidzon, Capak, Hasinger, Silverman, Pichon, Coupon, Aussel, Le~Borgne, Caputi, Cassata, Chang, Civano, Dunlop, Fynbo, Kartaltepe, Koekemoer, Le~F{\`{e}}vre, Le~Floc’h, Leauthaud, Lilly, Lin, Marchesi, Milvang-Jensen, Salvato, Sanders, Scoville, Smolcic, Stockmann, Taniguchi, Tasca, Toft, Vaccari, \& Zabl}]{cosmos_2015}
Laigle, C., McCracken, H.~J., Ilbert, O., {et~al.} 2016, The Astrophysical Journal Supplement Series, 224, 24, \dodoi{10.3847/0067-0049/224/2/24}

\bibitem[{Lange {et~al.}(2015)Lange, Driver, Robotham, Kelvin, Graham, Alpaslan, Andrews, Baldry, Bamford, Bland-Hawthorn, Brough, Cluver, Conselice, Davies, Haeussler, Konstantopoulos, Loveday, Moffett, Norberg, Phillipps, Taylor, L{\'{o}}pez-S{\'{a}}nchez, \& Wilkins}]{lange_15}
Lange, R., Driver, S.~P., Robotham, A.~S., {et~al.} 2015, Monthly Notices of the Royal Astronomical Society, 447, 2603, \dodoi{10.1093/MNRAS/STU2467}

\bibitem[{Lani {et~al.}(2013)Lani, Almaini, Hartley, Mortlock, H{\"{a}}u{\ss}ler, Chuter, Simpson, Van~der Wel, Gr{\"{u}}tzbauch, Conselice, Bradshaw, Cooper, Faber, Grogin, Kocevski, Koekemoer, \& Lai}]{Lani13}
Lani, C., Almaini, O., Hartley, W.~G., {et~al.} 2013, Monthly Notices of the Royal Astronomical Society, 435, 207, \dodoi{10.1093/MNRAS/STT1275}

\bibitem[{Lopes {et~al.}(2016)Lopes, Rembold, Ribeiro, Nascimento, \& Vajgel}]{Lopes16}
Lopes, P.~A., Rembold, S.~B., Ribeiro, A.~L., Nascimento, R.~S., \& Vajgel, B. 2016, Monthly Notices of the Royal Astronomical Society, 461, 2559, \dodoi{10.1093/MNRAS/STW1497}

\bibitem[{Mao {et~al.}(2018)Mao, Zentner, \& Wechsler}]{mao18}
Mao, Y.-Y., Zentner, A.~R., \& Wechsler, R.~H. 2018, Monthly Notices of the Royal Astronomical Society, 474, 5143, \dodoi{10.1093/mnras/stx3111}

\bibitem[{Matharu {et~al.}(2019)Matharu, Muzzin, Brammer, Van Der~Burg, Auger, Hewett, Van Der~Wel, Van~Dokkum, Balogh, Chan, Demarco, Marchesini, Nelson, Noble, Wilson, \& Yee}]{Matharu19}
Matharu, J., Muzzin, A., Brammer, G.~B., {et~al.} 2019, Monthly Notices of the Royal Astronomical Society, 484, 595, \dodoi{10.1093/MNRAS/STY3465}

\bibitem[{McKinney(2010)}]{pandas}
McKinney, W. 2010, in Proceedings of the 9th Python in Science Conference, 56--61, \dodoi{10.25080/Majora-92bf1922-00a}

\bibitem[{Miller {et~al.}(2019)Miller, Dokkum, Mowla, \& Wel}]{miller_19}
Miller, T.~B., Dokkum, P.~v., Mowla, L., \& Wel, A. v.~d. 2019, The Astrophysical Journal Letters, 872, L14, \dodoi{10.3847/2041-8213/AB0380}

\bibitem[{Miyazaki {et~al.}(2018)Miyazaki, Komiyama, Kawanomoto, Doi, Furusawa, Hamana, Hayashi, Ikeda, Kamata, Karoji, Koike, Kurakami, Miyama, Morokuma, Nakata, Namikawa, Nakaya, Nariai, Obuchi, Oishi, Okada, Okura, Tait, Takata, Tanaka, Tanaka, Terai, Tomono, Uraguchi, Usuda, Utsumi, Yamada, Yamanoi, Aihara, Fujimori, Mineo, Miyatake, Oguri, Uchida, Tanaka, Yasuda, Takada, Murayama, Nishizawa, Sugiyama, Chiba, Futamase, Wang, Chen, Ho, Liaw, Chiu, Ho, Lai, Lee, Jeng, Iwamura, Armstrong, Bickerton, Bosch, Gunn, Lupton, Loomis, Price, Smith, Strauss, Turner, Suzuki, Miyazaki, Muramatsu, Yamamoto, Endo, Ezaki, Ito, Kawaguchi, Sofuku, Taniike, Akutsu, Dojo, Kasumi, Matsuda, Imoto, Miwa, Suzuki, Takeshi, \& Yokota}]{ssp_1}
Miyazaki, S., Komiyama, Y., Kawanomoto, S., {et~al.} 2018, Publications of the Astronomical Society of Japan, 70, S1, \dodoi{10.1093/pasj/psx063}

\bibitem[{Mo {et~al.}(1998)Mo, Mao, \& White}]{mo98}
Mo, H.~J., Mao, S., \& White, S. D.~M. 1998, Monthly Notices of the Royal Astronomical Society, 295, 319, \dodoi{10.1046/j.1365-8711.1998.01227.x}

\bibitem[{Mo {et~al.}(2004)Mo, Yang, Van~den Bosch, \& Jing}]{mo04}
Mo, H.~J., Yang, X., Van~den Bosch, F.~C., \& Jing, Y.~P. 2004, Monthly Notices of the Royal Astronomical Society, 349, 205, \dodoi{10.1111/J.1365-2966.2004.07485.X}

\bibitem[{Mowla {et~al.}(2019)Mowla, Wel, Dokkum, \& Miller}]{mowla19}
Mowla, L., Wel, A. v.~d., Dokkum, P.~v., \& Miller, T.~B. 2019, The Astrophysical Journal, 872, L13, \dodoi{10.3847/2041-8213/ab0379}

\bibitem[{Nishizawa {et~al.}(2020)Nishizawa, Hsieh, Tanaka, \& Takata}]{photoz_hsc_pdr2}
Nishizawa, A.~J., Hsieh, B.-C., Tanaka, M., \& Takata, T. 2020, \dodoi{10.48550/arxiv.2003.01511}

\bibitem[{Park {et~al.}(2007)Park, Choi, Vogeley, Gott~III, \& Blanton}]{Park07}
Park, C., Choi, Y., Vogeley, M.~S., Gott~III, J.~R., \& Blanton, M.~R. 2007, The Astrophysical Journal, 658, 898, \dodoi{10.1086/511059}

\bibitem[{Paulino-Afonso {et~al.}(2019)Paulino-Afonso, Sobral, Darvish, Ribeiro, Van Der~Wel, Stott, Buitrago, Best, Stroe, \& Craig}]{Afonso19}
Paulino-Afonso, A., Sobral, D., Darvish, B., {et~al.} 2019, Astronomy {\&} Astrophysics, 630, A57, \dodoi{10.1051/0004-6361/201935137}

\bibitem[{Peebles(1969)}]{peebles_spin}
Peebles, P. J.~E. 1969, The Astrophysical Journal, 155, 393, \dodoi{10.1086/149876}

\bibitem[{Peng {et~al.}(2002)Peng, Ho, Impey, \& Rix}]{galfit}
Peng, C.~Y., Ho, L.~C., Impey, C.~D., \& Rix, H.-W. 2002, The Astronomical Journal, 124, 266, \dodoi{10.1086/340952}

\bibitem[{{Planck Collaboration} {et~al.}(2020){Planck Collaboration}, Aghanim, Akrami, Ashdown, Aumont, Baccigalupi, Ballardini, Banday, Barreiro, Bartolo, Basak, Battye, Benabed, Bernard, Bersanelli, Bielewicz, Bock, Bond, Borrill, Bouchet, Boulanger, Bucher, Burigana, Butler, Calabrese, Cardoso, Carron, Challinor, Chiang, Chluba, Colombo, Combet, Contreras, Crill, Cuttaia, de~Bernardis, de~Zotti, Delabrouille, Delouis, Di~Valentino, Diego, Dor{\'{e}}, Douspis, Ducout, Dupac, Dusini, Efstathiou, Elsner, En{\ss}lin, Eriksen, Fantaye, Farhang, Fergusson, Fernandez-Cobos, Finelli, Forastieri, Frailis, Fraisse, Franceschi, Frolov, Galeotta, Galli, Ganga, G{\'{e}}nova-Santos, Gerbino, Ghosh, Gonz{\'{a}}lez-Nuevo, G{\'{o}}rski, Gratton, Gruppuso, Gudmundsson, Hamann, Handley, Hansen, Herranz, Hildebrandt, Hivon, Huang, Jaffe, Jones, Karakci, Keih{\"{a}}nen, Keskitalo, Kiiveri, Kim, Kisner, Knox, Krachmalnicoff, Kunz, Kurki-Suonio, Lagache, Lamarre, Lasenby, Lattanzi, Lawrence, Le~Jeune, Lemos, Lesgourgues,
  Levrier, Lewis, Liguori, Lilje, Lilley, Lindholm, L{\'{o}}pez-Caniego, Lubin, Ma, Mac{\'{i}}as-P{\'{e}}rez, Maggio, Maino, Mandolesi, Mangilli, Marcos-Caballero, Maris, Martin, Martinelli, Mart{\'{i}}nez-Gonz{\'{a}}lez, Matarrese, Mauri, McEwen, Meinhold, Melchiorri, Mennella, Migliaccio, Millea, Mitra, Miville-Desch{\^{e}}nes, Molinari, Montier, Morgante, Moss, Natoli, N{\o}rgaard-Nielsen, Pagano, Paoletti, Partridge, Patanchon, Peiris, Perrotta, Pettorino, Piacentini, Polastri, Polenta, Puget, Rachen, Reinecke, Remazeilles, Renzi, Rocha, Rosset, Roudier, Rubi{\~{n}}o-Mart{\'{i}}n, Ruiz-Granados, Salvati, Sandri, Savelainen, Scott, Shellard, Sirignano, Sirri, Spencer, Sunyaev, Suur-Uski, Tauber, Tavagnacco, Tenti, Toffolatti, Tomasi, Trombetti, Valenziano, Valiviita, Van~Tent, Vibert, Vielva, Villa, Vittorio, Wandelt, Wehus, White, \& Zacchei}]{planck18}
{Planck Collaboration}, Aghanim, N., Akrami, Y., {et~al.} 2020, Astronomy {\&} Astrophysics, 641, A6, \dodoi{10.1051/0004-6361/201833910}

\bibitem[{Pozzetti {et~al.}(2010)Pozzetti, Bolzonella, Zucca, Zamorani, Lilly, Renzini, Moresco, Mignoli, Cassata, Tasca, Lamareille, Maier, Meneux, Halliday, Oesch, Vergani, Caputi, Kova{\v{c}}, Cimatti, Cucciati, Iovino, Peng, Carollo, Contini, Kneib, Le~F{\'{e}}vre, Mainieri, Scodeggio, Bardelli, Bongiorno, Coppa, De~La~Torre, De~Ravel, Franzetti, Garilli, Kampczyk, Knobel, Le~Borgne, Le~Brun, Pell{\`{o}}, Perez~Montero, Ricciardelli, Silverman, Tanaka, Tresse, Abbas, Bottini, Cappi, Guzzo, Koekemoer, Leauthaud, MacCagni, Marinoni, McCracken, Memeo, Porciani, Scaramella, Scarlata, \& Scoville}]{Pozzetti10}
Pozzetti, L., Bolzonella, M., Zucca, E., {et~al.} 2010, Astronomy {\&} Astrophysics, 523, A13, \dodoi{10.1051/0004-6361/200913020}

\bibitem[{Price-Whelan {et~al.}(2018)Price-Whelan, Sip{\H{o}}cz, G{\"{u}}nther, Lim, Crawford, Conseil, Shupe, Craig, Dencheva, Ginsburg, VanderPlas, Bradley, P{\'{e}}rez-Su{\'{a}}rez, de~Val-Borro, Aldcroft, Cruz, Robitaille, Tollerud, Ardelean, Babej, Bach, Bachetti, Bakanov, Bamford, Barentsen, Barmby, Baumbach, Berry, Biscani, Boquien, Bostroem, Bouma, Brammer, Bray, Breytenbach, Buddelmeijer, Burke, Calderone, Rodr{\'{i}}guez, Cara, Cardoso, Cheedella, Copin, Corrales, Crichton, D’Avella, Deil, Depagne, Dietrich, Donath, Droettboom, Earl, Erben, Fabbro, Ferreira, Finethy, Fox, Garrison, Gibbons, Goldstein, Gommers, Greco, Greenfield, Groener, Grollier, Hagen, Hirst, Homeier, Horton, Hosseinzadeh, Hu, Hunkeler, Ivezi{\'{c}}, Jain, Jenness, Kanarek, Kendrew, Kern, Kerzendorf, Khvalko, King, Kirkby, Kulkarni, Kumar, Lee, Lenz, Littlefair, Ma, Macleod, Mastropietro, McCully, Montagnac, Morris, Mueller, Mumford, Muna, Murphy, Nelson, Nguyen, Ninan, N{\"{o}}the, Ogaz, Oh, Parejko, Parley, Pascual, Patil,
  Patil, Plunkett, Prochaska, Rastogi, Janga, Sabater, Sakurikar, Seifert, Sherbert, Sherwood-Taylor, Shih, Sick, Silbiger, Singanamalla, Singer, Sladen, Sooley, Sornarajah, Streicher, Teuben, Thomas, Tremblay, Turner, Terr{\'{o}}n, Kerkwijk, de~la Vega, Watkins, Weaver, Whitmore, Woillez, \& Zabalza}]{astropy_2}
Price-Whelan, A.~M., Sip{\H{o}}cz, B.~M., G{\"{u}}nther, H.~M., {et~al.} 2018, The Astronomical Journal, 156, 123, \dodoi{10.3847/1538-3881/aabc4f}

\bibitem[{Racca {et~al.}(2016)Racca, Laureijs, Stagnaro, Salvignol, Lorenzo~Alvarez, Saavedra~Criado, Gaspar~Venancio, Short, Strada, B{\"{o}}nke, Colombo, Calvi, Maiorano, Piersanti, Prezelus, Rosato, Pinel, Rozemeijer, Lesna, Musi, Sias, Anselmi, Cazaubiel, Vaillon, Mellier, Amiaux, Berth{\'{e}}, Sauvage, Azzollini, Cropper, Pottinger, Jahnke, Ealet, Maciaszek, Pasian, Zacchei, Scaramella, Hoar, Kohley, Vavrek, Rudolph, \& Schmidt}]{euclid}
Racca, G.~D., Laureijs, R., Stagnaro, L., {et~al.} 2016, in Proceedings of SPIE Astronomical Telescopes + Instrumentation, ed. H.~A. MacEwen, G.~G. Fazio, M.~Lystrup, N.~Batalha, N.~Siegler, \& E.~C. Tong, 99040O, \dodoi{10.1117/12.2230762}

\bibitem[{Robitaille {et~al.}(2013)Robitaille, Tollerud, Greenfield, Droettboom, Bray, Aldcroft, Davis, Ginsburg, Price-Whelan, Kerzendorf, Conley, Crighton, Barbary, Muna, Ferguson, Grollier, Parikh, Nair, G{\"{u}}nther, Deil, Woillez, Conseil, Kramer, Turner, Singer, Fox, Weaver, Zabalza, Edwards, Azalee~Bostroem, Burke, Casey, Crawford, Dencheva, Ely, Jenness, Labrie, Lim, Pierfederici, Pontzen, Ptak, Refsdal, Servillat, \& Streicher}]{astropy_1}
Robitaille, T.~P., Tollerud, E.~J., Greenfield, P., {et~al.} 2013, Astronomy {\&} Astrophysics, 558, A33, \dodoi{10.1051/0004-6361/201322068}

\bibitem[{Shankar {et~al.}(2013)Shankar, Marulli, Bernardi, Mei, Meert, \& Vikram}]{shankar13}
Shankar, F., Marulli, F., Bernardi, M., {et~al.} 2013, Monthly Notices of the Royal Astronomical Society, 428, 109, \dodoi{10.1093/mnras/sts001}

\bibitem[{Shen {et~al.}(2003)Shen, Mo, White, Blanton, Kauffmann, Voges, Brinkmann, \& Csabai}]{shen03}
Shen, S., Mo, H.~J., White, S. D.~M., {et~al.} 2003, Monthly Notices of the Royal Astronomical Society, 343, 978, \dodoi{10.1046/j.1365-8711.2003.06740.x}

\bibitem[{Shimakawa {et~al.}(2021{\natexlab{a}})Shimakawa, Tanaka, Toshikage, \& Tanaka}]{hsc_morph_den}
Shimakawa, R., Tanaka, T.~S., Toshikage, S., \& Tanaka, M. 2021{\natexlab{a}}, Publications of the Astronomical Society of Japan, 73, 1575, \dodoi{10.1093/pasj/psab097}

\bibitem[{Shimakawa {et~al.}(2021{\natexlab{b}})Shimakawa, Higuchi, Shirasaki, Tanaka, Lin, Hayashi, Momose, Lee, Kusakabe, Kodama, \& Yamamoto}]{hsc_den}
Shimakawa, R., Higuchi, Y., Shirasaki, M., {et~al.} 2021{\natexlab{b}}, Monthly Notices of the Royal Astronomical Society, 503, 3896, \dodoi{10.1093/mnras/stab713}

\bibitem[{Simonyan \& Zisserman(2014)}]{vgg}
Simonyan, K., \& Zisserman, A. 2014, 3rd International Conference on Learning Representations, ICLR 2015 - Conference Track Proceedings.
\newblock \url{http://arxiv.org/abs/1409.1556}

\bibitem[{Siudek {et~al.}(2022)Siudek, Ma{\l}ek, Pollo, Iovino, Haines, Bolzonella, Cucciati, Gargiulo, Granett, Krywult, Moutard, \& Scodeggio}]{Siudek22}
Siudek, M., Ma{\l}ek, K., Pollo, A., {et~al.} 2022, A{\&}A, 666, 131, \dodoi{10.1051/0004-6361/202243613}

\bibitem[{Somerville {et~al.}(2018)Somerville, Behroozi, Pandya, Dekel, Faber, Fontana, Koekemoer, Koo, P{\'{e}}rez-Gonz{\'{a}}lez, Primack, Santini, Taylor, \& van~der Wel}]{somerville18}
Somerville, R.~S., Behroozi, P., Pandya, V., {et~al.} 2018, Monthly Notices of the Royal Astronomical Society, 473, 2714, \dodoi{10.1093/mnras/stx2040}

\bibitem[{Spearman(1904)}]{spearman_original}
Spearman, C. 1904, American Journal of Psychology, 15, 72

\bibitem[{Spergel {et~al.}(2013)Spergel, Gehrels, Breckinridge, Donahue, Dressler, Gaudi, Greene, Guyon, Hirata, Kalirai, Kasdin, Moos, Perlmutter, Postman, Rauscher, Rhodes, Wang, Weinberg, Centrella, Traub, Baltay, Colbert, Bennett, Kiessling, Macintosh, Merten, Mortonson, Penny, Rozo, Savransky, Stapelfeldt, Zu, Baker, Cheng, Content, Dooley, Foote, Goullioud, Grady, Jackson, Kruk, Levine, Melton, Peddie, Ruffa, \& Shaklan}]{ngrst}
Spergel, D., Gehrels, N., Breckinridge, J., {et~al.} 2013.
\newblock \url{http://arxiv.org/abs/1305.5422}

\bibitem[{Takada {et~al.}(2014)Takada, Ellis, Chiba, Greene, Aihara, Arimoto, Bundy, Cohen, Dor{\'{e}}, Graves, Gunn, Heckman, Hirata, Ho, Kneib, F{\`{e}}vre, Lin, More, Murayama, Nagao, Ouchi, Seiffert, Silverman, Sodr{\'{e}}, Spergel, Strauss, Sugai, Suto, Takami, \& Wyse}]{pfs}
Takada, M., Ellis, R.~S., Chiba, M., {et~al.} 2014, Publications of the Astronomical Society of Japan, 66, R1, \dodoi{10.1093/pasj/pst019}

\bibitem[{Tanaka(2015)}]{mizuki}
Tanaka, M. 2015, The Astrophysical Journal, 801, 20, \dodoi{10.1088/0004-637X/801/1/20}

\bibitem[{Tanaka {et~al.}(2018)Tanaka, Coupon, Hsieh, Mineo, Nishizawa, Speagle, Furusawa, Miyazaki, \& Murayama}]{hsc_photoz_pdr1}
Tanaka, M., Coupon, J., Hsieh, B.~C., {et~al.} 2018, Publications of the Astronomical Society of Japan, 70, 9, \dodoi{10.1093/PASJ/PSX077}

\bibitem[{Tasca {et~al.}(2009)Tasca, Kneib, Iovino, Le~F{\`{e}}vre, Kova{\v{c}}, Bolzonella, Lilly, Abraham, Cassata, Cucciati, Guzzo, Tresse, Zamorani, Capak, Garilli, Scodeggio, Sheth, Zucca, Carollo, Contini, Mainieri, Renzini, Bardelli, Bongiorno, Caputi, Coppa, De~La~Torre, De~Ravel, Franzetti, Kampczyk, Knobel, Koekemoer, Lamareille, Le~Borgne, Le~Brun, Maier, Mignoli, Pello, Peng, Perez~Montero, Ricciardelli, Silverman, Vergani, Tanaka, Abbas, Bottini, Cappi, Cimatti, Ilbert, Leauthaud, MacCagni, Marinoni, McCracken, Memeo, Meneux, Oesch, Porciani, Pozzetti, Scaramella, \& Scarlata}]{Tasca09}
Tasca, L.~A., Kneib, J.~P., Iovino, A., {et~al.} 2009, Astronomy {\&} Astrophysics, 503, 379, \dodoi{10.1051/0004-6361/200912213}

\bibitem[{van~der Wel(2008)}]{vdw08}
van~der Wel, A. 2008, The Astrophysical Journal, 675, L13, \dodoi{10.1086/529432/FULLTEXT/}

\bibitem[{Van Der~Wel {et~al.}(2014)Van Der~Wel, Franx, Van~Dokkum, Skelton, Momcheva, Whitaker, Brammer, Bell, Rix, Wuyts, Ferguson, Holden, Barro, Koekemoer, Chang, McGrath, H{\"{a}}ussler, Dekel, Behroozi, Fumagalli, Leja, Lundgren, Maseda, Nelson, Wake, Patel, Labb{\'{e}}, Faber, Grogin, \& Kocevski}]{vdw_14}
Van Der~Wel, A., Franx, M., Van~Dokkum, P.~G., {et~al.} 2014, The Astrophysical Journal, 788, 28, \dodoi{10.1088/0004-637X/788/1/28}

\bibitem[{van Dokkum {et~al.}(2014)van Dokkum, Bezanson, van~der Wel, Nelson, Momcheva, Skelton, Whitaker, Brammer, Conroy, Schreiber, Fumagalli, Kriek, Labb{\'{e}}, Leja, Marchesini, Muzzin, Oesch, \& Wuyts}]{dokkum_14}
van Dokkum, P.~G., Bezanson, R., van~der Wel, A., {et~al.} 2014, The Astrophysical Journal, 791, 45, \dodoi{10.1088/0004-637X/791/1/45}

\bibitem[{Virtanen {et~al.}(2020)Virtanen, Gommers, Oliphant, Haberland, Reddy, Cournapeau, Burovski, Peterson, Weckesser, Bright, van~der Walt, Brett, Wilson, Millman, Mayorov, Nelson, Jones, Kern, Larson, Carey, Polat, Feng, Moore, VanderPlas, Laxalde, Perktold, Cimrman, Henriksen, Quintero, Harris, Archibald, Ribeiro, Pedregosa, van Mulbregt, Vijaykumar, Bardelli, Rothberg, Hilboll, Kloeckner, Scopatz, Lee, Rokem, Woods, Fulton, Masson, H{\"{a}}ggstr{\"{o}}m, Fitzgerald, Nicholson, Hagen, Pasechnik, Olivetti, Martin, Wieser, Silva, Lenders, Wilhelm, Young, Price, Ingold, Allen, Lee, Audren, Probst, Dietrich, Silterra, Webber, Slavi{\v{c}}, Nothman, Buchner, Kulick, Sch{\"{o}}nberger, de~Miranda~Cardoso, Reimer, Harrington, Rodr{\'{i}}guez, Nunez-Iglesias, Kuczynski, Tritz, Thoma, Newville, K{\"{u}}mmerer, Bolingbroke, Tartre, Pak, Smith, Nowaczyk, Shebanov, Pavlyk, Brodtkorb, Lee, McGibbon, Feldbauer, Lewis, Tygier, Sievert, Vigna, Peterson, More, Pudlik, Oshima, Pingel, Robitaille, Spura, Jones, Cera,
  Leslie, Zito, Krauss, Upadhyay, Halchenko, \& V{\'{a}}zquez-Baeza}]{Virtanen2020SciPyPython}
Virtanen, P., Gommers, R., Oliphant, T.~E., {et~al.} 2020, Nature Methods, 17, 261, \dodoi{10.1038/s41592-019-0686-2}

\bibitem[{Wechsler {et~al.}(2002)Wechsler, Bullock, Primack, Kravtsov, \& Dekel}]{wechsler02}
Wechsler, R.~H., Bullock, J.~S., Primack, J.~R., Kravtsov, A.~V., \& Dekel, A. 2002, The Astrophysical Journal, 568, 52, \dodoi{10.1086/338765}

\bibitem[{Wechsler \& Tinker(2018)}]{wechsler_tinker}
Wechsler, R.~H., \& Tinker, J.~L. 2018, Annual Review of Astronomy and Astrophysics, 56, 435, \dodoi{10.1146/annurev-astro-081817-051756}

\bibitem[{Wechsler {et~al.}(2006)Wechsler, Zentner, Bullock, Kravtsov, \& Allgood}]{wechsler06}
Wechsler, R.~H., Zentner, A.~R., Bullock, J.~S., Kravtsov, A.~V., \& Allgood, B. 2006, The Astrophysical Journal, 652, 71, \dodoi{10.1086/507120}

\bibitem[{Weigel {et~al.}(2016)Weigel, Schawinski, \& Bruderer}]{Weigel16}
Weigel, A.~K., Schawinski, K., \& Bruderer, C. 2016, Monthly Notices of the Royal Astronomical Society, 459, 2150, \dodoi{10.1093/MNRAS/STW756}

\bibitem[{Whitaker {et~al.}(2011)Whitaker, Labb{\'{e}}, Van~Dokkum, Brammer, Kriek, Marchesini, Quadri, Franx, Muzzin, Williams, Bezanson, Illingworth, Lee, Lundgren, Nelson, Rudnick, Tal, \& Wake}]{newfirm}
Whitaker, K.~E., Labb{\'{e}}, I., Van~Dokkum, P.~G., {et~al.} 2011, The Astrophysical Journal, 735, 86, \dodoi{10.1088/0004-637X/735/2/86}

\bibitem[{Yang {et~al.}(2023)Yang, Gao, Frenk, Grand, Guo, Liao, \& Shao}]{yang21}
Yang, H., Gao, L., Frenk, C.~S., {et~al.} 2023, Monthly Notices of the Royal Astronomical Society, 518, 5253, \dodoi{10.1093/mnras/stac3335}

\bibitem[{Yang {et~al.}(2013)Yang, Mo, van~den Bosch, Bonaca, Li, Lu, Lu, \& Lu}]{yang13}
Yang, X., Mo, H.~J., van~den Bosch, F.~C., {et~al.} 2013, The Astrophysical Journal, 770, 115, \dodoi{10.1088/0004-637X/770/2/115}

\bibitem[{Yoon {et~al.}(2017)Yoon, Im, \& Kim}]{Yoon17}
Yoon, Y., Im, M., \& Kim, J.-W. 2017, The Astrophysical Journal, 834, 73, \dodoi{10.3847/1538-4357/834/1/73}

\end{thebibliography}
